\documentclass[pdflatex,sn-mathphys-num]{sn-jnl}% 
\usepackage{graphicx}%
\usepackage{multirow}%
\usepackage{amsmath,amssymb,amsfonts}%
\usepackage{amsthm}%
\usepackage{mathrsfs}%
\usepackage[title]{appendix}%
\usepackage{xcolor}%
\usepackage{textcomp}%
\usepackage{manyfoot}%
\usepackage{booktabs}%
\usepackage{algorithm}%
\usepackage{algorithmicx}%
\usepackage{algpseudocode}%
\usepackage{listings}%
\usepackage{graphicx}      
\usepackage{subcaption}     
\usepackage{booktabs}       
\usepackage{xcolor}
\usepackage{geometry}
\theoremstyle{thmstyleone}%

\theoremstyle{thmstyletwo}%

\theoremstyle{thmstylethree}%

\begin{document}

\title{A simple experiment for observing clustering and dynamics of coalescing particles in air turbulence}

\author[1]{\fnm{Linli} \sur{Fu}}

\author[1]{\fnm{Jun} \sur{Feng}}

\author[1]{\fnm{Yating} \sur{Chen}}

\author[1]{\fnm{Fanxi} \sur{Gong}}

\author[1]{\fnm{Xiaohui} \sur{Meng}}

\author*[1]{\fnm{Ewe-Wei} \sur{Saw}}\email{ewsaw3@gmail.com}

\affil[1]{\orgdiv{School of Atmospheric Sciences}, \orgname{Sun Yat-sen University}, \city{Zhuhai}, \postcode{51900}, 
\state{Guangdong}, \country{China}}

\abstract{A novel experimental platform is developed with a focus on simplicity to investigate the dynamics of inertial particles (micro-droplets) in air turbulence. The ultimate goal of the effort is the observation of  particle collision and coalescence in turbulent flows, while the immediate focus is on the influence of collision-coalescence process on other dynamical metrics such as the radial distribution function (RDF) and relative velocity statistics. The main observation tool is a three-dimensional Lagrangian particle tracking (LPT) system, designed to achieve high-resolution measurements at high particle number density and at deep sub-Kolmogorov scales in turbulent flow. The system leverages the simplicity of LED illumination combined with high-speed spinning-disk atomizers, enabling reliable tracking of particles with diameters of approximately $10~\mu\mathrm{m}$ and larger under controlled turbulent conditions. A minimum resolvable particle separation of $r/\eta \approx 0.1$ is achieved, thereby extending the accessible measurement regime for small inertial particles. A central contribution of this work lies in the systematic identification and mitigation of three dominant sources of spurious particles: false stereo-matching induced spurious particle (FMIS), interpolation induced spurious particles (IIS), and threshold induced fragmentation (TIF). An angle-based geometric filtering criterion is introduced, through which the artifact due to FMIS on RDF are strongly suppressed. Together, these procedures establish a validated workflow for obtaining reliable small-scale statistics. Using this framework, the radial distribution function (RDF) and a normalized pseudo-collision rate are measured at near-contact separations for particles with Stokes numbers $St \approx 0.2$--$1.0$. A clear enhancement of sub-Kolmogorov clustering with increasing Stokes number is observed, and consistent near-contact statistics are ensured through the proposed filtering strategy. The present study therefore extends the operational limits of LPT and provides a reliable experimental methodology for investigating inertial-particle dynamics at previously inaccessible spatial scales.}

\keywords{Inertial particle clustering; Lagrangian particle tracking; Spurious particle elimination; Radial distribution function; Particle-pair statistics}

\maketitle

\section{Introduction}\label{sec1}

The dynamics and spatial distribution of inertial particles in turbulence are important topics in multiphase flow research. In many natural and engineering systems, for example cloud droplet growth \cite{Shaw2003}, atmospheric aerosol deposition \citep{Wang1993}, and spray combustion \citep{Elghobashi1994}—particles display highly non-uniform spatial distributions manifested as clustering. This clustering primarily results from the finite response time of the particles, which prevents them from faithfully following rapid fluid motions; consequently, particles are expelled from vortical regions and preferentially accumulate in strain-dominated zones \cite{Maxey1987,Squires1991}. This preferential concentration produces interstitial multiscale structures that substantially modify momentum, heat, and mass transport \cite{Eaton1994}. Therefore, elucidating clustering mechanisms is both theoretically important and practically necessary for accurate modeling and optimization of multiphase systems.

Several dimensionless parameters influence the clustering of inertial particles, among which the Stokes number (\textit{St}) is predominant. St quantifies the ratio of the particle response time to a characteristic turbulent timescale; near \textit{St} $\mathrm{\approx}$ 1 particle–flow interactions peak, producing maximal clustering \cite{Bec2007}. In this regime, particle distributions exhibit strong intermittency and multifractal statistics that reflect complex multiscale coupling with local turbulent structures \cite{Bec2003,Biferale2004}.

Despite advances in theory and DNS, experimental investigations of inertial particle clustering at very small scales remain relatively scarce. DNS provides valuable mechanistic detail, but typically relies on idealized assumptions and faces prohibitive cost at high Reynolds numbers or for polydisperse systems. Laboratory experiments complement simulations by capturing realistic flow–particle coupling, optical non-idealities, and realistic turbulence \cite{Maas1993,Mali1993}; they deliver repeatable datasets for validating particle dynamics metrics such as radial distribution functions \cite{Saw2008, Monchaux2010}, and Lagrangian statistics \cite{Xu2007}. Thus, careful experiments are indispensable for inspiring and guiding theories and models.

However, experimental endeavors face significant challenges: in many cases, particles of practical interest are much smaller than the characteristic turbulent scales, so high-resolution observation strains conventional optical techniques \cite{Saw2014}. Imaging noise, particle occlusion, and correspondence ambiguities can yield spurious detections and false trajectories that bias particle statistics \cite{Warhaft2002,Bodenschatz2010}. These issues are exacerbated at high seeding densities. or under limited background illumination.

Advanced reconstruction methods have mitigated many multi-camera limitations: techniques such as Shake-the-Box (\textsc{STB}) \cite{Schanz2016} and volumetric matching \cite{Wieneke2012} increase matching density and temporal continuity, enabling long trajectories even at elevated seeding. Nevertheless, particle-image overlap, parallax distortion, and near-degenerate geometric configurations still yield ambiguous correspondences at high concentrations, increasing false-match rates and corrupting statistics \cite{Tan2020,Dabiri2019}. Therefore, suppressing spurious detections and removing false trajectories is essential to improve experimental reliability.

Here we present an experimental platform that combines LED illumination, spinning-disks, and a high-speed Lagrangian particle tracking (LPT) system, specifically designed to observe particle dynamics in turbulence while minimizing spurious artifacts. Turbulence is produced by a pair of counter rotating disks in a closed acrylic chamber. A multi-camera high-speed imaging system together with customized image-processing and matching algorithms enables 3D trajectory reconstruction at high spatial and temporal resolution. Building on established LPT techniques, we introduce a geometric filter based on the
orientation angle of particle-pair separation vectors; this angle-based criterion suppresses false pairings arising from projection degeneracy, thereby improving trajectory fidelity of particle statistics.

The subsequent sections describe the experimental apparatus, including the design of the turbulence-box, the configuration of the spinning-disk, and the calibration of the system (Sec.~\ref{sec2}). Section~\ref{sec3} addresses typical tracking errors and associated filtering strategies, followed by results pertaining to particle pseudo-collision (to be defined) and clustering statistics under various conditions in Section \ref{sec4}. 

\section{Methodology}\label{sec2}
\subsection{Experimental Setup}\label{subsec2.1}

\begin{figure}[htbp]
	\centering
	\begin{subfigure}[c]{0.48\textwidth} 
		\includegraphics[width=\textwidth]{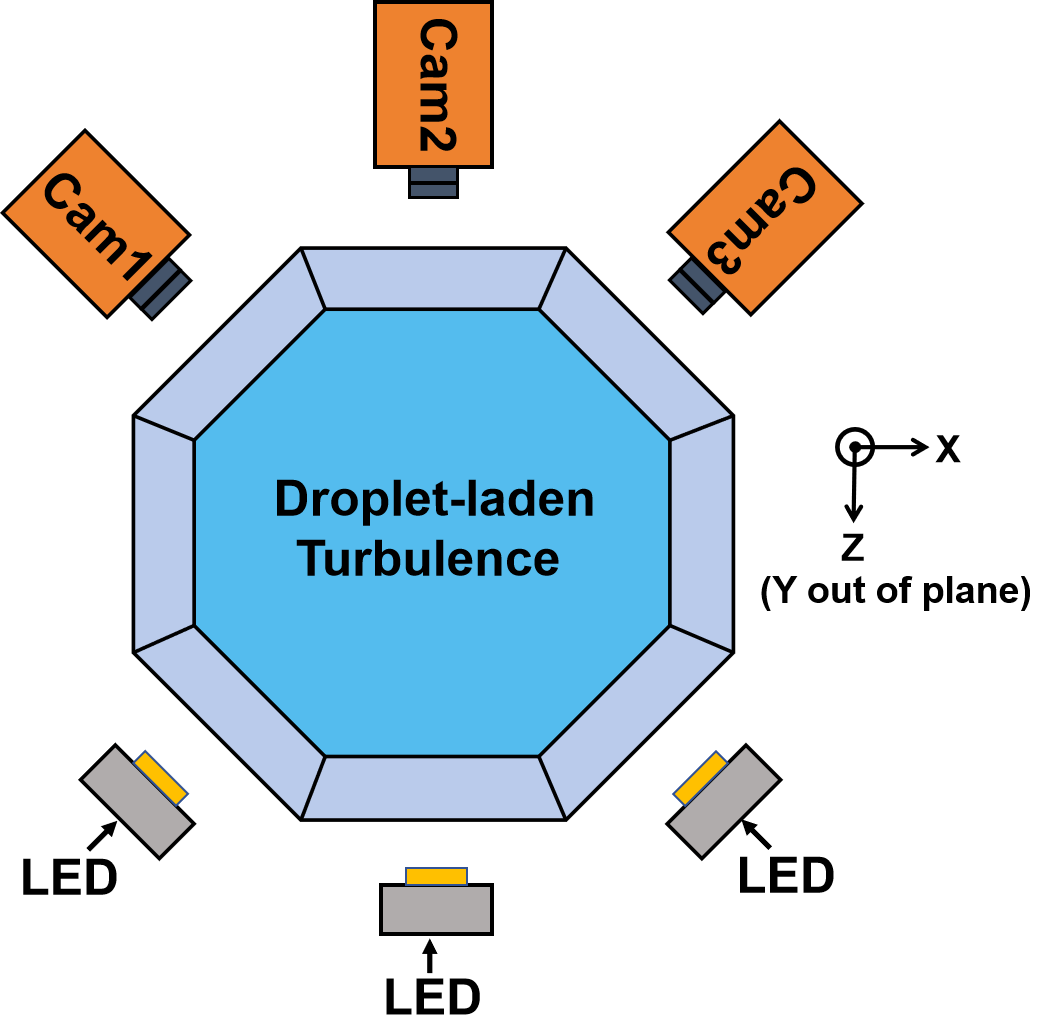} 
		\caption{Top view}
		\label{fig:ptv_top}
	\end{subfigure}
	\hspace{0.02\textwidth} 
	\begin{subfigure}[c]{0.48\textwidth}  
		\raisebox{0.3em}{\includegraphics[width=\textwidth]{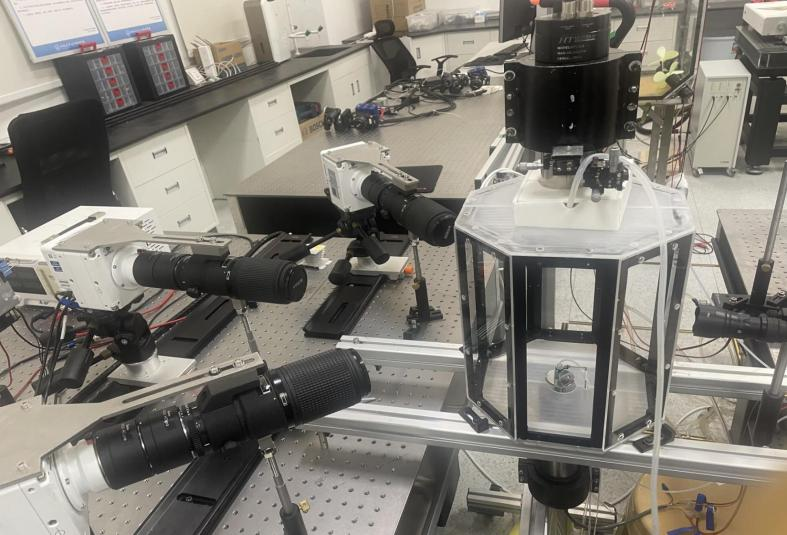}}
		\caption{Side view}
		\label{fig:ptv_side}
	\end{subfigure}
	
    \caption{Schematic and picture of the LPT experimental system. Panel (a) illustrates the optical arrangement, including the octagonal turbulence chamber (height 300 mm, flat-to-flat distance 250 mm). Three high-speed cameras are arranged around the chamber at approximately 45° intervals for stereoscopic particle tracking. Illumination is provided by high-power LED arrays positioned opposite to each camera and aligned with the optical axis. Panel (b) shows the three cameras mounted on adjustable mechanical stages for precise alignment, while the turbulence chamber is fixed between two optical tables.
    }
\label{fig:3D_LPT_setup}
\end{figure}

To investigate the dynamical behavior of small inertial particles in turbulent flows, a laboratory-scale multiphase turbulence facility was developed. The core of the apparatus is a turbulence chamber constructed from transparent acrylic panels with an octagonal-prism geometry (height $300\,\mathrm{mm}$, width $250\,\mathrm{mm}$). This configuration provides multiple optical access paths and enables the generation of a flow field that is moderately homogeneous and isotropic within the central region. A stainless-steel exterior frame reinforces the acrylic structure and minimizes deformation under operation. The top and bottom plates incorporate mountings for the particle-generation units while mechanically isolating the driver motors from the chamber walls to suppress vibration-induced imaging artefacts. The overall configuration is illustrated in Fig.~\ref{fig:3D_LPT_setup}.

A three-camera Lagrangian Particle Tracking (LPT) system was implemented to capture the three-dimensional trajectories of inertial droplets. The system consists of  three high-speed digital cameras (Phantom VEO 640L), high-brightness LED  volumetric illumination (Luminus SFT-40-W) and telephoto macro lenses. Cameras are arranged in a horizontal plane (X–Z), focused on the chamber center, with Camera 1 and 3 symmetrically positioned around Camera 2 at approximately 45° separation. All cameras are aligned normal to the acrylic windows to minimize refractive distortion. 

 Lenses used are Nikon AF Micro-Nikkor 200 mm lenses extended with teleconverters. The effective physical resolution at the view volume is $4~\mu\mathrm{m}/\mathrm{pixel}$, yielding a view volume of approximately 3 × 3 × 2 mm³. The Kolmogorov time scale in the facility satisfies \(\tau_\eta \sim
10^{-4}~\mathrm{s}\); accordingly, the imaging frequency was set to 10k~fps and exposure time to $1~\mu s$ to resolve rapid small-scale particle motion. High-power LED arrays were selected instead of laser illumination to eliminate speckle noise and yield more uniform grayscale gradients, significantly improving particle-edge localization and sub-pixel centroid fitting. Camera calibration was performed using Tsai’s method  ~\cite{Tsai1987,Ouellette2006,Xu2007} with a precision 3D calibration target. Images are downloaded from the cameras via 10-Gigabit Ethernet links.

\subsection{Spinning Disk}\label{subsec2.2}
Small droplets, playing the role of small inertial particles, were generated using a spinning-disk atomization mechanism \citep{Philipson1973}. Each unit consists of a high-speed motor, a controlled liquid-supply system, and a horizontally mounted stainless-steel disk with a diameter of $4\,\mathrm{cm}$. Liquid delivered to the disk center via a syringe needle spreads over the rotating disk and is flung to the edge by centrifugal force; there it breaks up into droplets via Rayleigh-Taylor instability. Previous studies have shown that such a setup generates a characteristic droplet size distribution featuring both primary and smaller satellite droplets, leading to a bimodal size spectrum \cite{Ahmed2012,Sahoo2021,Chen2024}. 

\begin{figure}[htbp]
	\centering
	\begin{subfigure}[b]{0.48\textwidth}
		\includegraphics[width=\textwidth]{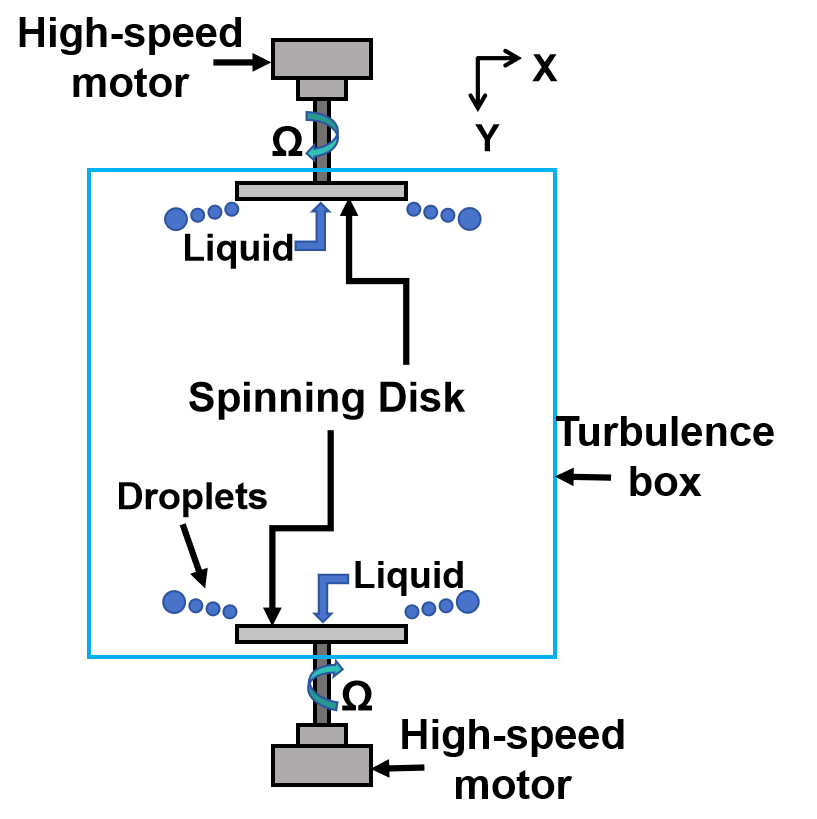}
		\caption{Principle of particle generation}
		\label{fig:gen_principle}
	\end{subfigure}
	\hfill
	\begin{subfigure}[b]{0.48\textwidth}
		\includegraphics[width=\textwidth,trim=0.10in 1.00in 0.07in 0.00in,clip]{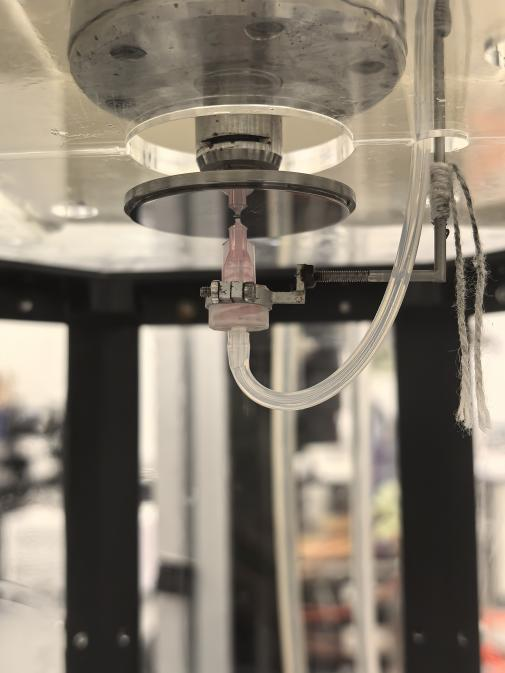}
		\caption{Experimental implementation}
		\label{fig:gen_experiment}
	\end{subfigure}
	\caption{Particle generation system based on spinning disk. Panel (a) two counter-rotating disks (diameter $4\,\mathrm{cm}$) are positioned at the upper and lower boundaries of the turbulence chamber. Liquid is supplied to both and is atomized into fine droplets by the strong centrifugal force produced by high-speed rotation. The droplets are subsequently mixed into the turbulence chamber. Panel (b) the disks are mounted coaxially and driven by high-speed motors. Droplet sizes are controlled by the speed of the spinning disks.}
	\label{fig:particle_generation}
\end{figure}

To reduce injection anisotropy, two counter-rotating disks are positioned at the upper and lower boundaries of the chamber, simultaneously supplying droplets toward the central mixing region. The motors are driven independently and adjustments of disk speeds allows for fine control of droplet size distribution. The configuration of the particle-generation system is shown in Figs.~\ref{fig:particle_generation}. The working fluid is a mixture of distilled water and 3 \% nonionic surfactant (X100), with a resulting density of $\sim 1000~\mathrm{kg/m^3}$. This ensures droplets remain sufficiently heavy relative to air, validating the use of linear Stokes drag for particle–fluid coupling \cite{Maxey1983}.

Droplet diameters were measured under the same experimental conditions as the turbulence measurements, using short-exposure images to minimize motion blur. This approach ensures that the measured size distribution accurately reflects the droplets within the flow (Sec.~\ref{subsec2.3}). In addition to droplet generation, the rotating disks also induce a strong shear-dominant airflow, which serves as the primary source of turbulence within the measurement region.

\subsection{Particle size characterization}\label{subsec2.3}

To obtain accurate particle diameters, a physically grounded intensity-thresholding scheme was applied. In this kind of imaging setup, which some would call "shadow imaging", droplets appear darker than the background in the raw images (12-bits/pixel), a lower intensity corresponds to a more focused particle. The mean background pixel intensity $\langle I_{\rm b} \rangle$ was estimated 
by analyzing the pixel intensities of particle free regions in each frame. In this section, $\langle \cdot \rangle$ denotes a spatial average over particle free pixels within a single frame. These regions were identified in a two-step procedure. First, a preliminary standard deviation $\sigma_{\mathrm{0}}$ was estimated directly from the raw intensity field of the image. 
At this stage, the statistics include both background illumination fluctuations and particle-induced intensity deficits. Pixels with intensity below $3 \sigma_{\mathrm{0}}$ were excluded to remove particle-containing pixels. 
Based on the remaining particle-free pixels, the background intensity fluctuations (standard deviation) $\sigma_b$ were subsequently evaluated. The resulting
$\sigma_b$, was estimated to be
$ 0.014$, $0.013$, and $0.012$ for experiment with spinning-disk speed of 42,000 rpm, 30,000 rpm, and 18,000 rpm, respectively. Here, the image intensity $I$ was normalized to the range $[0,1]$.
Incorporating these noise estimates, a particle detection threshold $T_{\rm 1}$
was defined as
\begin{equation} \label{eq:T_loose}
T_{\rm 1} = \langle I_{\rm b} \rangle - 6\sigma_b,
\end{equation}
such that any pixels with intensity lower that $T_{\rm 1}$ is considered to be a part of a particle.
{$T_{\rm 1}$ is considered a lenient threshold that may lead to a slight overestimation of droplet sizes.}

Another strict threshold could be define as
\begin{equation} \label{eq:T_strict}
T_{\rm 2} = \langle I_{\rm p} \rangle + 3\sigma_p,
\end{equation}
where $I_{\rm p}$ is the minimum intensity of the sharply focused droplets within each dataset.
 
The corresponding standard deviation of particle brightness, denoted as $\sigma_p$, was measured as $0.048$, $0.037$, and $0.038$ for experiments at 42,000 rpm, 30,000 rpm, and 18,000 rpm, respectively. 
This definition is considered to be a strict threshold with potential of underestimating droplet size. The variability $\sigma_p$ primarily reflects optical defocusing and the size-dependent scattering characteristics of the droplets.

\begin{table}[htbp]
\centering
\caption{Particle brightness distribution and threshold parameters at different disk speeds. $\langle I_{\rm b} \rangle$: mean background intensity (average gray level of background noise);
$I_{\rm p}$: mean intensity of the best-focused particles; $\sigma_b$: standard deviation of background noise; $\sigma_p$: standard deviation of particle intensity; $T_{\rm 1}$ and $T_{\rm 2}$: permissive and strict thresholds.}
\begin{tabular}{ccccccc}
\hline
Disk speed (rpm) & $\langle I_{\rm b} \rangle$ & $I_{\rm p}$ & $\sigma_b$ & $\sigma_p$ & $T_{\rm 1}$ & $T_{\rm 2}$ \\
\hline
42{,}000 & 0.72 & 0.31 & 0.014 & 0.048 & 0.64 & 0.45 \\
30{,}000 & 0.66 & 0.32 & 0.013 & 0.037 & 0.58 & 0.39 \\
18{,}000 & 0.89 & 0.52 & 0.012 & 0.038 & 0.82 & 0.64 \\
\hline
\end{tabular}
\label{tab:brightness_threshold}
\end{table}

The two threshold criteria defined above produce two sets particle-size estimates ($d_1,\ d_2$), where $d_{\rm 1}$ tends to overestimate true particle sizes and vici-versa for $d_{\rm 2}$. Also, the permissive threshold $T_{\rm 1}$ tends to include mildly broadened particle profiles, whereas the strict threshold $T_{\rm 2}$ detects only the most well-resolved particle images. As a result, the two thresholds offer complementary upper- and lower-bound measurements of the particle size. To obtain a best estimate the final particle diameter was defined as the arithmetic mean of the two measurements:
\begin{equation} \label{eq:D}
  d = \frac{d_{\rm 1} + d_{\rm 2}}{2}.
\end{equation}

By varying the rotational speed of the spinning disk atomizer (18,000 -- 42,000 rpm), droplet sizes ranging from $10~\mu\mathrm{m}$ to $80~\mu\mathrm{m}$ could be reliably generated. To systematically assess the droplet generation performance, a series of experiments were conducted. 

\begin{figure}[htbp]
	\centering
	\includegraphics[width=2.8in, height=2.0in]{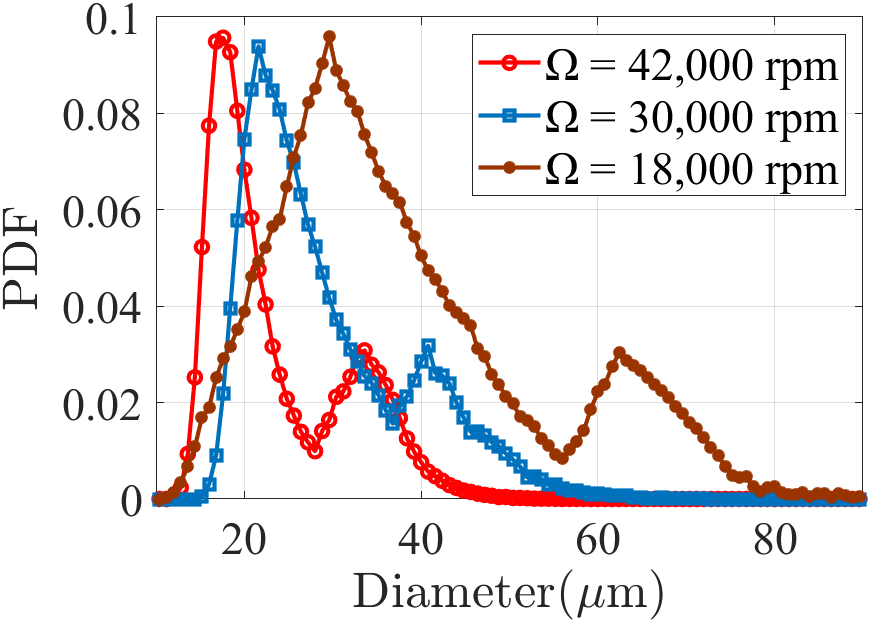}
	\caption{Particle size distribution. The x-axis represents the particle diameter ($\mu\mathrm{m}$), while the y-axis shows the probability density function (PDF) of the particle size. The red circular curve corresponds to 42,000 rpm, the blue rectangular curve to 30,000 rpm, and the brown star-shaped curve to 18,000 rpm.}
	\label{fig:particle_distribution}
\end{figure}

Fig.~\ref{fig:particle_distribution} illustrates the probability distributions of particle sizes under three different rotational speeds. Under all conditions, a bimodal distribution is observed, indicating the coexistence of two distinct particle populations. This trend is consistent with the mechanisms of primary droplet breakup and secondary satellite droplet formation in rotary atomization processes \cite{Hinze1955}. The number of larger particles is approximately one-fourth of the smaller particles, aligning well with earlier findings.

\begin{table}[h!]
\centering
\caption{Peak positions and widths of particle-size distributions under different disk rotation speeds.}
\begin{tabular}{cccc}
\toprule
Rotation speed (rpm) & Peak 1 ($\mu\pm\sigma$, $\mu$m) & Peak 2 ($\mu\pm\sigma$, $\mu$m) \\
\midrule
42,000  & $17.6 \pm 3.1$ & $33.6 \pm 4.1$ \\
30,000  & $21.6 \pm 4.8$ & $40.8 \pm 3.7$ \\
18,000  & $29.6 \pm 8.5$ & $62.4 \pm 5.4$ \\
\bottomrule
\end{tabular}
\label{tab:peak_values}
\begin{tablenotes}
\footnotesize
\item[$\mu$] mean particle diameter
\item[$\sigma$] standard deviation
\end{tablenotes}
\end{table}

At a rotational speed of 42,000 rpm, the particle-size distribution becomes narrow and well-defined, exhibiting two dominant peaks at $17.6 \pm 3.1~\mu\mathrm{m}$ and $33.6 \pm 4.1~\mu\mathrm{m}$. At 30,000 rpm, the distribution broadens, showing a primary peak at $21.6 \pm 4.8~\mu\mathrm{m}$ and a secondary peak at $40.8 \pm 3.7~\mu\mathrm{m}$. The overall size range also expands, with droplets exceeding $60~\mu\mathrm{m}$, indicating increased polydispersity at intermediate rotation speeds. At the lowest rotation speed of 18,000 rpm, the distribution becomes substantially broader and flatter, characterized by two peaks located at $29.6 \pm 8.5~\mu\mathrm{m}$ and $62.4 \pm 5.4~\mu\mathrm{m}$, with maximum droplet sizes approaching $70~\mu\mathrm{m}$. This behavior reflects enhanced instability in the droplet formation process under low rotational speeds, resulting in a markedly more dispersed size distribution.

\subsection{Turbulence characterization and particle dynamic parameters}\label{subsec2.4}
 The turbulent flow in the measurement region is characterized by its velocity fluctuations, Reynolds number, energy dissipation rate, and the associated Kolmogorov scales. The turbulence intensity is quantified by the root-mean-square velocity fluctuation $u_{\mathrm{rms}}$, from which the Reynolds number $Re$ is evaluated.
\begin{equation}\label{Re}
Re = \frac{u_{\mathrm{rms}}\, L}{\nu},
\end{equation}
where $\nu$ is the kinematic viscosity of air, and $L$ is the characteristic length scale of the flow. In the present experiments, $L = 0.1~\mathrm{m}$ is adopted based on empirical considerations (i.e., it is taken as half of the chamber width). Here, $Re$ is introduced as a nominal Reynolds number and is used solely for comparing different operating conditions rather than for formal theoretical analysis.

The turbulent velocity fluctuation \(u_{\mathrm{rms}}\) for each operating condition was obtained using all the tracks of the smaller particle group (Peak 1 in Table~\ref{tab:peak_values}). At the spinning-disk speed of 42{,}000~rpm, the result is $u_{\mathrm{rms}} = 0.550\,\mathrm{m\,s^{-1}} $, while at 30{,}000~rpm and 18{,}000~rpm, they are $u_{\mathrm{rms}} = 0.438\,\mathrm{m\,s^{-1}}$ and $u_{\mathrm{rms}} = 0.339\,\mathrm{m\,s^{-1}}$ respectively.

To characterize the small-scale properties of the turbulence that are relevant to particle dynamics, the energy dissipation rate $\varepsilon$ and the associated Kolmogorov scales are required. The local Kolmogorov time scale \(\tau_\eta\) and length scale $\eta$ are defined as \cite{Kolmogorov1941}:
\begin{equation} \label{eq:tau_eta}
	\tau_\eta = \left( \frac{\nu}{\varepsilon} \right)^{1/2},
\end{equation}
\begin{equation} \label{eq:eta}
	\eta = \left(\frac{\nu^3}{\varepsilon}\right)^{1/4},
\end{equation}

Once $\tau_\eta$ is determined, the Stokes number (\(St\)) is defined as the ratio of the particle dynamic response time (\(\tau_p\)) to the Kolmogorov time scale (\(\tau_\eta\)):
\begin{equation} \label{eq:Stokes}
	St = \frac{\tau_p}{\tau_\eta},
\end{equation}
\(\tau_p\) reflects the time required for the particle's velocity to adjust the changes in the fluid's velocity. For spherical particles with small diameters and mass densities much greater than the carrier fluid density, under low  particle Reynolds number (Stokes flow) conditio the response time can be estimated \cite{Maxey1983}:
\begin{equation} \label{eq:tau_p}
	\tau_p = \frac{1}{18 \nu} \left( \frac{\rho_d}{\rho} \right) d^2,
\end{equation}
where in this work, \(\rho_d = 1000~\mathrm{kg/m^3}\) is the particle mass density, 
\(\rho = 1.2~\mathrm{kg/m^3}\) is the fluid density and \(d\) is the particle diameter. 

In our experiments, the ambient temperature varied between 18~°C and 27~°C. The temperature dependence of the fluid density is negligible within this range: the variation in air density is less than 3\%, and the densities of water and solid particles remain essentially constant. Therefore, only the temperature-induced change in air kinematic viscosity is corrected. Taking $\nu_{20} = 1.48 \times 10^{-5}\,\mathrm{m^2/s}$ as the reference value at 20~°C, the corresponding viscosity variation is $\nu = (1.51 \pm 0.06) \times 10^{-5}\,\mathrm{m^2/s}$, which is  accounted in the evaluation of uncertainty in $\tau_p$ and $St$.

The turbulence energy dissipation rate $\varepsilon$ is estimated using two complementary approaches.The primary estimate of the turbulence energy dissipation rate was obtained from the second-order particle velocity structure function $S_2(r)$. Assuming locally homogeneous and isotropic turbulence, the relation $S_2(r)=\varepsilon r^2/(15\nu)$ holds for separations within the viscous dissipative scales, enabling the dissipation rate to be evaluated from the measured $S_2(r)$.

Reliable computation of $S_2(r)$ requires continuous and temporally resolved particle trajectories. In the present experiments, the 3D particle positions were recorded simultaneously from three high-speed cameras. Due to the rapid motion of droplets in the highly turbulent flow, some trajectories were intermittently lost or fragmented, which may introduce uncertainty and/or bias into the velocity estimates. To mitigate trajectory fragmentation, isolated particle detections were conservatively reconnected in a post-processing step based on short-time velocity consistency assumption. This procedure supplements the primary tracking results without introducing interpolated or synthetic particle positions (details to be presented in a separate publication). After trajectory reconnection, the particle velocities were computed, and the structure function $S_2(r)$ was evaluated across a range of separations. The resulting $S_2$–$r$ relation yielded a direct estimate of the dissipation rate $\varepsilon$ for each operating condition.

Separately, at small separations, the radial distribution function of inertial particles has been shown to follow an approximate power-law form for monodisperse suspensions, i.e., 
$g(r) \sim c_0 \left( r/\eta \right)^{-c_1}$,
where $c_1$ is sometimes called the clustering exponent and can be interpreted as a metric of small-scale clustering intensity. \cite{Saw2012,Bec2007}.
This behavior reflects enhanced particle accumulation at sub-Kolmogorov scales and arises from the balance between inward inertial drift and turbulent diffusion \cite{Chun2005}. It is established that $c_1$ depends on St, and  
in the weak-inertia and dilute-particle limit ($St \ll 1$), perturbative expansion analyses predict that $c_1$ scales approximately with $St^2$ (see e.g., \cite{Chun2005}). Based on this theoretical background, the RDF may be used to indirectly estimate the dissipation rate ($\epsilon$).  The definition and calculation of the RDF will be presented in the sequel. The dissipation rate can be estimated {by finding the correct value for $\epsilon$ such that the particle Stokes number calculated using Eq.~\ref{eq:Stokes} and Eq.~\ref{eq:tau_p} matches with the St-value that could be independently inferred from the measured clustering exponent ($c_1$) based on the known $c_1$-$St$ relation curated in Fig. 4 in \cite{Saw2012}. 

Together, the two approaches provide complementary estimates of the turbulence energy dissipation rate in the measurement region. However, in order to avoid confusion due to multiple $\epsilon$ estimates, The RDF-based estimate is used in this manuscript solely as a tool for consistency check. Unless otherwise stated, the dissipation rate and the corresponding Kolmogorov scales used in the following analysis are those obtained from the more direct and orthodox structure-function method.

The particle properties and flow parameters for all experiments are summarized in Table~\ref{tab:all_parameters}, including the particle diameter $d$, turbulent velocity fluctuation $u_{\mathrm{rms}}$, Reynolds number $Re$, and
particle response time $\tau_p$. Based on these measurements, turbulence parameters inferred from both the RDF-based method and the second-order velocity structure function ($S_2$) method are listed, including the dissipation rate $\varepsilon$, Kolmogorov length scale $\eta$, and the corresponding
Stokes number $St$.}

\begin{table*}[t]
	\centering
	\caption{
		Summary of particle and turbulence parameters for all experiments. Turbulence dissipation rate, Kolmogorov scales, and Stokes numbers are estimated independently using the RDF-based method and the second-order velocity structure function method.
	}
	\label{tab:all_parameters}
	\renewcommand{\arraystretch}{1.2}
	\setlength{\tabcolsep}{4.5pt}
	\footnotesize 
	\begin{tabular*}{\textwidth}{@{\extracolsep{\fill}} c c c c c c c c c c c c @{}}
		\toprule
		Disk Speed & $d$ & $u_{\mathrm{rms}}$ & $Re$ & $\tau_p$ 
		& $\varepsilon_{\scriptscriptstyle\mathrm{R}}$ 
		& $\varepsilon_{\scriptscriptstyle\mathrm{S}}$ 
		& $\eta_{\scriptscriptstyle\mathrm{R}}$ & $\eta_{\scriptscriptstyle\mathrm{S}}$ 
		& $St_{\scriptscriptstyle\mathrm{R}}$ & $St_{\scriptscriptstyle\mathrm{S}}$ \\
        (rpm) 
        & ($\mu$m) 
        & ($\mathrm{m\,s^{-1}}$)
        &  
        & (ms) 
        & \multicolumn{2}{c}{($\mathrm{m^2\,s^{-3}}$)} 
        & \multicolumn{2}{c}{($\mu$m)} 
        &  
        &  \\
		\midrule
		42k   & $17.6 \pm 3.1$ & 0.550 & 3642 & $0.95 \pm 0.05$ & 0.43 & 1.36 & 299 & 224 & 0.16 & 0.28 \\
		42k  & $33.6 \pm 4.1$ & 0.530 & 3510 & $3.50 \pm 0.05$ & 0.43 & 1.36 & 299 & 224 & 0.58 & 1.04 \\
		\addlinespace[3pt]
		30k  & $21.6 \pm 4.8$ & 0.438 & 2914 & $1.40 \pm 0.07$ & 0.09 & 0.07 & 443 & 471 & 0.11 & 0.10 \\
		30k  & $40.8 \pm 3.7$ & 0.380 & 2517 & $5.10 \pm 0.04$ & 0.09 & 0.07 & 443 & 471 & 0.39 & 0.35 \\
		\addlinespace[3pt]
		18k &  $29.6 \pm 8.5$ & 0.339 & 2252 & $2.70 \pm 0.22$ & 0.07 & 0.05 & 475 & 512 & 0.18 & 0.15 \\
		18k &  $62.4 \pm 5.4$ & 0.264 & 1722 & $11.9 \pm 0.09$ & 0.07 & 0.05 & 475 & 512 & 0.80 & 0.69 \\
		\bottomrule
	\end{tabular*}
	\footnotesize
	\begin{flushleft}
		$\varepsilon_{\scriptscriptstyle\mathrm{R}}$, 
		$\eta_{\scriptscriptstyle\mathrm{R}}$, 
		$St_{\scriptscriptstyle\mathrm{R}}$
		are obtained from the RDF-based method;
		$\varepsilon_{\scriptscriptstyle\mathrm{S}}$, 
		$\eta_{\scriptscriptstyle\mathrm{S}}$,  
		$St_{\scriptscriptstyle\mathrm{S}}$
		are estimated based on the second-order velocity structure function.
	\end{flushleft}
\end{table*}

\subsection{Computing the Radial Distribution Function}\label{subsec2.5}
The RDF represents the of probability of finding another particle at a distance $r$ from a reference particle, relative to that expected for a uniform random distribution. 
Following the {work} in \cite{Saw2012}, the RDF can be calculated by
\begin{equation} \label{eq:RDF}
g(r) =
\dfrac{\psi(r)/N}{(N-1)\, \delta V_r / V}
\end{equation}
where $\psi(r)$ represents the summed number of particles located at a separation $r$ from each trial particle (the trial particles are taken to be the entire population of $N$ particles, double counting is allowed). The quantity $\delta V_r/V$ denotes the ratio of a small sampling volume at distance $r$ (i.e., a spherical shell of thickness $\delta r$) to the total measurement volume $V$.

To minimize the effects of frame-to-frame fluctuations in particle number, the raw RDF for each frame was computed using Eq.~\eqref{eq:RDF} and averaged to obtain $g_{\mathrm{raw}}(r)$. To mitigate the artifact due to variations in the interrogation volume near the view volume boundaries (finite view volume effect), a reference RDF, $g_{\mathrm{v}}(r)$, was obtained by considering all detected particles throughout the entire experiment, as if they were simultaneously present in the same volume. $g_{\mathrm{v}}(r)$ essentially captures the RDF-signature of the "shape" of the view-volume. Following that, the raw RDF was divided by $g_{\mathrm{v}}(r)$ to yield the final, inhomogeneity-corrected RDF, $g_{\text{eff}}(r)$.
\begin{equation} \label{eq:RDF_norm}
     g_{\text{eff}}(r) = \frac{g_{\text{raw}}(r)}{g_{\text{v}}(r)}
\end{equation}
This procedure suppresses the artifacts caused by any large scale inhomogeneity in particle density and view volume boundary effects\footnote{The latter (finite view volume effect) can be interpreted as a special case of the former (inhomogeneous density) in the sense that the particle density changes from finite inside the view volume to zero outside of it.}, ensuring that $g_{\text{eff}}(r)$ reflects the intrinsic spatial clustering of the particles. This method follows (with generalization) from the methods described in \cite{Saw2008, Saw2008phd} (see also the opened review (by second reviewer, RC2) of \cite{Dodson2019RC2}). The effectiveness of this method will be demonstrated later in the results section pertaining to RDF (Fig.~\ref{fig:rdf}).

\subsection{Calibration and Image Processing}\label{subsec2.6}
Accurate knowledge of the imaging geometry is essential for three-dimensional particle tracking. In the present system, three high-speed cameras were arranged around the turbulence chamber (see Sec.~\ref{subsec2.1}) to provide overlapping views. Each camera was calibrated using a calibration plate with known marker coordinates placed inside the measurement region. The calibration {followed the Tsai method based on pin-hole camera model} \cite{Tsai1987}: multiple images of the plate were acquired at different orientations, and nonlinear optimization was applied to determine intrinsic parameters (focal length, principal point location, radial distortion) and extrinsic parameters (rotation matrix and translation vector). These calibrated parameters form the geometric basis for computing three-dimensional particle positions.

After calibration, particle images were processed using a combination of grayscale thresholding to separate particles from the background and sub-pixel localization. The thresholds used for particle tracking is not necessary the same as the thresholds for particle sizing described in Sec.~\ref{subsec2.3}. The former is optimized for maximizing particle detection and minimizing false positive due to optical noise, without concern over the accuracy of the resulting sizes of the accepted particles.
Subsequently, the centroid method was applied to compute sub-pixel coordinates.

Three-dimensional particle positions reconstruction may be heuristically understood as a multi-view ray-based
triangulation approach. For each particle candidate detected in the reference view (e.g., of Camera~1), a back-projection ray was constructed from the camera through the image coordinate. Candidate image points in Cameras~2 and~3 were treated in the same manner, forming corresponding back-projection rays in three-dimensional space. Multi-view matching was performed by evaluating the geometric consistency among these rays \cite{Xu2007}. Specifically, candidate particle (a combination of one point per each camera) were accepted when the minimum distance between their back-projection rays fell below a prescribed matching tolerance. In the present study, the tolerance was set to 3~pixels. This value was chosen to be smaller than the nominal particle diameter (approximately 4~pixels), such that ideally only rays originating from the same physical particle would be associated. The accepted ray "intersections" were then used to determine the three-dimensional particle positions for each individual frame in the recorded movie. The same tolerance criterion is applied in the subsequent analysis described in Sec~\ref{subsec3.1}

Temporal particle trajectories were then obtained by linking the three-dimensional positions across successive frames of the recorded movie. A three-frame predictor-corrector method was used, which is equivalent to applying a minimum-acceleration criterion. Particle's next position is predicted based on last known velocity. When only a single candidate particle was found inside the predicted search window (space within a fixed range from predicted position), the temporal link was assigned directly. When multiple candidates existed, the candidate yielding the smallest acceleration was selected. This method requires sufficiently high temporal resolution; the cameras operated at $10$k\,fps, corresponding to a sampling interval of approximately $1/40 \times \tau_{\eta}$ for most demanding experiment (42\,k\,rpm), which is adequate for resolving particle motion in the present work.

\begin{figure}[t]
    \centering
    \begin{subfigure}[b]{0.48\textwidth}
        \centering
        \includegraphics[width=\textwidth]{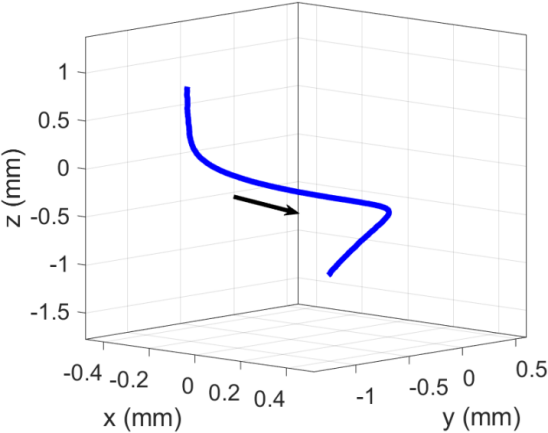}
        \caption{The trajectory of droplet A}
        \label{fig:dropletA}
    \end{subfigure}
    \hfill
    \begin{subfigure}[b]{0.48\textwidth}
        \centering
        \includegraphics[width=\textwidth]{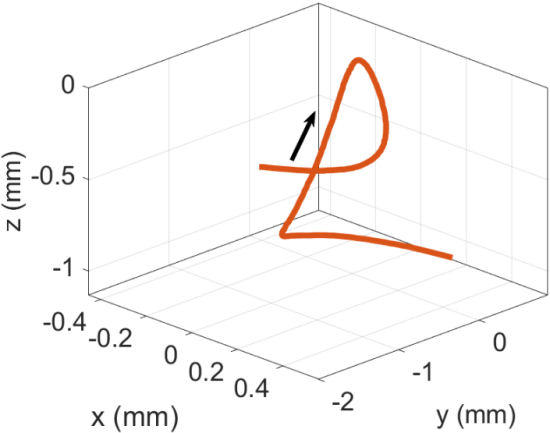}
        \caption{The trajectory of droplet B}
        \label{fig:dropletB}
    \end{subfigure}
	\caption{Single-particle trajectory. The X-, Y-, and Z-axes denote the horizontal, vertical (gravity-aligned), and imaging-depth directions, respectively. The black arrow indicates the particle’s direction of motion.}
	\label{fig:single_particle_trajectory}
\end{figure}

\begin{figure}
	\centering
	\includegraphics[width=0.5\textwidth,height=2.10in]{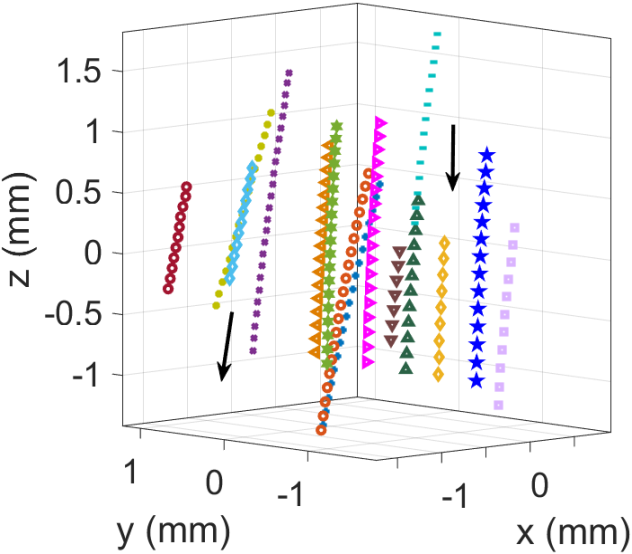}
	\caption{Particle Tracks in 3D Space during a Short Time Interval (0.0034s). Each curve represents the trajectory of a single particle moving in three-dimensional space over time, with different colors and marker styles used to distinguish individual droplets. No tracks from different times have been combined; all trajectories are measured simultaneously. The black arrow indicates the direction of movement of the droplets.}
	\label{fig:particle_tracks_3d}
\end{figure}

From experiment case of $42$\,k\,rpm, representative trajectories were selected to illustrate the measured particle motion. Fig.~\ref{fig:single_particle_trajectory} shows the three-dimensional trajectory of a single particle, and Fig.~\ref{fig:particle_tracks_3d} presents several particle trajectories over a short interval of $0.0034$\,s. The X-axis corresponds to the horizontal (left--right) direction, the Y-axis represents the vertical direction aligned with gravity, and the Z-axis denotes the imaging depth direction.

As shown in Figs.~\ref{fig:single_particle_trajectory} and \ref{fig:particle_tracks_3d}, the measured trajectories maintain good spatial continuity and capture small-scale structures of the turbulent flow. Some trajectories exhibit looping or swirling behavior, demonstrating that the present methodology is capable of resolving complex inertial-particle dynamics in turbulence.

\section{Misidentification and Spurious Particle Events}\label{sec3}
Our interest in small, near collisional, scale dynamics combined with high seeding concentration,
introduces substantial challenges for both temporal tracking and multi-view stereo matching. During data processing, three representative types of artifacts or errors were identified. These errors can produce spurious particle and trajectories in the tracking results, ultimately affecting the reliability of particle clustering and dynamical statistics and analyses.

\subsection{False Stereo-matching-induced Spurious Particles (FMIS)}\label{subsec3.1}
\begin{figure}[htbp]
	\centering
	\includegraphics[width=0.6\textwidth,height=1.5in]{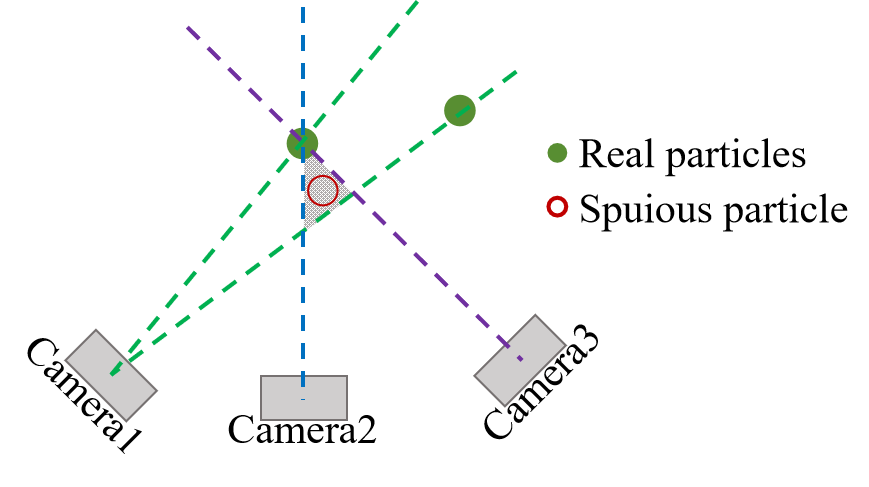}
	\caption{Spurious particle from projection ambiguity (three cameras arranged in the same plane). The matching process between the three camera views and the actual particle is illustrated. Each detected particle in the image plane defines a back-projection ray originating from the camera passing through the centroid of the particle's 2D location in the image plane. The green dashed line denotes the back-projection ray from Camera\,1, while the blue and purple dashed lines correspond to the back-projection rays from Camera\,2 and Camera\,3, respectively. The gray shadows represent areas where false particles could form. The red hollow circle indicates an incorrect matching result caused by projection ambiguity when rays corresponding to different particles come within distances smaller than a prescribed matching-tolerance-distance.}
	\label{fig:spurious_particle}
\end{figure}

In the 3D stereo-matching stage, a type of spurious particle can arise due to projection ambiguity. This occurs when two particles are closely spaced in the image plane of one camera, even though their actual 3D-position in the measurement volume are not necessarily close. Each detected spot defines a back-projection ray, and when rays from closely spaced spots in  e.g., Camera~1 encounter rays from the other two cameras, multiple intersections may occur. This is illustrated in Fig.~\ref{fig:spurious_particle}. Specifically, a single particle detected in Camera~2 and Camera~3 may intersect with both rays from Camera~1, forming a small triangular region in which an additional false intersection is likely, which would then lead to occurrence of a phantom particle. In practice, an “intersect” is assumed when two projection rays reach a distance smaller than a prescribed matching tolerance as detailed in Sec~\ref{subsec2.6}. A phantom particle may then be reconstructed within this region.

\begin{figure}[htbp]
	\centering
	\begin{subfigure}[b]{0.32\textwidth}
		\centering
		\includegraphics[width=\textwidth]{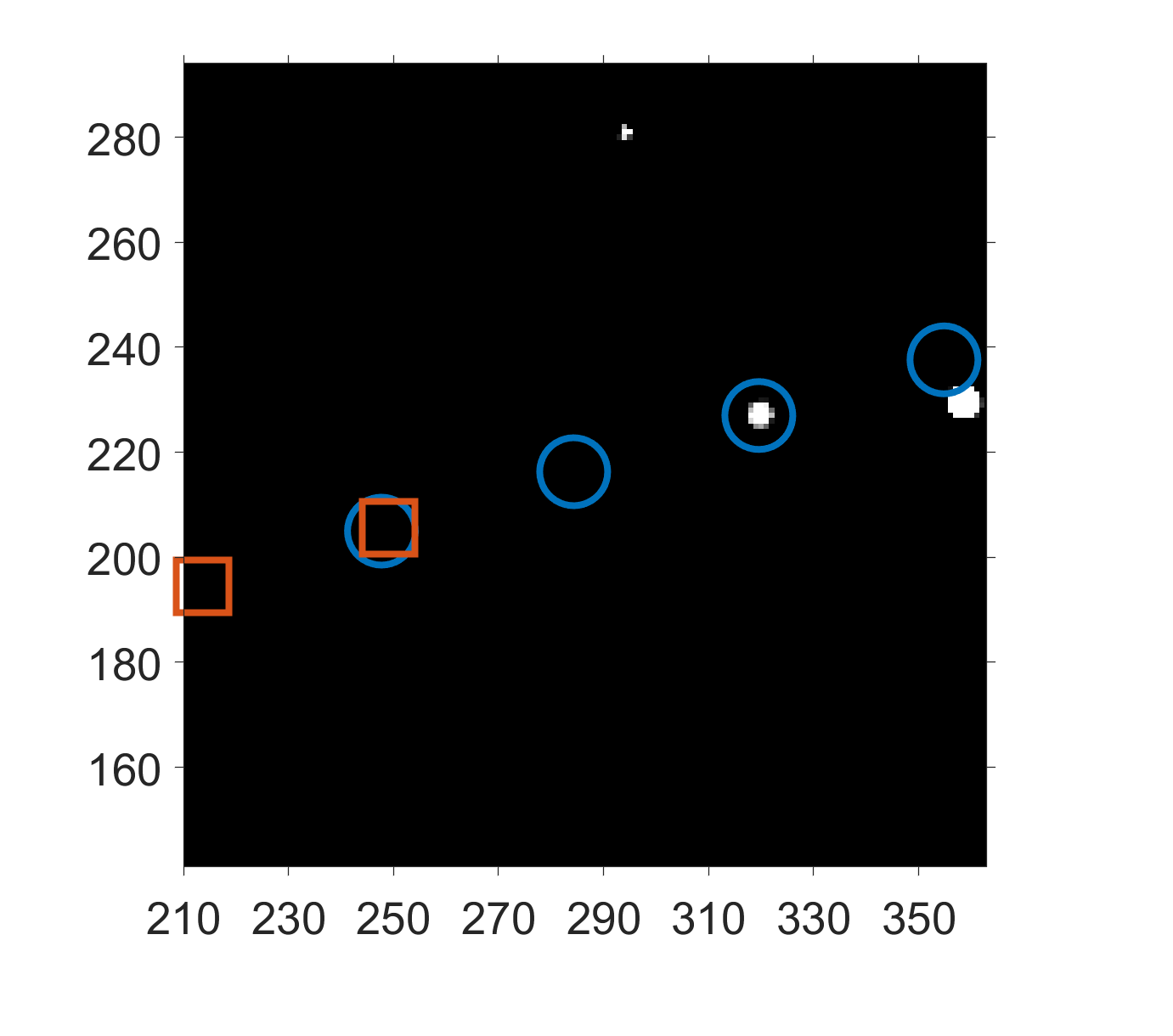}
		\caption{Camera\,1 at $T - 2\Delta t$}
		\label{fig:Camera1 at T - 2t_1993}
	\end{subfigure}
	\hfill
	\begin{subfigure}[b]{0.32\textwidth}
		\centering
		\includegraphics[width=\textwidth]{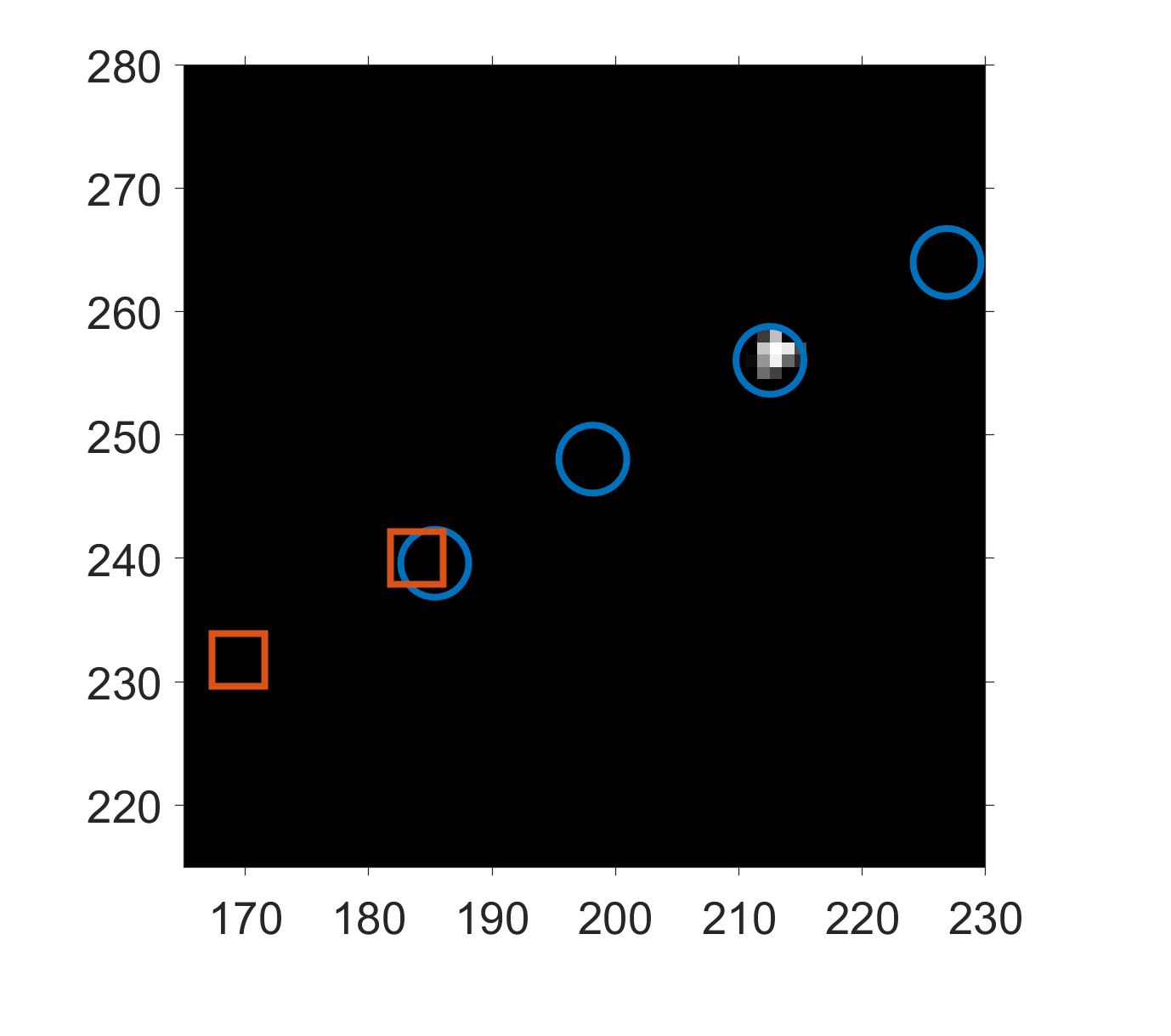}
		\caption{Camera\,2 at $T - 2\Delta t$}
		\label{fig:Camera2 at T - 2t_1993}
	\end{subfigure}
	\hfill
	\begin{subfigure}[b]{0.32\textwidth}
		\centering
		\includegraphics[width=\textwidth]{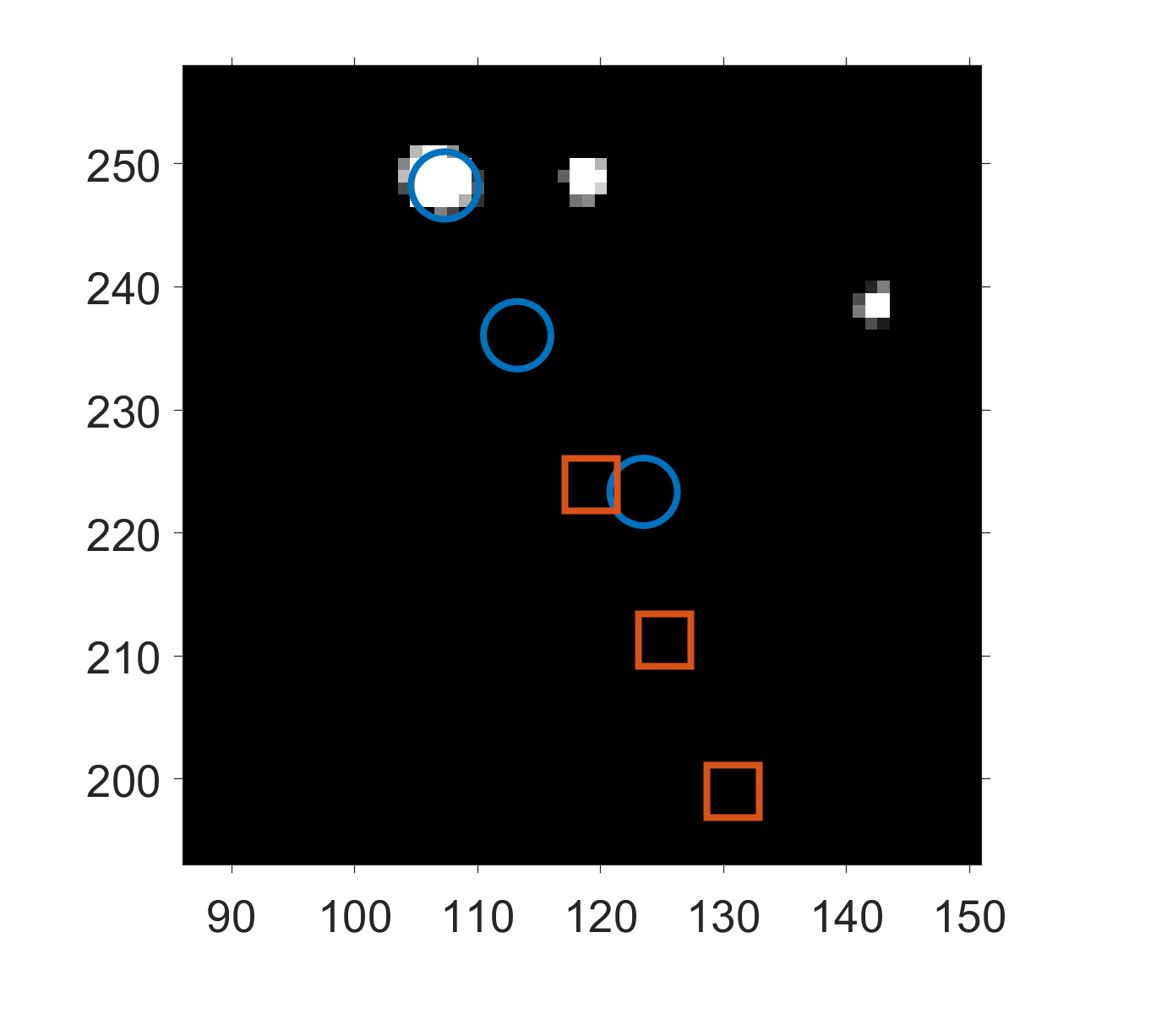}
		\caption{Camera\,3 at $T - 2\Delta t$}
		\label{fig:Camera3 at T - 2t_1993}
	\end{subfigure}
	\hspace{0.05\textwidth}
	\begin{subfigure}[b]{0.32\textwidth}
		\centering
		\includegraphics[width=\textwidth]{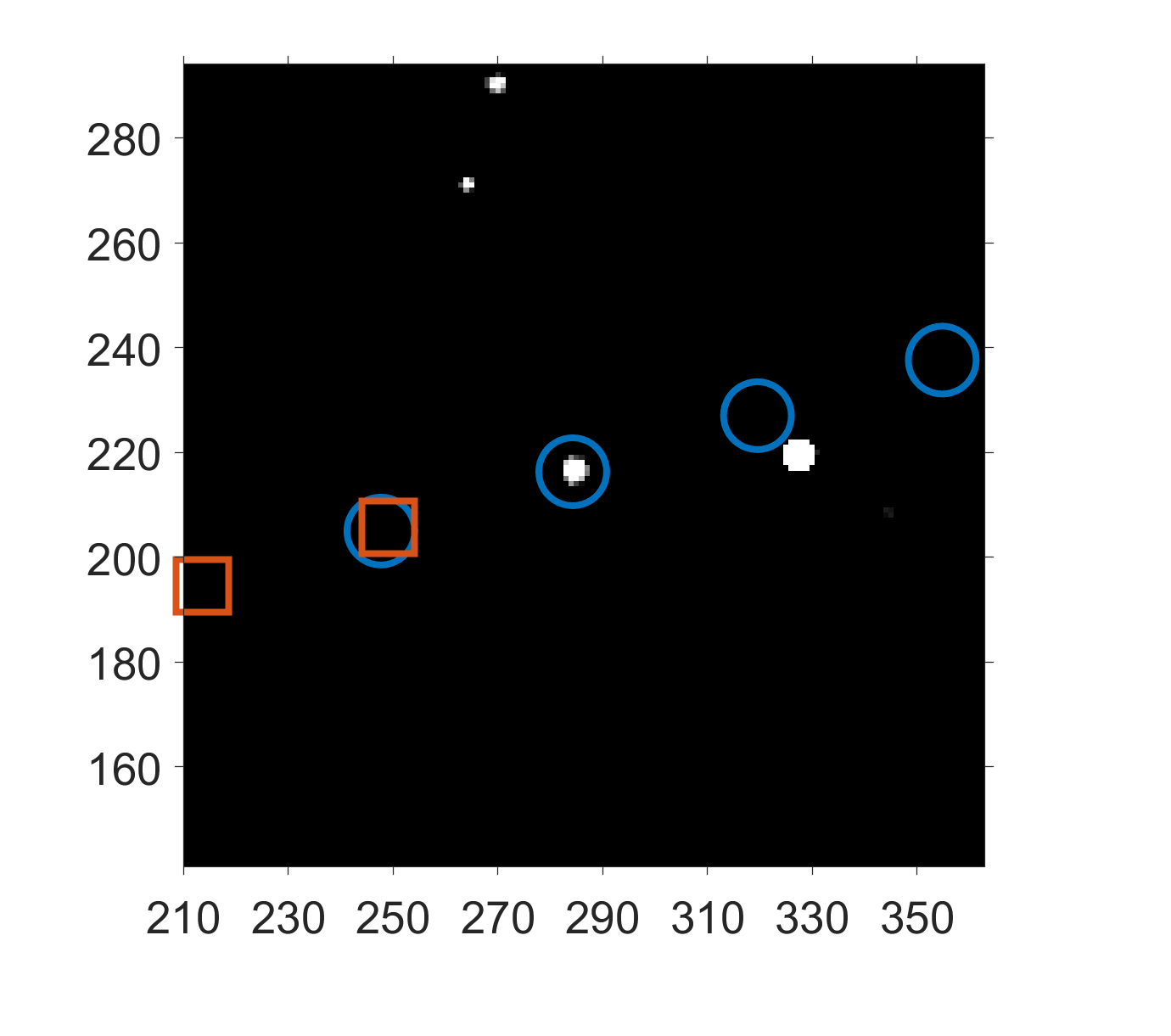}
		\caption{Camera\,1 at $T - \Delta t$}
		\label{fig:Camera1 at T - 1t_1993}
	\end{subfigure}
	\hfill
	\begin{subfigure}[b]{0.32\textwidth}
		\centering
		\includegraphics[width=\textwidth]{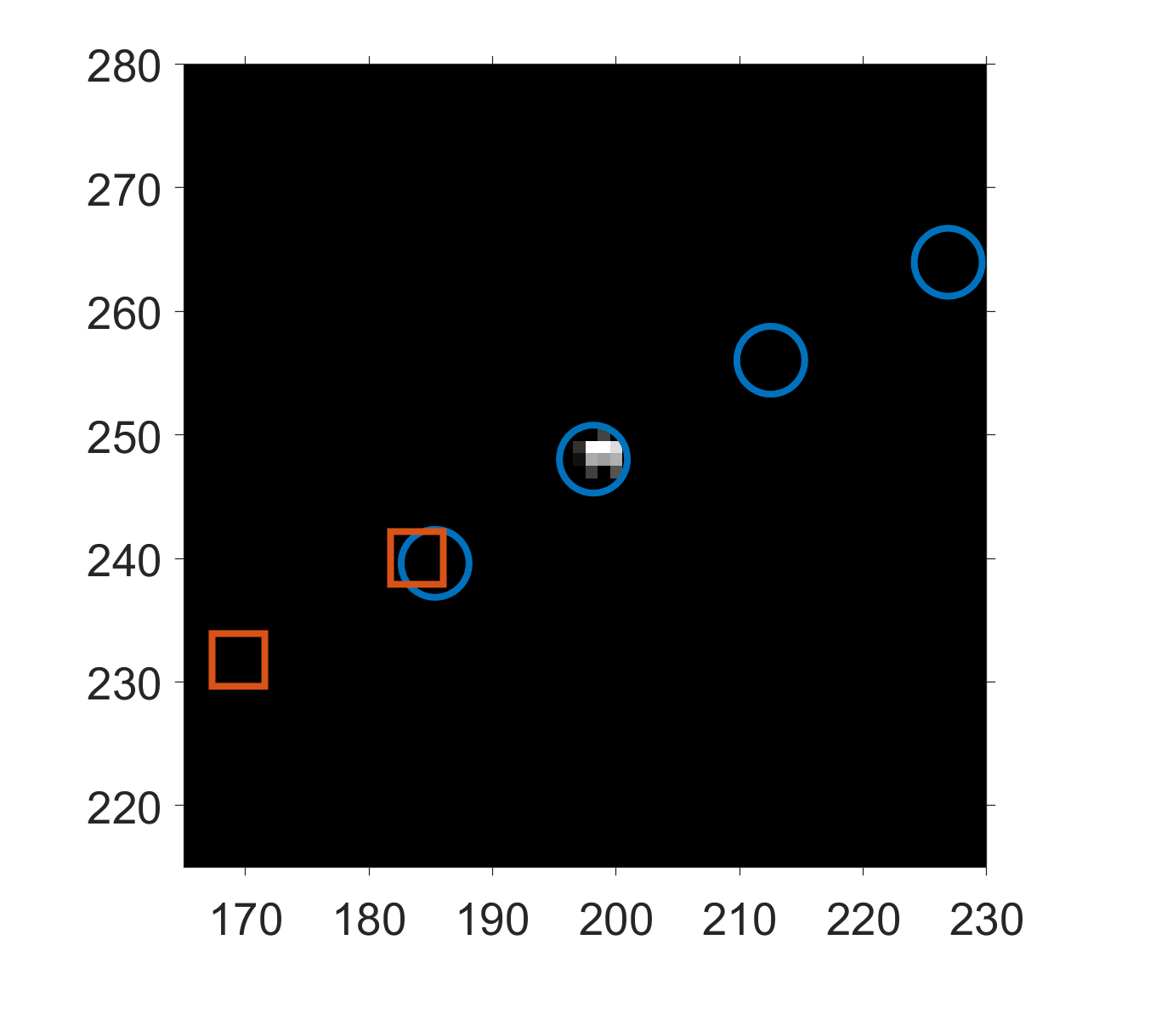}
		\caption{Camera\,2 at $T - \Delta t$}
		\label{fig:Camera2 at T - 1t_1993}
	\end{subfigure}
	\hfill
	\begin{subfigure}[b]{0.32\textwidth}
		\centering
		\includegraphics[width=\textwidth]{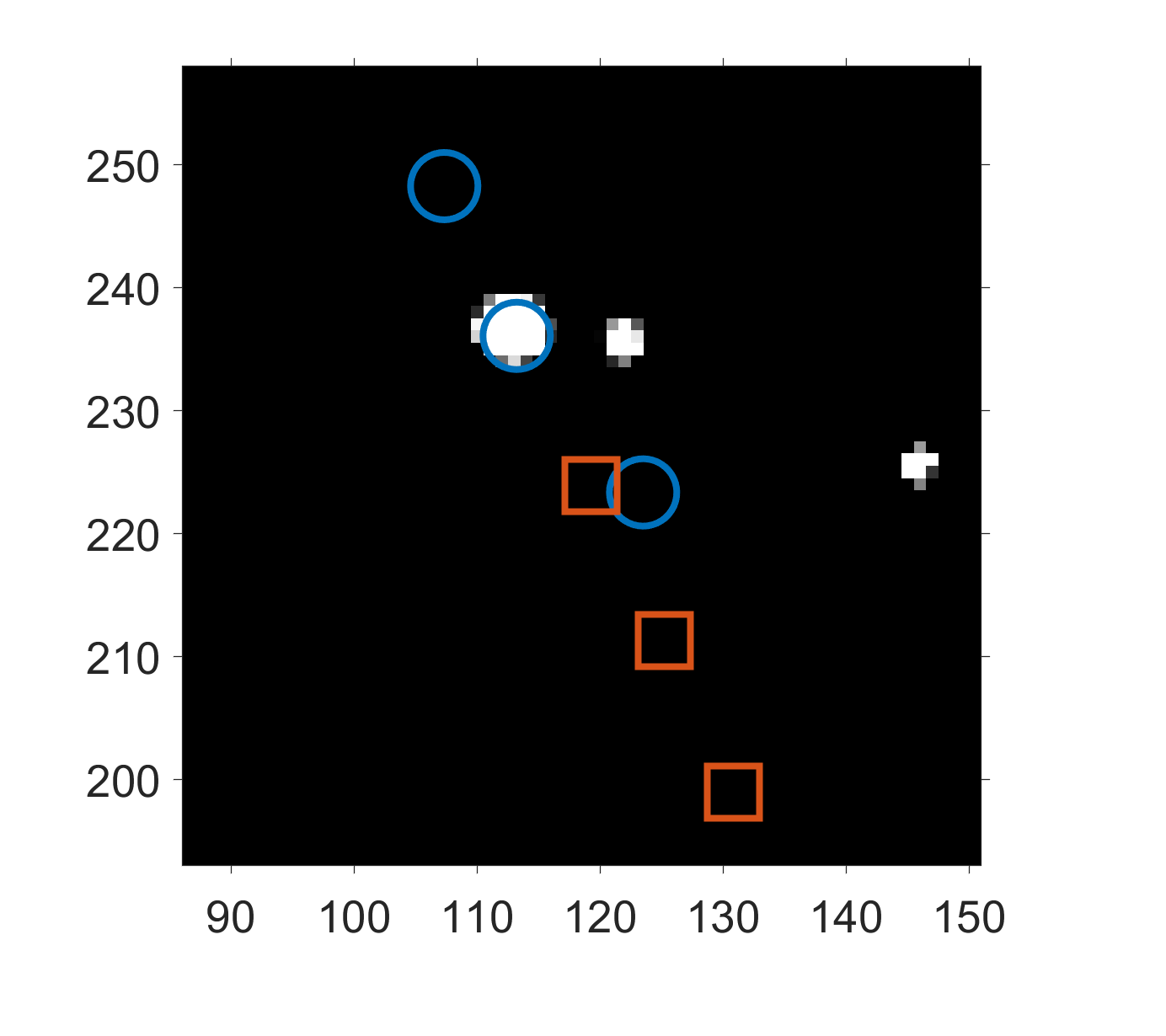}
		\caption{Camera\,3 at $T - \Delta t$}
		\label{fig:Camera3 at T - 1t_1993}
	\end{subfigure}
	\hspace{0.05\textwidth}
	\begin{subfigure}[b]{0.32\textwidth}
		\centering
		\includegraphics[width=\textwidth]{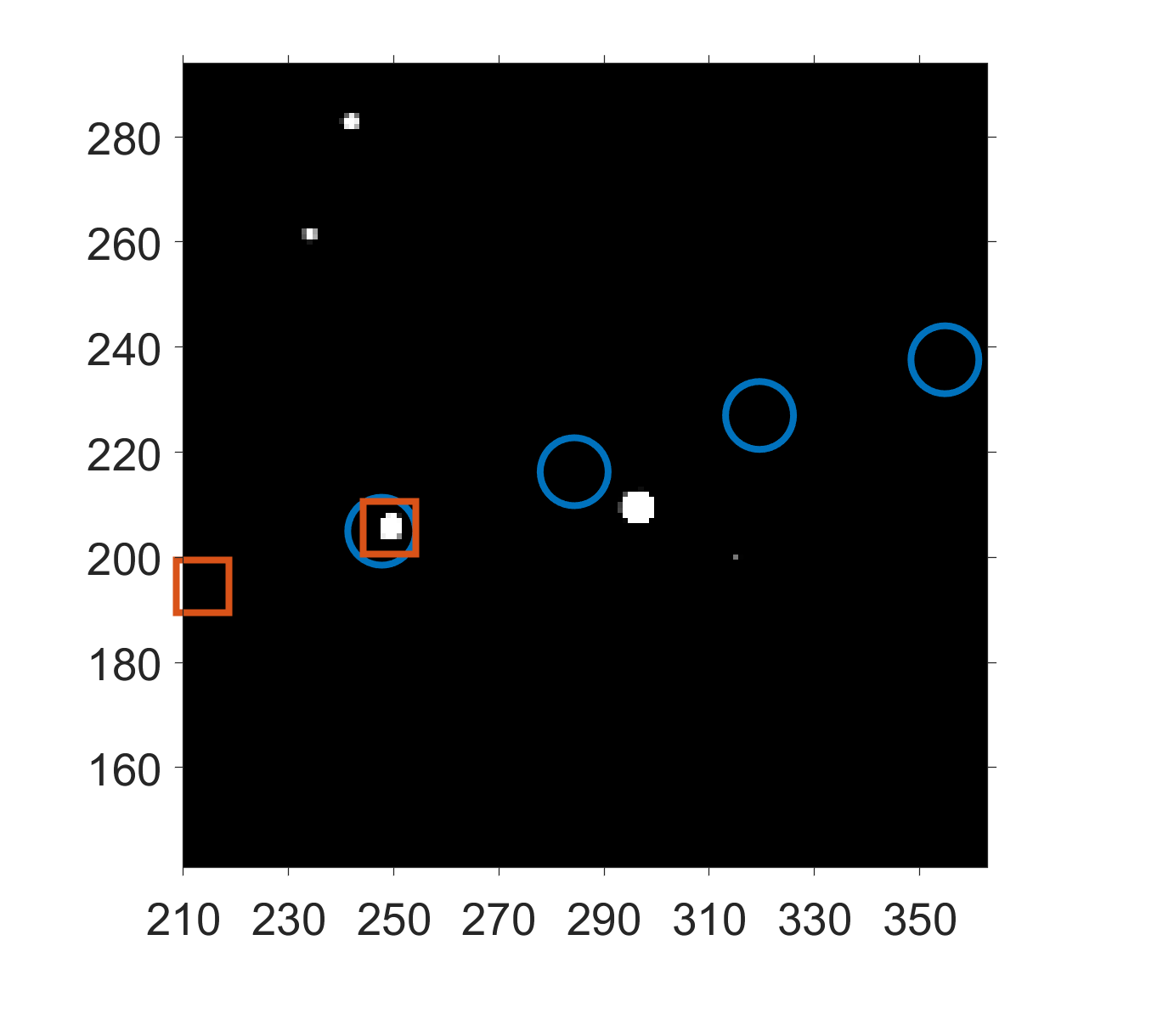}
		\caption{Camera\,1 at $T$}
		\label{fig:Camera1 at T_1993}
	\end{subfigure}
	\hfill
	\begin{subfigure}[b]{0.32\textwidth}
		\centering
		\includegraphics[width=\textwidth]{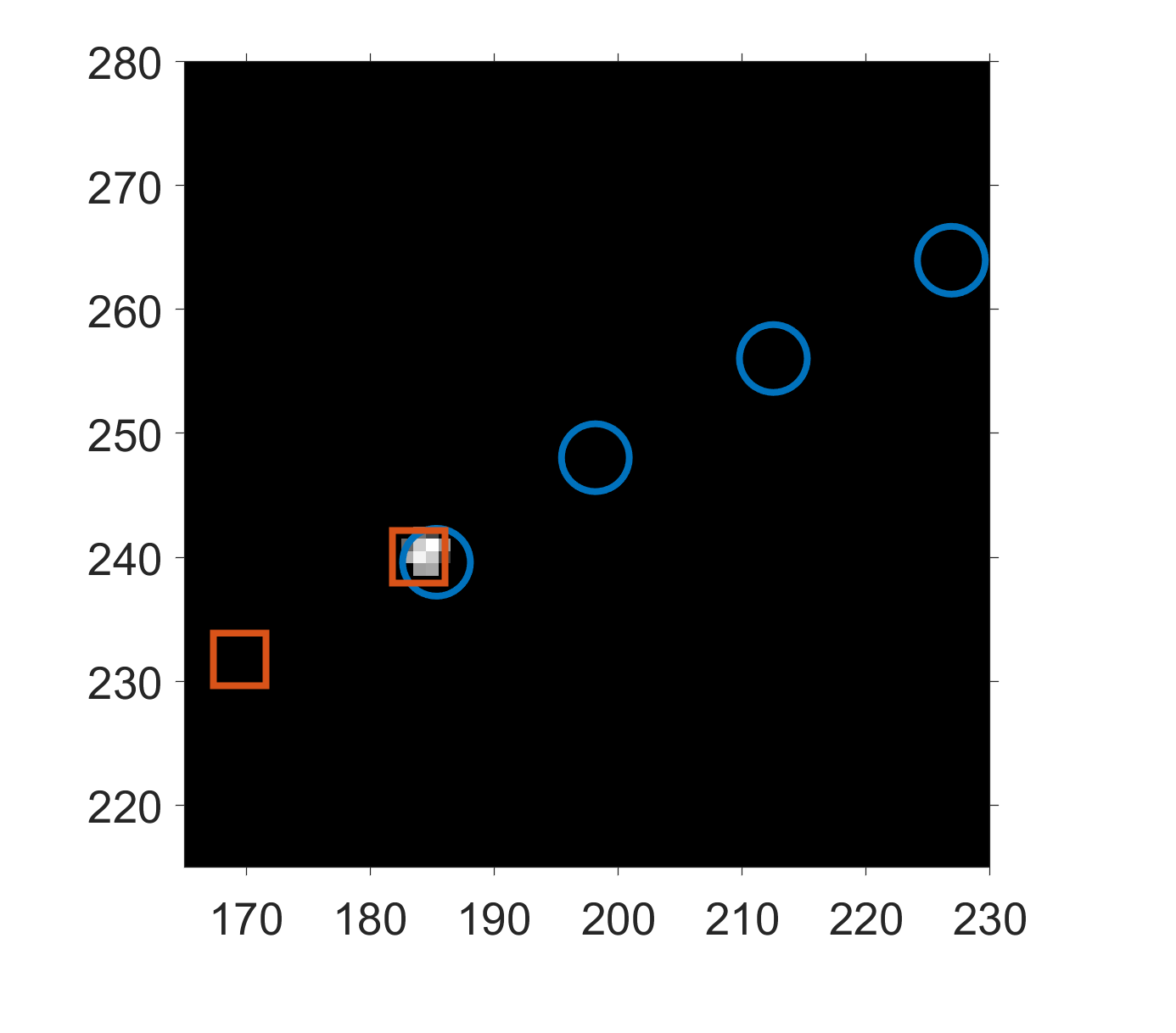}
		\caption{Camera\,2 at $T$}
		\label{fig:Camera2 at T_1993}
	\end{subfigure}
	\hfill
	\begin{subfigure}[b]{0.32\textwidth}
		\centering
		\includegraphics[width=\textwidth]{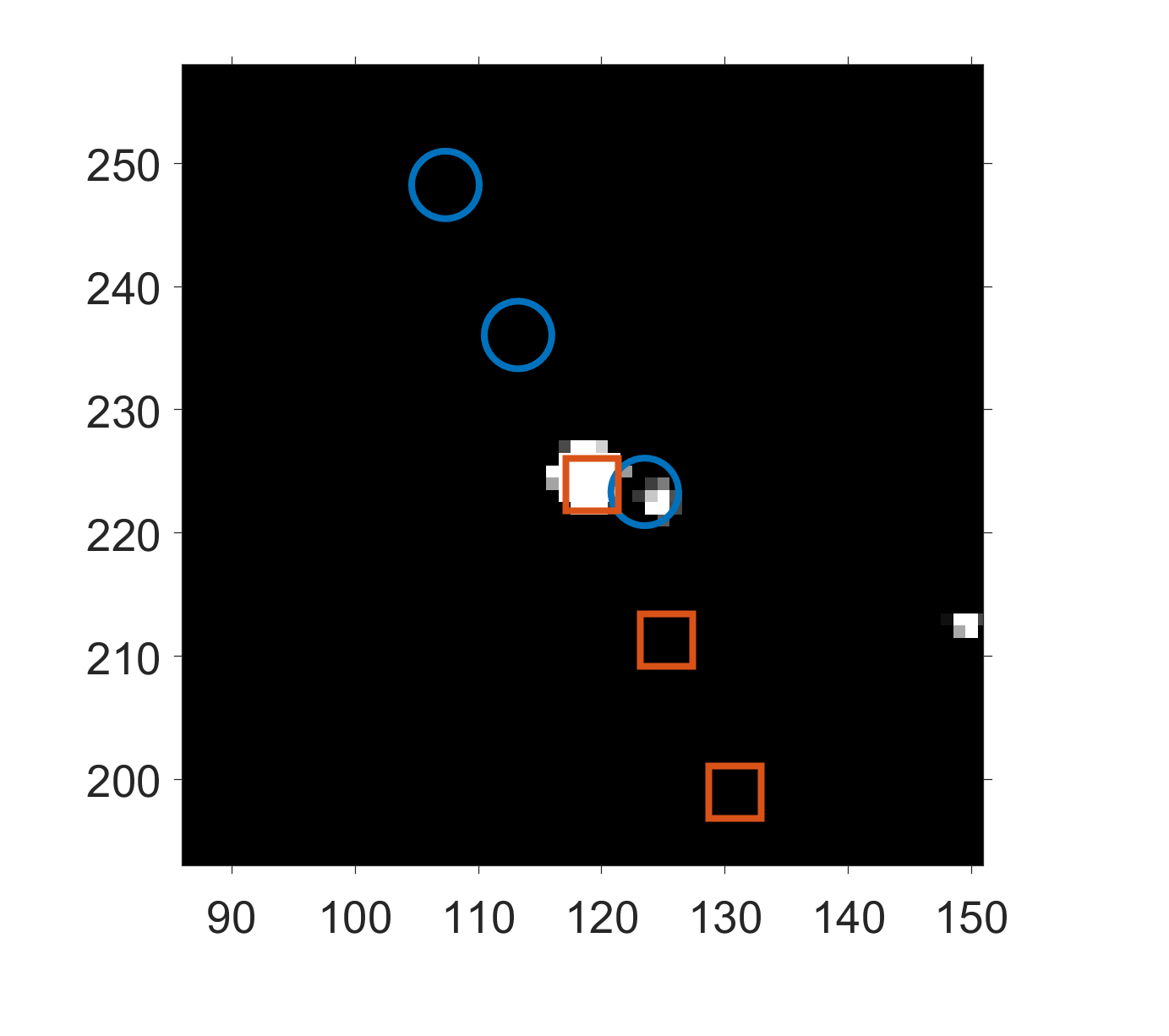}
		\caption{Camera\,3 at $T$}
		\label{fig:Camera3 at T_1993}
	\end{subfigure}
	\hfill
	\begin{subfigure}[b]{0.32\textwidth}
		\centering
		\includegraphics[width=\textwidth]{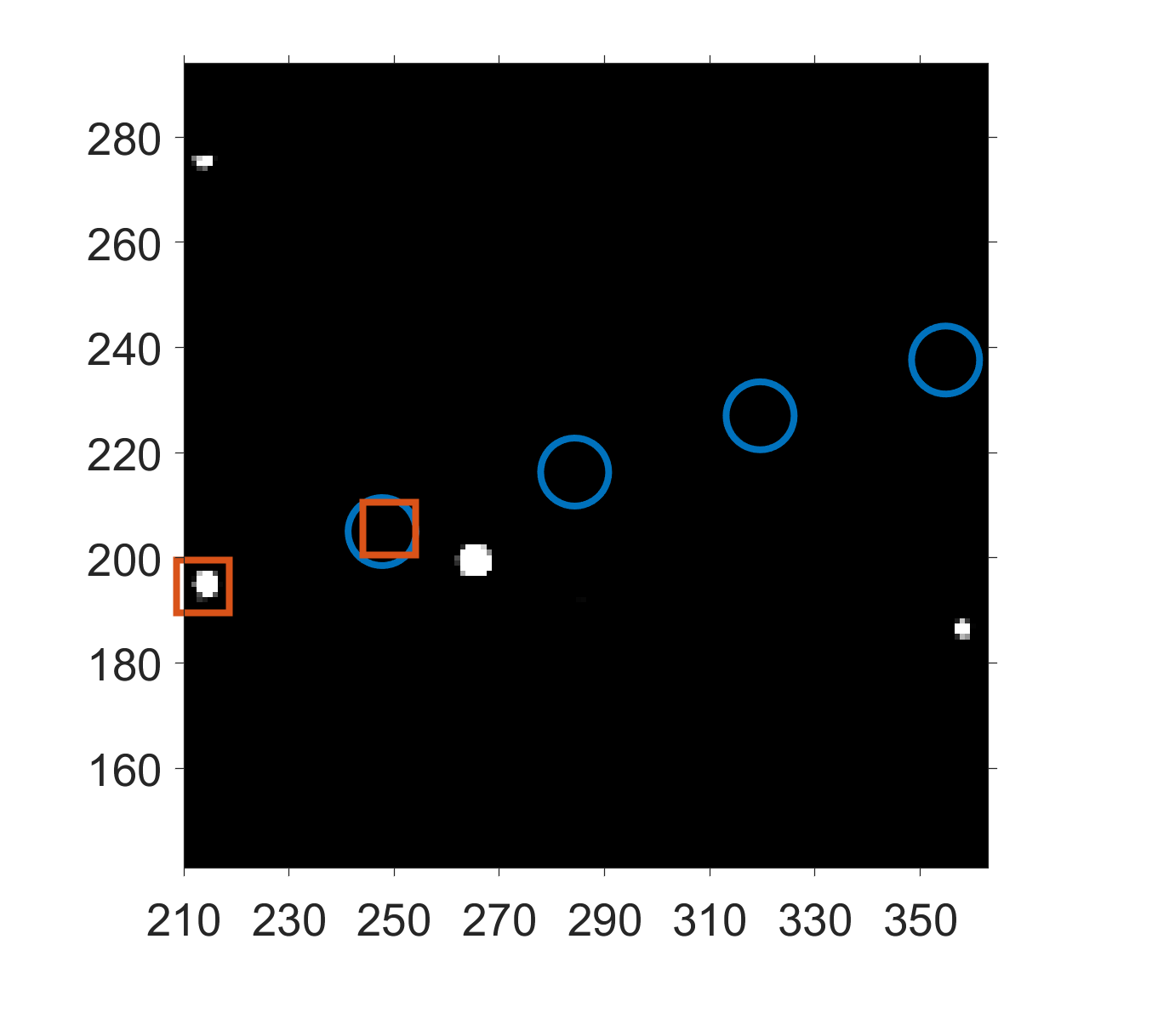}
		\caption{Camera\,1 at $T + \Delta t$}
		\label{fig:Camera0 at T + 1t_1993}
	\end{subfigure}
	\hfill
	\begin{subfigure}[b]{0.32\textwidth}
		\centering
		\includegraphics[width=\textwidth]{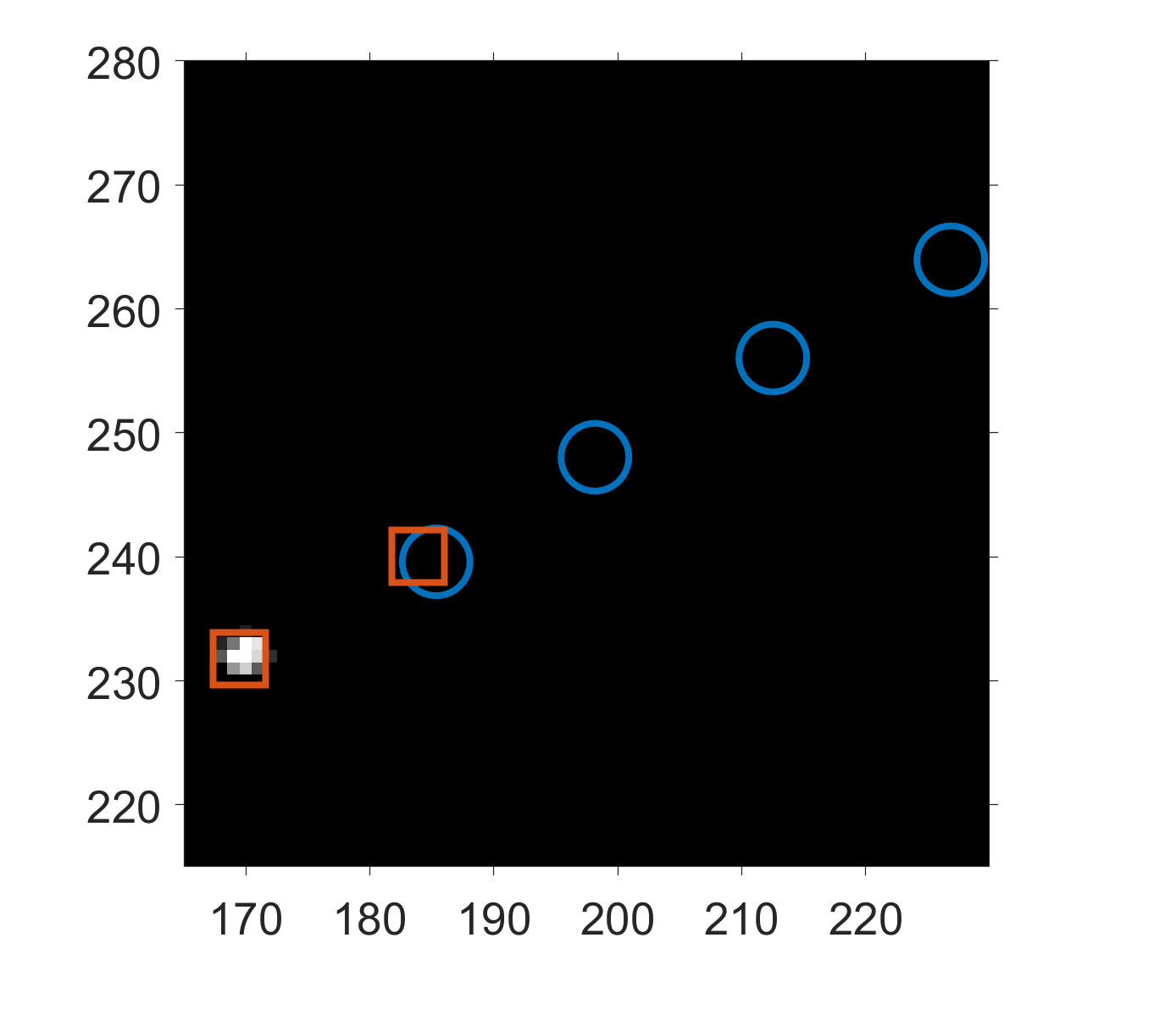}
		\caption{Camera\,2 at $T + \Delta t$}
		\label{fig:Camera2 at T + 1t_1993}
	\end{subfigure}
	\hfill
	\begin{subfigure}[b]{0.32\textwidth}
		\centering
		\includegraphics[width=\textwidth]{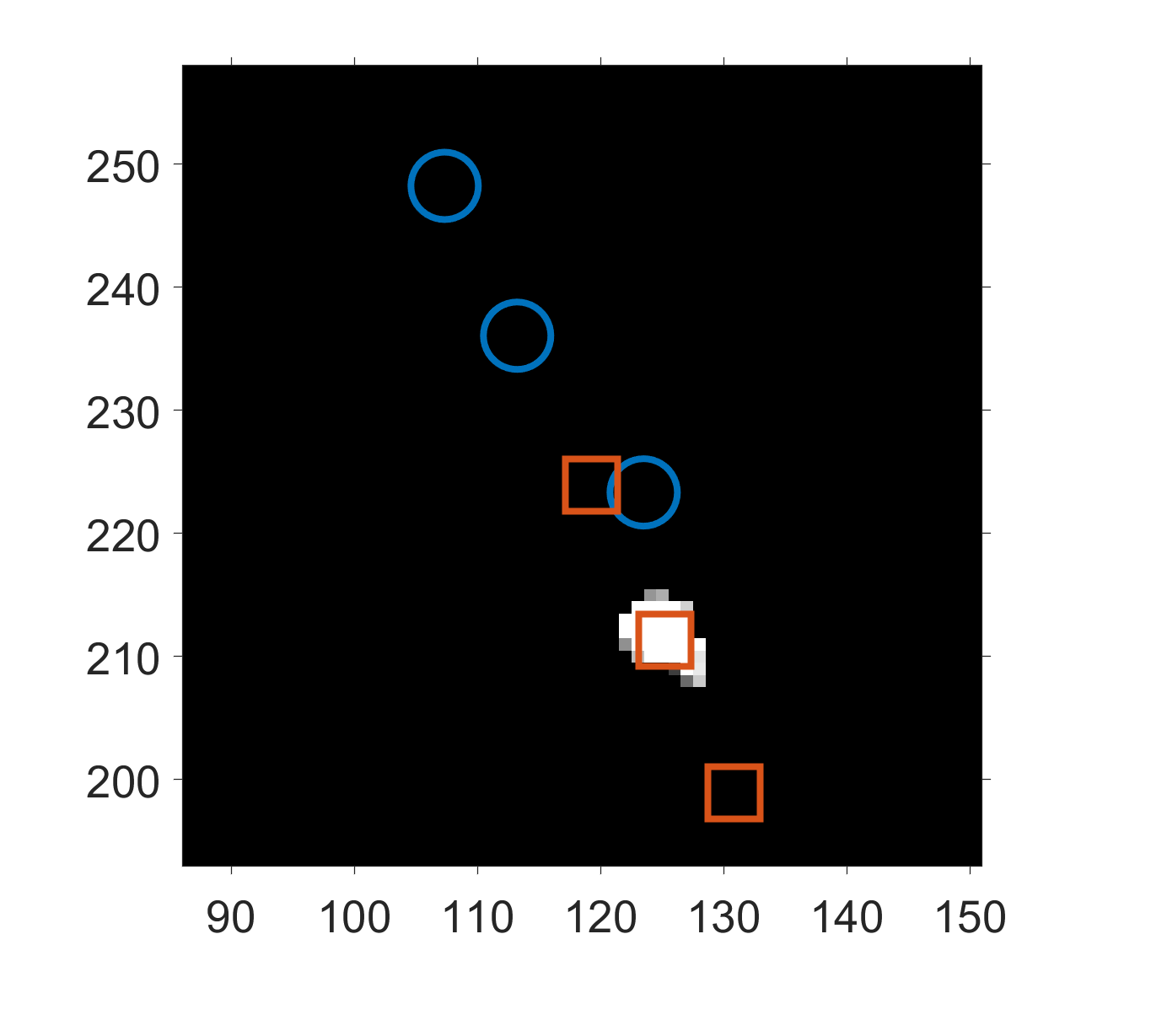}
		\caption{Camera\,3 at $T + \Delta t$}
		\label{fig:Camera3 at T + 1t_1993}
	\end{subfigure}
	\caption{Confirmed example of false stereo-matching induced spurious particles (FMIS) illustrated using synchronized image sequences from three cameras at four consecutive frames (time steps). Panels (a)--(c), (d)--(f), (g)--(i), and (j)--(l) correspond to $T-2\Delta t$, $T-1\Delta t$, $T$, and $T+1\Delta t$, respectively. Blue circles and orange squares denote two reconstructed particle trajectories (overlaid on all images for reference, irrespective of the specific frame at which each point was detected). At $T-2\Delta t$ and $T-1\Delta t$, the two particles are clearly separated in all camera views. At $T$, the particles become closely spaced in Camera\,3, resulting in two nearly overlapping reconstructed trajectories, leading to spurious closely separated particle pair.} 
	\label{fig:tracking_error_demo}
\end{figure}

Fig.~\ref{fig:tracking_error_demo} illustrates a confirmed example of particle tracking error arising from projection overlap. The figure shows synchronized image-plane trajectories from all three cameras for the same particle pair over four consecutive frames. At $T+\Delta t$ and $T+2\Delta t$, both particles are clearly resolved in at least two cameras (in this case Camera\,1 and 3), leading to unambiguous stereo matching and reconstructed trajectory. 
(In this specific example, Camera\,2 sees only one of the particles, with the other particle likely located outside its view.)
At $T+3\Delta t$, however, the two particles become extremely close in Camera\,3, while remaining distinguishable in Camera\,1. This projection overlap results in ambiguous ray intersections and causes the stereo-matching algorithm to erroneously reconstruct two nearly coincidental three-dimensional trajectories. At the subsequent time step ($T+4\Delta t$), the matching ambiguity is resolved and only one trajectory is retained. Consequently, a supposedly continuous particle track becomes artificially split, and at one instance, a spurious particle is generated at very close vicinity to a real particle. Such events will artificially raise the magnitude of RDF at small (near collisional) scales.
More examples of occurrence of this type of error are shown in Appendix~\ref{secA1}.

When particle concentration is high, this type of error will occur more frequently, causing RDF values at small scales to be overestimated. This effect could be quite severe considering that its occurrence is tied to the probability for projections of two particles to coincide in the view of any single camera which is much larger than probability for genuine close 3D encounters of turbulently advected particles. This type error or artifact is hereafter referred to as \textbf{FMIS}, denoting false stereo-matching induced spurious particles. We must emphasize that this artifact is not specific to the way our cameras are configured and is expected for any configuration of multi-cameras LPT system. Thus, the implication of these findings and its mitigation methodology (Sec.~\ref{sec4}) for general LPT or particle tracking velocimetry (PTV) applications should not be underestimated.

\subsection{Thresholding Induced Particle Fragmentation (TIF)}\label{subsec3.2}
Particles that are slightly out of focused could appears as diffused expanded dark spots. While these particles are typically recognized as single objects under basic threshold settings, background noise suppression often necessitates raising the threshold value. This adjustment can cause regions of varying brightness within a single particle to be misclassified as multiple smaller particles, leading to fragmentation. This introduces spurious particles with small inter-particle distances that is detrimental for accurate RDF measurement. While increasing the threshold helps suppress false positives, excessive threshold may fail to detect actual in-focus particles due to brightness fluctuation. Therefore, selecting an optimal background threshold involves a trade-off between minimizing false detections and maximizing particle identification accuracy---making it a critical parameter in image preprocessing. 

\begin{figure}[htbp]
    \centering
    \begin{subfigure}[b]{0.3\textwidth}
        \includegraphics[width=\linewidth,height=1.56in]{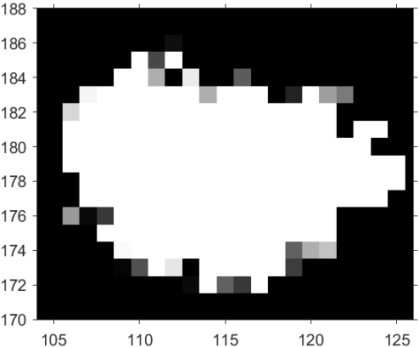}
        \caption{$T$}
        \label{fig:7a}
    \end{subfigure}
    \hfill
    \begin{subfigure}[b]{0.3\textwidth}
        \includegraphics[width=\linewidth,height=1.56in]{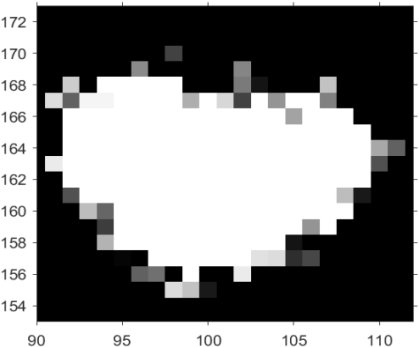}
        \caption{$T + \Delta t$}
        \label{fig:7b}
    \end{subfigure}
    \hfill
    \begin{subfigure}[b]{0.3\textwidth}
        \includegraphics[width=\linewidth,height=1.56in]{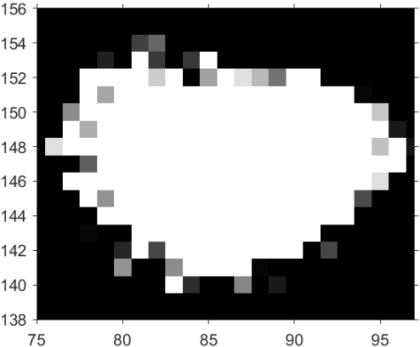}
        \caption{$T + 2\Delta t$}
        \label{fig:7c}
    \end{subfigure}

    \vspace{1em}

    \begin{subfigure}[b]{0.3\textwidth}
        \includegraphics[width=\linewidth,height=1.56in]{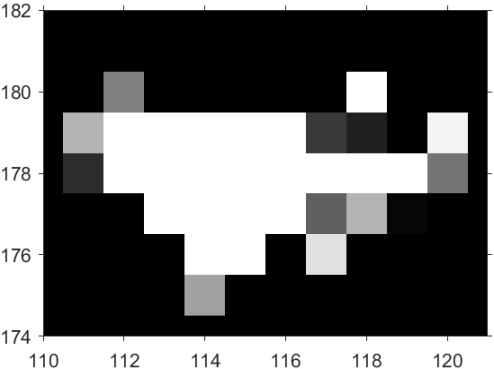}
        \caption{$T$}
        \label{fig:7d}
    \end{subfigure}
    \hfill
    \begin{subfigure}[b]{0.3\textwidth}
        \includegraphics[width=\linewidth,height=1.56in]{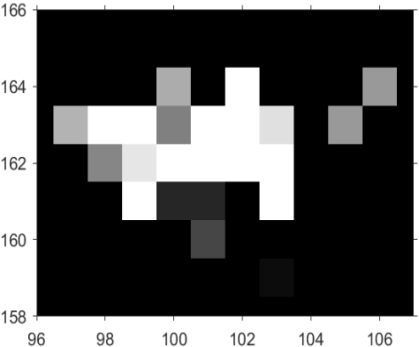}
        \caption{$T + \Delta t$}
        \label{fig:7e}
    \end{subfigure}
    \hfill
    \begin{subfigure}[b]{0.3\textwidth}
        \includegraphics[width=\linewidth,height=1.56in]{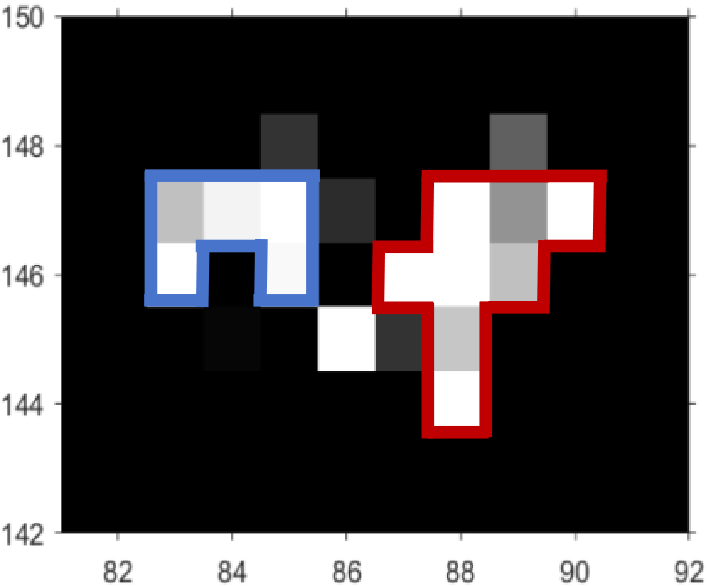}
        \caption{$T + 2\Delta t$}
        \label{fig:7f}
    \end{subfigure}

    \caption{Particle splitting under different background thresholds.  
    Top row (a--c): baseline threshold; bottom row (d--f): elevated threshold.  
    Columns correspond to consecutive time steps $T$, $T + \Delta t$, $T + 2\Delta t$.  
    In panel (f), one particle is erroneously split into two due to the higher threshold, causing a mismatch.}
    \label{fig:particle_splitting}
\end{figure}

A typical case of such threshold-induced particle splitting is shown in Fig.~\ref{fig:particle_splitting}. In this work, the particle detection algorithm defines a valid particle as a cluster of pixels connected to the group via at least one of its four sides (the 4-connected neighborhood criterion). Although the two bright regions in Fig.~\ref{fig:7f} appear visually contiguous, they share only a single corner pixel, which does not satisfy the 4-connected requirement. Consequently, the algorithm recognizes them as two distinct particles. This example demonstrates how increasing the threshold can artificially break pixel connectivity in blurred particle images, resulting in false particle splitting. Such mismatches arising from non-ideal thresholding are hereafter referred to as \textbf{TIF}, denoting thresholding-induced particle fragmentation.

\begin{figure}[htbp]
	\centering
	\includegraphics[width=2.65in, height=1.97in]{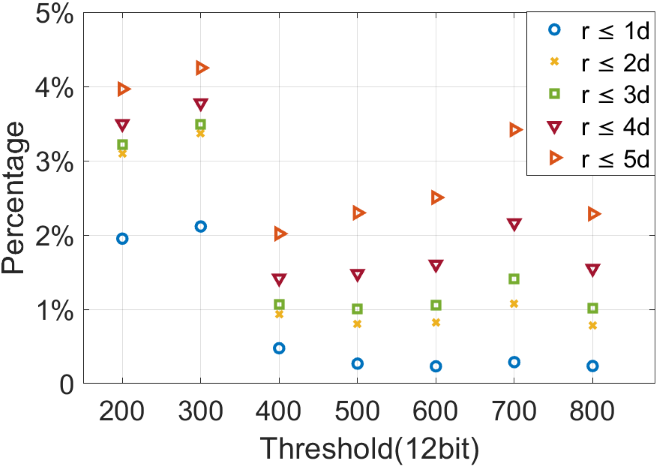}
	\caption{Effect of threshold on particle pair distribution.Percentage of particle pairs within different distance as a function of the brightness threshold (12-bit). Circles: $r \leq 1d$; crosses: $r \leq 2d$; squares: $r \leq 3d$; triangles: $r \leq 4d$; upward triangles: $r \leq 5d$. Here, $r$ denotes the three-dimensional distance between two particles. The results show that the proportion decreases as the threshold increases and exhibits two local minima, occurring around threshold values of 600 and 800.}
	\label{fig:threshold}
\end{figure}

To mitigate particle splitting, optimizing the background intensity threshold is essential. Local adjustment of the gray-level threshold can significantly reduce such events. In this study, the dataset from Exp.~1 was processed using thresholds of 200, 300, 400, 500, 600, 700, and 800, and the proportion of particle pairs at different inter-particle distances was calculated. The results are presented in Fig.~\ref{fig:threshold}. It can be observed that, as the threshold increases, the proportion of particle pairs with distance $\leq 2d$ (where $d$ is the particle diameter) exhibits a distinct trend: when the threshold is too low, weak gray regions are misidentified as independent particles, leading to a higher proportion of false splitting; when the threshold is too high, the original particle boundaries are truncated, which also introduces additional splitting. In contrast, within an intermediate threshold range (approximately 600--800), the proportion reaches its minimum, indicating the most stable particle detection. 

Considering the experimental conditions, a threshold of 20\%--30\% of the average background intensity is adopted (for the Exp~1, the average background intensity is about 2300, corresponding to thresholds of 600--800). This strategy effectively reduces false segmentation, thereby providing a solid basis for subsequent analysis. It is important to note that the threshold discussed here refers to the threshold used for particle tracking procedure, which is distinct and independent of the dual-threshold used in determination of the particle size (Sec~\ref{subsec2.3}). This distinction is deliberate as each of these tasks has different requirements and constraints.

\subsection{Interpolation Induced Spurious Particle (IIS)}\label{subsec3.3}
During the particle-tracking stage, a particle may disappear in a single frame due to occlusion, or detection failure. To maintain temporal continuity, most tracking algorithms interpolate the particle position between two consecutive detections if only one or a few is missing. Specifically, the instantaneous velocity is estimated from two successive detections at $t_1$ and $t_2$ :
\begin{equation} \label{eq:velocity_estimate}
	\mathbf{v}(t_2) = \frac{\mathbf{x}(t_2) - \mathbf{x}(t_1)}{t_2 - t_1},
\end{equation}
and the predicted position for the missing frame at $t_3$ is computed as:
\begin{equation} \label{eq:predicted_position}
	\mathbf{x}_{\mathrm{pred}}(t_3) = \mathbf{x}(t_2) + \mathbf{v}(t_2)\,(t_3 - t_2),
\end{equation}

\begin{figure}[htbp]
	\centering
	\includegraphics[width=4.0in,height=0.7in]{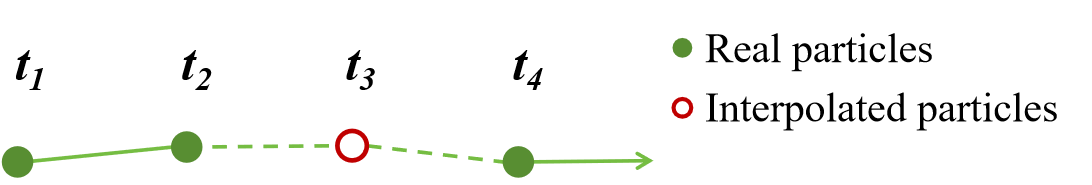}
	\caption{Interpolated particle. Correctly matched particles are shown as green filled circles. Red hollow circles denote inferred particles that were artificially introduced to maintain the continuity of the trajectories. \textit{$t_{2}-t_{1} = t_{3}-t_{2} = t_{4}-t_{3} = \Delta t$}.}
	\label{fig:interpolated_particle}
\end{figure}

This procedure preserves the temporal continuity of particle trajectories. A typical example is shown in Fig.~\ref{fig:interpolated_particle}. Possible error arising from trajectory interpolation are hereafter referred to as \textbf{IIS}, denoting interpolation-induced spurious particles.

Each interpolated point is explicitly flagged in the output with \(\texttt{interpolated}=1\), indicating a ``suspected spurious particle''. The advantage of this method is that it effectively prevents the fragmentation of particle trajectories caused by short-term detection loss \citep{Ouellette2006}, thereby preserving the temporal coherence of the data. However, it should be emphasized that these interpolated particles are not treated as real measurements. Instead, they serve as placeholders that allow the trajectory to be reconstructed smoothly. As the interpolation flag is preserved during the tracking stage, the flagged particles can be readily excluded from subsequent analyses to avoid introducing non-physical artifacts.

\section{Analysis and Mitigation of False Matching Induced Spurious Particles}\label{sec4}
As discussed above, spurious particles may arise due to projection ambiguities during the stereo-matching process. To quantify this effect, 20 datasets were randomly selected from the representative case (42\,k~rpm, small particles). The focus was placed on spurious particles generation of the FMIS type. Here, spurious particles are identified at the trajectory level rather than from single-frame image. A particle is classified as spurious if it satisfies at least one of the following criteria: (i) it exhibits abrupt appearance or disappearance without physically plausible
motion, or (ii) it undergoes artificial merging or splitting behavior inconsistent across the three camera views. Statistical results show that such error are most frequent when the inter-particle distance satisfies $1.5d<r<2.5d$, with approximately 26\% of detected particles in this range identified as spurious. For $r<1.5d$, most spurious particles originate from TIF. IIS since particle image segmentation is restricted to such small scale regime.

In order to mitigate adverse effect of FMIS and at the same time further reveal its spatial characteristics, we calculated the angles between the particle pair displacement vectors and the principal coordinate planes (X-Z, X-Y, and Y-Z planes). Specifically, given two particles $P_i = (x_1, y_1, z_1)$ and $P_j = (x_2, y_2, z_2)$, their displacement vector is defined as
\begin{equation} \label{eq:displacement_vector}
	\vec{r}_{ij} = P_i - P_j
\end{equation}
To evaluate the orientation of this vector relative to the coordinate planes, we make use of the plane-normal vectors
$n_{\scriptscriptstyle XZ} = (0,1,0)$, 
$n_{\scriptscriptstyle XY} = (0,0,1)$, and 
$n_{\scriptscriptstyle YZ} = (1,0,0)$, corresponding to the XZ, XY, and YZ planes, respectively. Below we define $\theta$ to denote the angle between the particle-pair displacement vector and the corresponding coordinate plane, evaluated via its normal vector such that:
\begin{equation} \label{eq:angle}
	\theta = 90^{o} - \cos^{-1}\!\left( 
	\frac{\vec{r}_{ij} \cdot n}{\|\vec{r}_{ij}\| \, \|n\|} 
	\right).
\end{equation}

\begin{figure}[htbp]
	\centering
	\includegraphics[width=2.65in, height=1.95in]{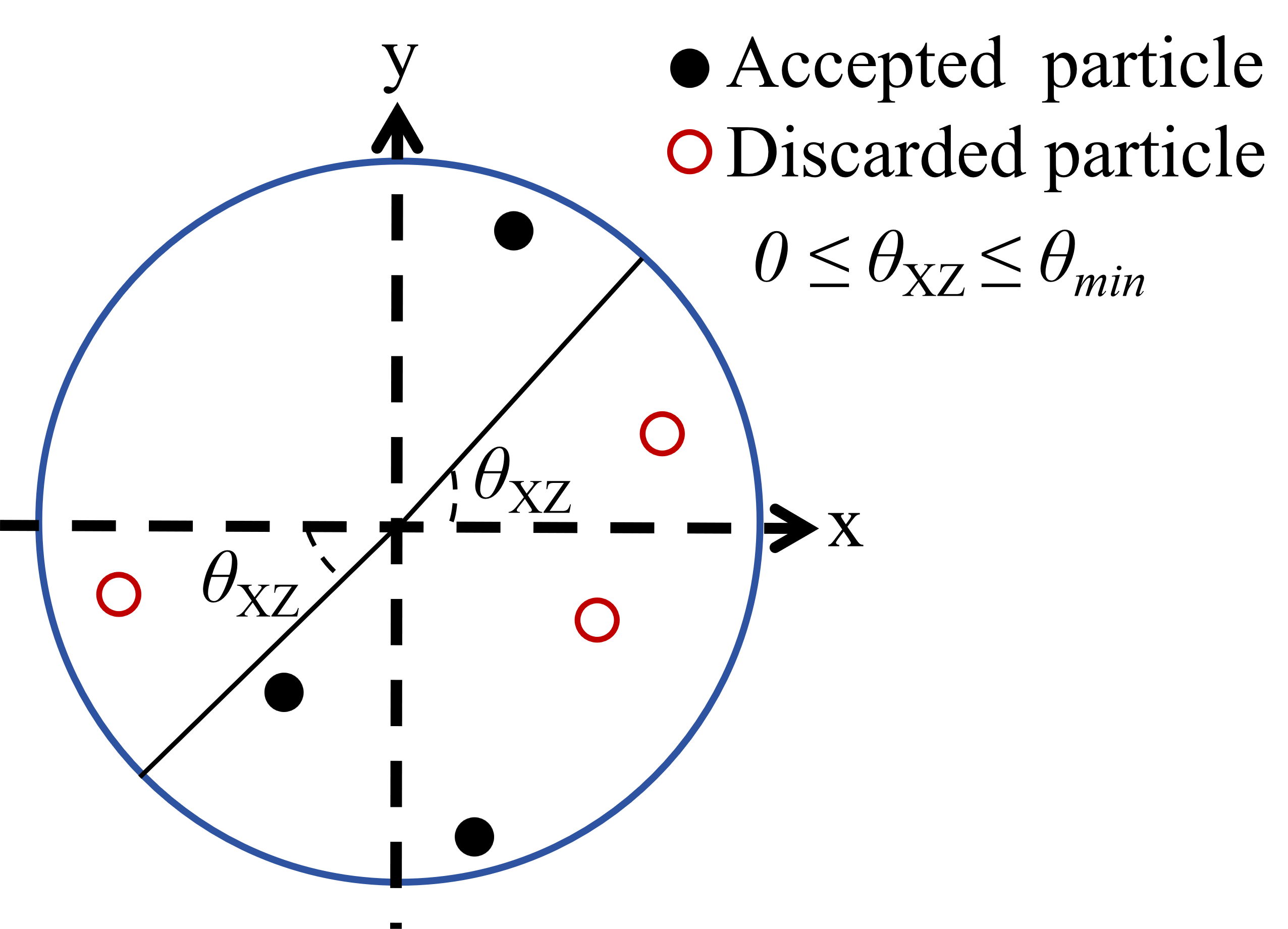}
	\caption{Particle pairs are filtered according to their orientation angle $\theta$ relative to the X-Z plane. Only particle pairs with \textit{$\theta_{XZ}$} exceeding a predefined threshold ($\theta_{XZ}$ $>$ $\theta_{\min}$) are accepted (black dots), while particles with \textit{$\theta_{XZ} \le \theta_{\min}$} are discarded (red hollow circles).}
	\label{fig:orientation_filtering}
\end{figure}

Owing to the specific configuration of the cameras in our setup, we focus on the  orientation angle $\theta_{XZ}$, i.e., the angle measured from the horizontal plane on which our camera constellation lies. We make use of this angle to define a criterion that would allow us to filter out the effect of the FMIS error from the data. Here, we demonstrate the criterion's efficacy for the case of RDF calculation. An illustration of the angle definition and the subsequent filtering criterion is shown in Fig.~\ref{fig:orientation_filtering}. Specifically, the filtering criterion is define as: particle pairs whose orientation angle ~$\theta_{XZ}$ is smaller than a prescribed threshold~$\theta_{\min}$ are considered unfavorable and are therefore discarded, whereas pairs with $\theta_{XZ} > \theta_{\min}$ are retained. The rationale for this criterion is based on two points. Firstly, due to the co-planar configuration of our cameras on the horizontal (XZ) plane, a FMIS-type spurious particle have a strong tendency to appear at positions such that its displacement relative to a nearby real particle is small (of order of $d$) and at a small angle from the XZ-plane. Secondly, owing to isotropy of small scale turbulence, the resultant clustering of inertial particles is also known to be isotropic (i.e., without preferred orientation) when external forces with symmetry-breaking tendency are absent. By sacrificing small angles pair statistics, this method allows us to filtered out the FMIS-type artifacts from the RDF result.

\begin{figure}[htbp]
	\centering
	\includegraphics[width=3.0in,height=2.3in]{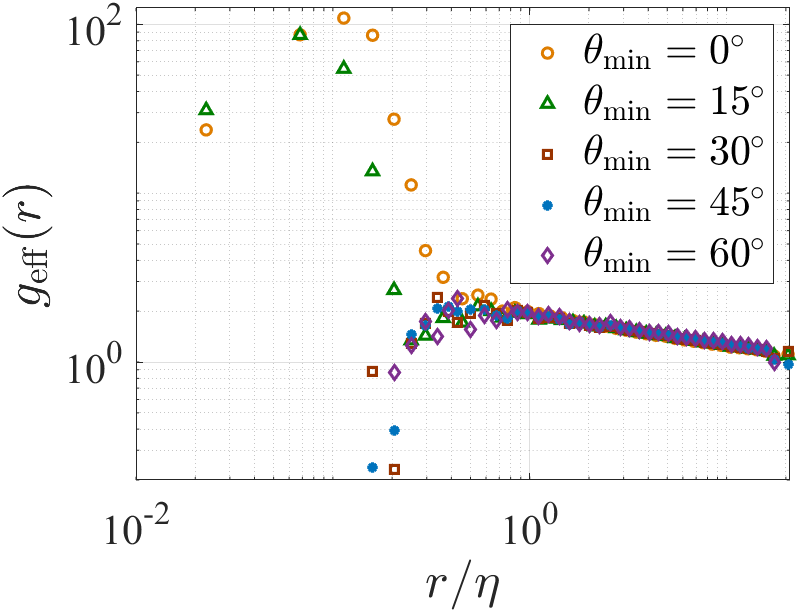}
	\caption{Effect of orientation angle filtering on the radial distribution function $g_{\mathrm{eff}}(r)$ of particle pairs from Exp~1. The RDF is plotted for particle pairs filtered by different minimum orientation angle thresholds $\theta_{\min}$, including $0^\circ$ (yellow circles), $15^\circ$ (green triangles), $30^\circ$ (brown rectangles), $45^\circ$ (blue stars), and $60^\circ$ (purple diamonds).}
	\label{fig:rdf_angle_filtering}
\end{figure}

Several angular threshold $\theta_{\min}$ ($0^\circ$, $15^\circ$, $30^\circ$, $45^\circ$, $60^\circ$) were applied to the data of 42\,k~rpm (small particles), thereby allowing us to systematically examine the effect of increasing angular thresholds on the particle RDF. A lower threshold ($0^\circ$ or $15^\circ$) serves as a baseline without strong geometric exclusion, while higher thresholds ($30^\circ$--$60^\circ$) progressively remove particle pairs with orientations close to the X-Z plane, which are more likely to be affected by the FMIS type error owing to the chosen configuration of our cameras. Fig.~\ref{fig:rdf_angle_filtering} presents the radial distribution function $g_{\mathrm{eff}}(r)$, obtained under these angular thresholds. Several key observations can be made:

\begin{enumerate}
    \item[1] \textbf{Pronounced small-scale clustering at low thresholds:} Without filtering (0°) or with a low threshold (15°), $g_{\mathrm{eff}}(r)$ exhibits a sharp rise at small separations (\textit{r/$\eta$} = 0.01--0.1), with peak values exceeding $10^2$. Such extreme peaks likely reflect spurious clustering caused by false stereo-matching, i.e., the FMIS type error, as defined in \ref{subsec3.1}.

    \item[2] \textbf{Effectiveness of angular filtering:} Increasing the threshold to 30°, 45°, and 60° significantly reduces $g_{\mathrm{eff}}(r)$ in the small-scale region, suppressing the extreme peak, {while has little to no effect on the RDF values at larger separations.} As discussed above, since turbulent-induced inertial clustering of  particle is known to be isotropic (orientation independent) in the absence of any symmetry-breaking external forces, this observation strongly suggests that particle pairs with small separations and orientations close to the X–Z plane are more likely artifacts of FMIS, and the angular filtering effectively mitigates the effects of these artifacts, recovering the true physical RDF signature of particles.

    \item[3] \textbf{Negligible impact on large scales:} For $r \gg d$, RDF curves under all thresholds converge, consistent with the understanding that (FMIS) spurious particles primarily affect small-scale statistics and have limited impact on larger scales.
\end{enumerate}

These results demonstrate that particle pairs with $\theta_{XZ}$ less than 30° are most susceptible to FMIS spurious particle effects. Accordingly, applying an angular threshold of at least 30° improves the systematic accuracy of the RDF. In this study, a 45° threshold is adopted for all subsequent analyses to balance the removal of spurious contributions while retaining sufficient data for statistical convergence.

\section{Measured Particle Dynamical Statistics}\label{sec5}
In order to explore the effectiveness of the current experimental methodology, we experimentally investigates the particle radial distribution function (RDF) and the normalized pseudo collision rate under different experimental conditions. 

\subsection{Particle pseudo collision rate}\label{subsec5.1}
To quantify the frequency of close-contact events between inertial particles, we introduce a pseudo-collision rate, $R_{pc}$. Instead of measuring how often particle physically collide, $R_{pc}$ measures how often particle pairs come within a distance closer than a prescribed threshold. The pseudo-collision rate was computed in two steps. First, we define the raw pseudo-collision rate:
\begin{equation} \label{eq:Rraw}
	R_{\text{raw}} = \frac{N_{c,i}}{T}
\end{equation}
where $N_{c,i}$ denotes the number of close-contact events below a set threshold, and $T$ represents the total sampling duration ($T = 11.94s$ in this study). 
Second, the normalized pseudo-collision rate is obtained as:
\begin{equation} \label{eq:Rpc}
	R_{pc} = \frac{R_{\text{raw}}}{N_p^2} = \frac{N_{c,i}}{T \times N_p^2}
\end{equation}
where $N_p$ is the average number of observed particles per frame. Normalization by $N_p^2$ allows comparison across datasets with different particle concentrations, ensuring that $R_{pc}$ reflects the intrinsic tendency for particle pairs to approach each other. This normalization follows common practice for pair-based statistics and is adopted here as an operational measure to facilitate comparison. To make the pseudo-collision rate dimensionless, we multiply by $\tau_\eta$, and introduce the dimensionless pseudo-collision rate:
\begin{equation} \label{eq:Rstar}
	R_{pc}^* = R_{pc} \, \tau_\eta.
\end{equation}

Only post-processed particle trajectories reconstructed using the velocity-prediction-based post-tracking algorithm were considered (see Sec.~\ref{subsec2.4}). Particles identified as interpolation-induced (IIS; Sec.~\ref{subsec3.3}) were excluded, and additional distance- and orientation-based filters were applied to suppress spurious proximity events related to false-stereo-matching (details in Sec.~\ref{sec4}). Within any 20-frame temporal window, only the first approach between any particle pair was counted as a valid event.

\begin{figure}[htbp]
	\centering
	\begin{subfigure}[c]{0.48\textwidth}
	   \includegraphics[width=\textwidth]{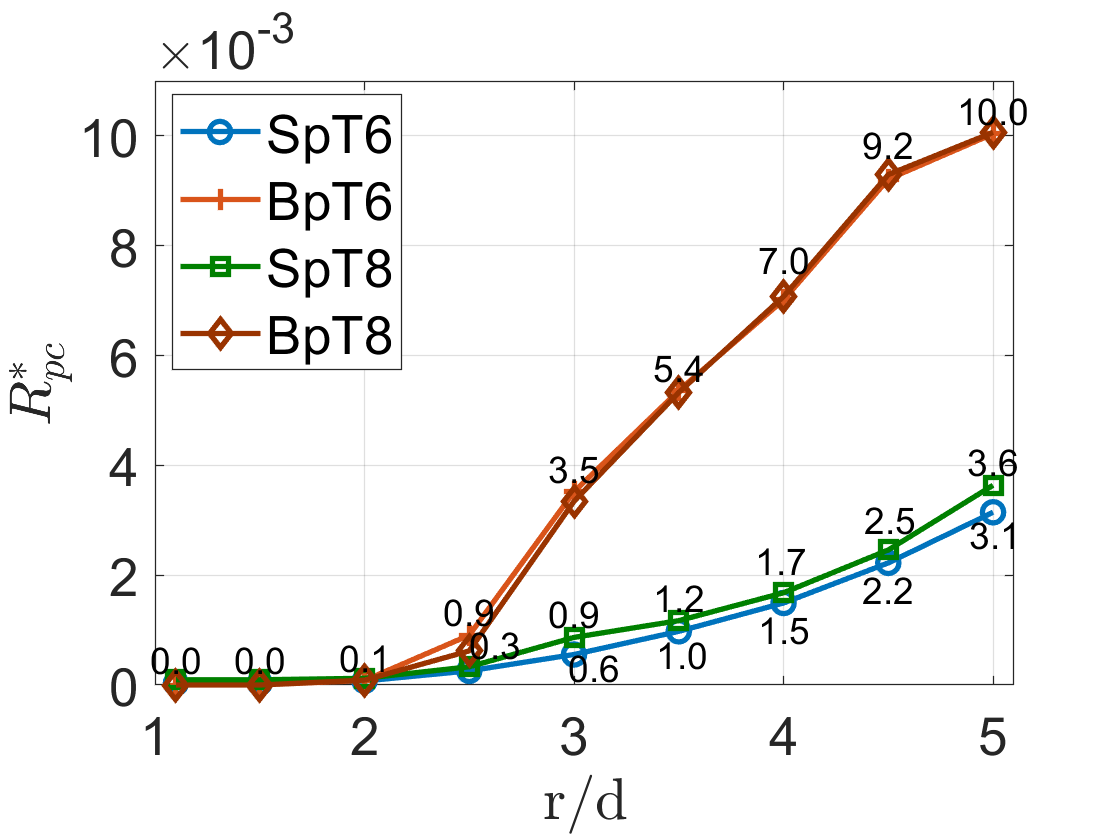}
	   \caption{$R_{pc}^*$}
   \end{subfigure}
	\hfill
	\begin{subfigure}[c]{0.48\textwidth}
		\includegraphics[width=\textwidth]{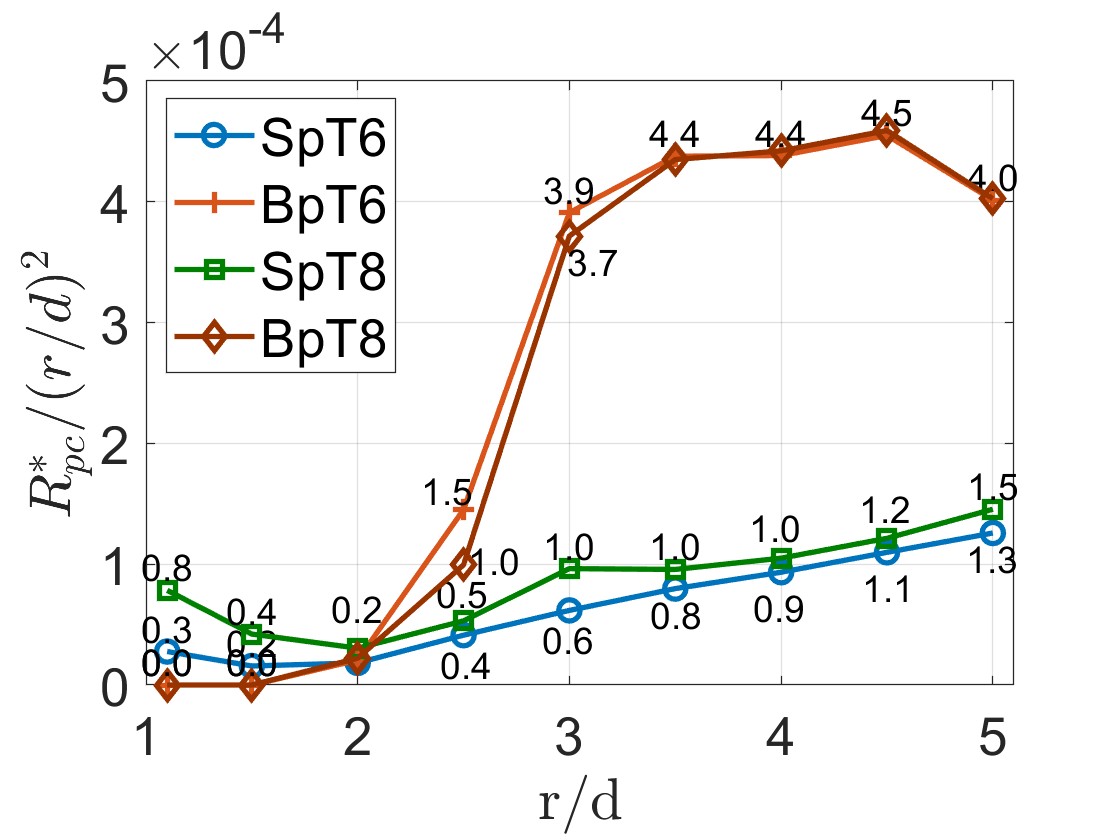}
		\caption{$R_{pc}^*/(r/d)^2$}
	\end{subfigure}
	\caption{Panel (a), Normalized pseudo-collision rate $R_{ pc}^*$. Panel (b), geometrically corrected counterparts, $R_{pc}^*/(r/d)^2$. These are plotted as functions of dimensionless separation $r/d$ for small particles ($17.6~\mu$m, Sp) and large particles ($33.6~\mu$m, Bp), processed using image binarization thresholds of 600 (T6) and 800 (T8); such that the combined labels SpT6, SpT8, BpT6, and BpT8 denote the corresponding particle size–threshold combinations. Results are shown after applying post-processing filters to remove IIS and FMIS types spurious particles. }
   	\label{fig:Rpc}
\end{figure}

Fig.~\ref{fig:Rpc} presents the normalized pseudo-collision rate $R_{pc}^*$ and its geometrically compensated form, $R_{pc}^*/(r/d)^2$, as a function of the dimensionless separation distance $r/d$. The two representations are shown to separate geometric effects from clustering-induced enhancements. The analysis compares two particle sizes and two image binarization thresholds. Four representative cases (SpT6, BpT6, SpT8, BpT8) correspond to small particles ($17.6\mu\mathrm{m}$) and large particles ($33.6\mu\mathrm{m}$), processed under image binarization thresholds of 600 and 800. As discussed in Sec.~\ref{sec3}, we see that only TIF type spurious particles are directly affected by the binarization threshold; (IIS) and (FMIS) particles remain unaffected.

As shown in Fig.~\ref{fig:Rpc}(a), $R_{pc}^*$ increases monotonically with inter-particle separation $r/d$ for all cases, primarily due to the geometric growth of the cross-section area of the imaginary spherical shell associated with the pseudo-collision. A clear segregation based on particle size is observed: the large particles (BpT6, BpT8) exhibit higher $R_{pc}^*$ values than the small particles (SpT6, SpT8) across the measured range. In contrast, variations in the image binarization threshold produce only minor changes, indicating that $R_{pc}^*$ is relatively insensitive to TIF spurious particles under the present processing conditions.

To reduce the influence of geometric cross-section growth, $R_{pc}^*$ is normalized by $(r/d)^2$, and the result is shown in Fig.~\ref{fig:Rpc}(b). At small separations ($r/d \lesssim 2.5$), the normalized radial distribution functions for all particle groups remain relatively low and similar, which may be attributed to direct particle interactions (such as collisions or repulsion) and the strong correlation of the flow at this scale, both of which limit random clustering. Beyond $r/d \approx 3$, the compensated rates seem to approach a plateau, indicating random, kinetic theory like, relative trajectories.

\subsection{Radial Distribution Function (RDF)}\label{sec5.2}

Here we present experimentally measured (RDFs) of inertial particles over a range of Stokes numbers, with particular emphasis on near collisional scale clustering and its statistical convergence. The definition of radial distribution function has been introduced earlier. 

To facilitate comparison between different particle sizes, the Stokes number is evaluated using the Kolmogorov scales derived from the energy dissipation rate estimated using the second-order velocity structure function (see Sec.~\ref{subsec2.4}). Six cases were considered with the corresponding particle sizes and Stokes numbers given in  Table~\ref{tab:st_c1}.

As $r$ decreases, the sampling volume associated with each separation bin diminishes proportionally to $r^2 \Delta r$, which inevitably reduces the number of particle pairs available for statistics and limits the convergence of the RDF at small separations. As a result, RDF estimates at small $r/\eta$ are subject to increased statistical uncertainty. The uncertainty is quantified using the standard error of the mean (SEM),
\begin{equation} \label{eq:SEM}
		\mathrm{SEM}(r) = \frac{\sigma_s(r)}{\sqrt{N_{\mathrm{s}}}}
	\end{equation}
where $\sigma_s(r)$ denotes the standard deviation of the RDF values across independent samples (i.e., the different runs conducted for each experiment), at a particular separation ($r$) bin and $N_{\mathrm{s}}$ is the number of samples.  
This definition follows standard statistical practice for ensemble-averaged quantities \cite{Bendat2011}. 

\begin{figure}[htbp]
	\centering
	\subfloat[$42\,\mathrm{k\,rpm}$]{
		\includegraphics
		[width=0.48\linewidth, height=0.4\linewidth]{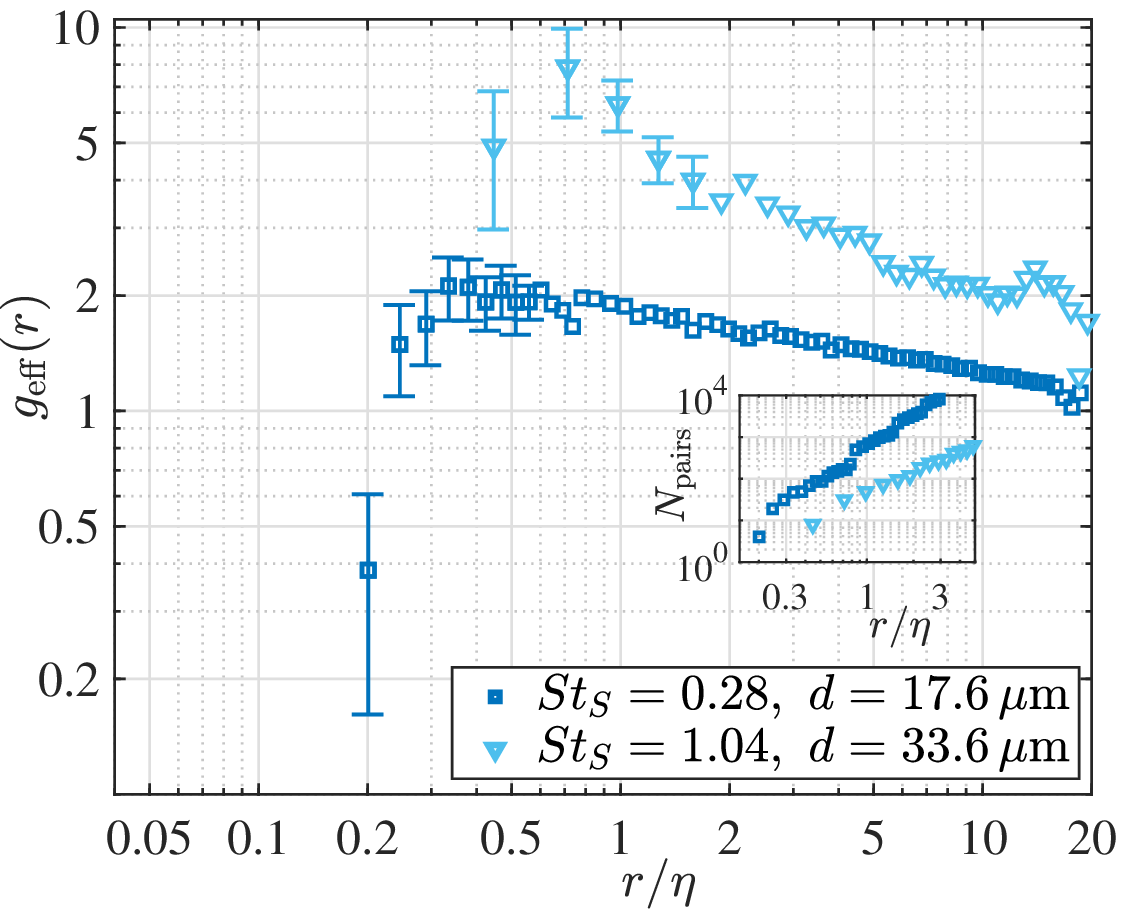}
		\label{fig:rdf_42k}
	}
	\hspace{-1em}
	\subfloat[$30\,\mathrm{k\,rpm}$]{
		\includegraphics[width=0.48\linewidth, height=0.4\linewidth]{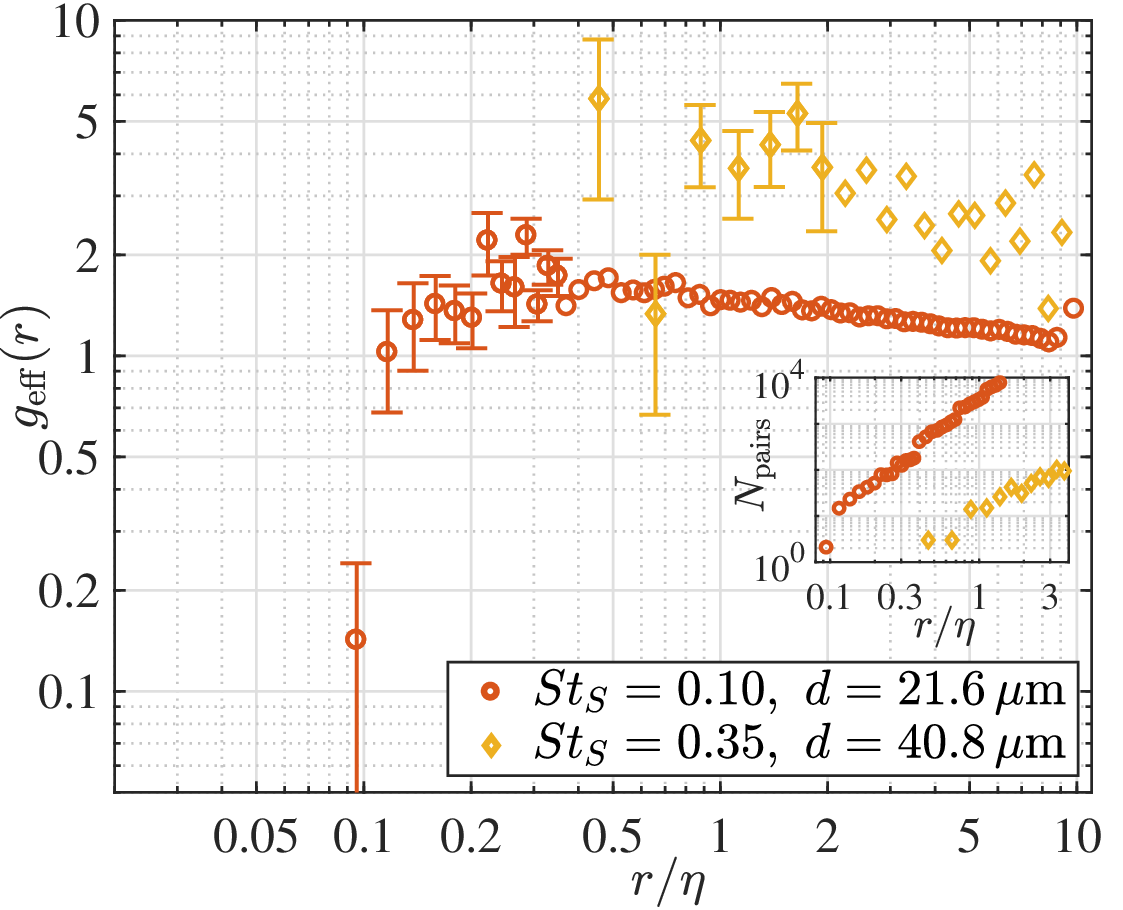}
		\label{fig:rdf_30k}
	}
	\vspace{0.6em}
	\subfloat[$18\,\mathrm{k\,rpm}$]{
		\includegraphics[width=0.48\linewidth, height=0.4\linewidth]{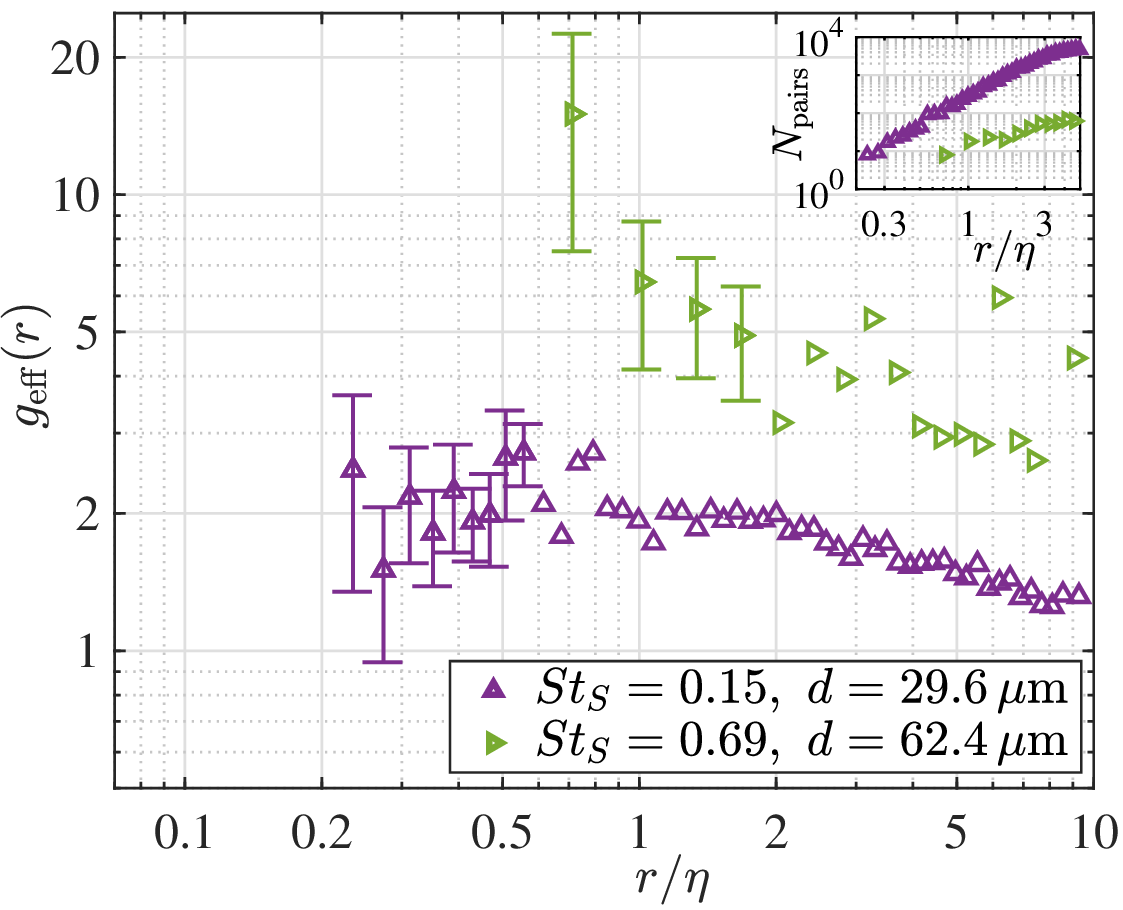}
		\label{fig:rdf_18k}
	}
	\hspace{-1em}
	\subfloat[Revised RDF comparison]{
		\includegraphics[width=0.48\linewidth, height=0.4\linewidth]{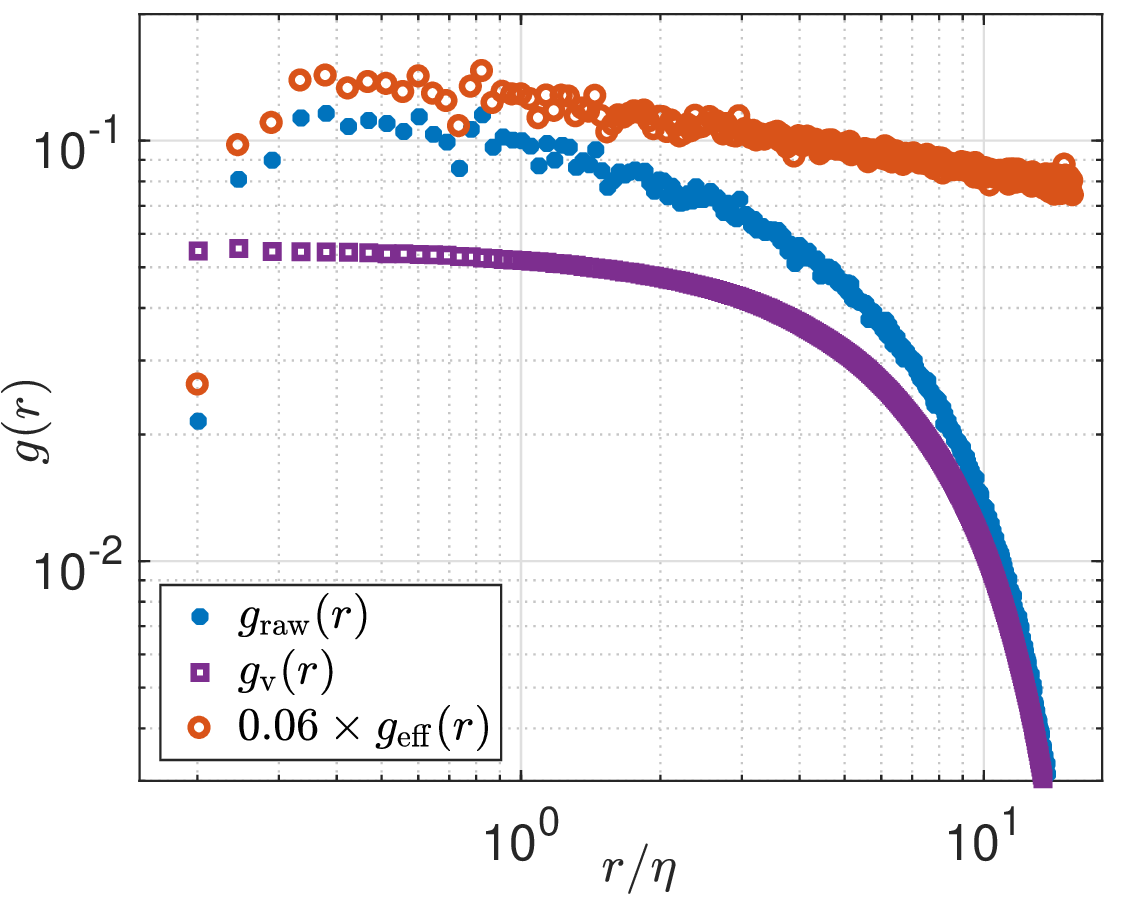}
		\label{fig:rdf_all}
	}
	
	\caption{Radial distribution functions $g_{\mathrm{eff}}(r)$ for inertial particles measured under different experimental conditions. Panels (a)--(c) correspond to experiments of 42k, 30k, and 18k\,rpm, respectively, each case includes two particle sizes representing different Stokes numbers. $d/\eta$ = (0.08, 0.15), (0.05, 0.09), (0.06, 0.12) for the (smaller, larger) particles in each of these panels respectively. Panel (d) shows the effectiveness of the method used to remove artifact (modulation) on the RDF due to finite observation volume (details in text) for the data from the 42k\,rpm experiment.}
	\label{fig:rdf}
\end{figure}

Fig.~\ref{fig:rdf} presents the measured RDFs normalized by the Kolmogorov length scale. To reduce cluttering in the final result, the RDF values across different $r$-bins were merged and averaged as $r$ increases. {Across all rotation speeds, the RDFs exhibit a distinct power-law scaling region, $g(r) \sim c_0 \left( r/\eta \right)^{-c_1}$, typically spanning $0.7 < r/\eta < 10$. where $c_1$ quantifies the intensity of particle inertial clustering.} {Panels (a)--(c) correspond to experiment of 42k, 30k, and 18k rpm, respectively, representing different turbulent forcing conditions. For each speed, two particle sizes are shown, resulting in different Stokes numbers under otherwise identical flow conditions. Here, $St_{\scriptscriptstyle\mathrm{S}}$ quantifies particle inertia based on flow statistics, whereas $c_1$ is an effective exponent obtained from RDF scaling; they are not operational derivatives of each other}.

For clarity, Table~\ref{tab:st_c1} summarizes all experimental cases discussed in this section, including the disk rotation speed, particle diameter, the Stokes number $St_{\scriptscriptstyle\mathrm{S}}$ estimated from the second-order velocity structure function (Sec.~\ref{subsec2.4}), and the clustering exponent $c_1$ obtained from the RDF scaling. 
The table provides a compact overview of how particle inertia and clustering intensity vary across the six cases and serves as a reference for the detailed discussion of the RDF behaviors shown in Fig.~\ref{fig:rdf}(a)--(d).

At highest turbulent intensity (42k\,rpm; Fig.~\ref{fig:rdf}(a)), the RDFs exhibit clear enhancement at small separations for both particle sizes, indicating strong clustering at dissipative scales. The larger particles ($St_{\scriptscriptstyle\mathrm{S}} = 1.04$, 33.6~\textmu m) display significantly higher RDF values and steeper decay with increasing $r/\eta$ than the smaller particles ($St_{\scriptscriptstyle\mathrm{S}} = 0.28$, 17.6~\textmu m), reflecting the increasing importance of inertial effects under strong turbulent forcing. For the smaller particles, at very small separations ($r/d \lesssim 3.8 $), a decrease from the peak is observed, while for the larger particles, the same trend is arguably also discernible but less compelling. This will be discussed in details in the Discussion section.

At the intermediate rotation speed (30k rpm; Fig.~\ref{fig:rdf}(b)), the overall clustering intensity is relatively reduced. Small particles ($St_{\scriptscriptstyle\mathrm{S}} = 0.10$, 21.6~\textmu m) exhibit more modest RDF enhancement around $\eta$ and the RDF also drops for particle separations below $r/d \approx 4.4$. The larger particles ($St_{\scriptscriptstyle\mathrm{S}} = 0.35$, 40.8~\textmu m) show similar with slightly steeper slope ($c_1$) but without a clear peak nor the subsequent decent as $r$ decreases towards smaller values.

Fig.~\ref{fig:rdf}(c) show the results of the 18k\,rpm experiment. We see that the RDF for the small particles ($St_{\scriptscriptstyle\mathrm{S}} = 0.15$, 29.6~\textmu m) also exhibit power-law scaling in the range $1 < r/\eta < 10$, but the trend is not clear for the larger particles due to significant statistical scatter. There is no clear sign of RDF drop off at the smallest $r$, which may be due to the poor statistical convergence (in this case, $d/\eta \approx 0.06 $).  It should be noted that at such low turbulent intensity. The flow is affected by the presence of a persistent, non-stationary large-scale vortex. Such a vortex may lead to biases in the estimates of dissipation rate $\epsilon$ and Stokes numbers. It may also affect the shape of the observed RDF.

Fig.~\ref{fig:rdf}(d) shows, from the case of 42k\,rpm, the raw observed RDF  $g_{\text{raw}}(r)$, RDF associated to the view volume $g_{\text{v}}(r)$ and the final effective RDF $g_{\text{eff}}(r)$ ($\equiv g_{\text{raw}}/g_{\text{v}}$) (all RDFs shown in the preceding panels are of this kind). We see that after division by the view-volume associated RDF, the final RDF shows a much cleaner power law trend. This figure showcases the effectiveness of this method for the removal of the finite-view-volume (or large-scale inhomogeneity) induced modulation on the RDF (further details in Sec.~\ref{subsec2.5}).

\begin{table}[htbp]
\centering
\caption{Summary of experimental conditions, particle Stokes numbers $St_{\scriptscriptstyle\mathrm{S}}$ estimated from the second-order velocity structure function, and the effective clustering exponent $c_1$ obtained from power-law fits of the RDFs.}
\label{tab:st_c1}
\begin{tabular}{cccccc}
\hline
Particle index & Disk speed (rpm) & $d$ ($\mu$m) & $d/\eta$ & $St_{\scriptscriptstyle\mathrm{S}}$ &  $c_1$ \\
\hline
1 & 42k & 17.6 & 0.08 & 0.28 & 0.24 \\
2 & 42k & 33.6 & 0.15 & 1.04 & 0.54 \\
3 & 30k & 21.6 & 0.05 & 0.10 & 0.18 \\
4 & 30k & 40.8 & 0.09 & 0.35 & 0.48 \\
5 & 18k & 29.6 & 0.06 & 0.15 & 0.31 \\
6 & 18k & 62.4 & 0.12 & 0.69 & 0.53 \\
\hline
\end{tabular}
\end{table}

\section{Discussion}\label{sec6}
In some experimental studies, an increase of the radial distribution function at separations in the order of tenths of particle diameter has been reported under specific (turbulent) flow and particle-property conditions \cite{Yavuz2018}. Such a feature is not observed in the present measurements. If the clustering enhancement reported in previous studies is assumed to be physically robust, one possible reason for its absent in our measurements could be electrostatic interactions between particles. This possibility cannot be completely excluded a priori, as particle charging has been shown to suppress particle clustering \cite{Lu2010}. However, several aspects of the present experimental configuration suggest that electrostatic effects are unlikely to play a dominant role here. Specifically, the particles are generated from a liquid atomization process in a humid air environment, where charge relaxation times are expected to be short, and no external electric fields are applied. Moreover, no anomalous long-range repulsion or aggregation behavior indicative of electrostatic interactions is observed in the measured particle trajectories. As particle charge was not directly measured in the present study, the role of electrostatic effects cannot be conclusively assessed. Addressing this question will require dedicated measurements and controlled charging conditions, which is beyond the scope of the present work.

Another mechanism that has been proposed as a cause of reduction of the radial distribution function at very small separations is the depletion of near-contact particle pairs caused by droplet coalescence \cite{Saw2022, Meng2023}. For coalescing droplets, a reduction in RDF is expected at separations below $r \approx 2.5d$ \cite{Saw2022}. Using the measured values of $d/\eta$ listed in Table~\ref{tab:st_c1}, the corresponding depletion scale can be estimated individually for each case according to the relation $r_{\text{de}}/\eta = 2.5\, d/\eta$. The resulting values are: 42k--small particles, $r_{\text{de}}/\eta = 0.20$;  42k--large particles, $r_{\text{de}}/\eta = 0.38$; 30k--small particles, $r_{\text{de}}/\eta = 0.13$;  30k--large particles, $r_{\text{de}}/\eta = 0.23$; 18k--small particles, $r_{\text{de}}/\eta = 0.15$;  and 18k--large particles, $r_{\text{de}}/\eta = 0.30$.

To enable a consistent comparison across different cases, we define the observed onset of RDF roll-off as the separation at which $g(r)$ systematically deviates from the observed intermediate-scale power-law scaling. 
In the 42k--small particles case, the deviation occurs at $r/\eta \approx 0.35$, while for 42k--large particles it occurs at $r/\eta \approx 0.70$ (the corresponding $r_{\text{de}}/\eta$ values are 0.20 and 0.38 respectively). In the 30k--small particles case, the roll-off starts at $r/\eta \approx 0.25$, whereas no statistically significant roll-off observed for the large particles case (the $r_{\text{de}}/\eta$ values are 0.13 and 0.23 respectively). For both 18k cases, a reliable roll-off cannot be established due to limited statistical convergence at the smallest separations. For all cases where a measurable deviation is observed, the separations at which the RDF decreases are significantly larger than the scale predicted from coalescence alone. For the cases where deviation is detectable, the observed roll-off scale is roughly twice as large as the predicted scale where depletion should start.  This suggests that coalescence-induced depletion may not be the sole mechanism responsible for the observed RDF reduction in the present experiments.

\section{Conclusions}\label{sec7}
This study addresses the experimental challenges of resolving inertial particle motion and statistics at near-contact separations in turbulent flows using three-dimensional Lagrangian particle tracking (LPT). The primary objective is to establish a reliable experimental methodology capable of producing physically consistent small-scale statistics. The key conclusions are summarized as follows:
\begin{itemize}
	\item[(1)] Three dominant sources of spurious particles relevant for high-resolution LPT were systematically identified: interpolation-induced artifacts (IIS), threshold-induced particle fragmentation (TIF), and spurious particle arising from false stereo matching (FMIS). While IIS and TIF have been discussed in previous LPT studies, the FMIS-type spurious particles are shown here to be intrinsically linked to the three-dimensional viewing geometry and found to be particularly severe when particle separations approach the particle triangulation limit. Importantly, these spurious detections are not random but exhibit structured spatial signatures that directly contaminate near-contact statistics such as the radial distribution function (RDF) and pseudo-collision rate.
	\item[(2)] To mitigate FMIS, an orientation-based geometric filtering criterion was introduced based on the co-planar(XZ) configuration of the three cameras. Owing to this viewing geometry, FMIS-induced spurious particles tend to form separation vectors that lie preferentially close to the XZ plane. An orientation angle $\theta_{XZ}$ was therefore defined to quantify this geometric bias, and particle pairs with $\theta_{XZ}$ below a prescribed threshold were excluded. This criterion is physically justified by the expected small-scale isotropy of inertial particle clustering in the absence of symmetry-breaking forces. Genuine particle pairs do not exhibit systematic orientation bias, whereas FMIS artifacts do. The filtering procedure thus selectively suppresses geometrically induced spurious pairs while preserving the statistically isotropic component of the true particle distribution. The effectiveness of the proposed methodology was validated through systematic filtering of the radial distribution function. %and 
	\item[(3)] Radial distribution function and normalized pseudo-collision rate over a range of Stokes numbers $St_{\scriptscriptstyle\mathrm{S}} \approx 0.2$--$1.0$ were measured. 
	Across all cases, the RDFs exhibit a well-defined intermediate-scale power-law scaling region, $g(r) \sim c_0 (r/\eta)^{-c_1}$, with a systematic dependence of the exponent $c_1$ on particle inertia. After suppression of spurious particles, no unphysical shoot up are observed at the smallest separations. At very small separations, a gradual roll-off of the RDF is observed in some cases. The characteristic scale of this deviation is found to be roughly twice the depletion zone scale previously found to be associated with droplet collision-coalescence.
	
\end{itemize}

This work demonstrates that the reliability of near-contact inertial particle statistics in high-density LPT measurements critically depends on the systematic identification and suppression of structured measurement biases. By explicitly addressing geometry-induced stereo ambiguities and resolution-related limitations, the present methodology establishes a physically consistent framework for extracting small-scale clustering statistics in turbulence. The resulting measurements exhibit robust scaling behavior and systematic dependence on particle inertia, while the deviations at the smallest separations suggest possible influence of coalescence induced depletion or weak electrostatic repulsion. More broadly, this study highlights the importance of coupling experimental design, geometric considerations, and statistical validation when probing collisional-scale particle dynamics. The proposed workflow provides a generalizable strategy for improving the fidelity of small-scale particle statistics in multiphase turbulence experiments. 

\section{Acknowledgements}
This work was mainly supported by the National Natural Science Foundation of China (Grant No. 11872382) and by the Thousand (Young) Talents Program of China.

\section*{Declarations}
\noindent\textbf{Funding}
This work was supported by the National Natural Science Foundation of China (Grant No. 11872382).

\noindent\textbf{Conflict of interest}
The authors declare that they have no conflict of interest.

\noindent\textbf{Ethics approval and consent to participate}
Not applicable.

\noindent\textbf{Consent for publication}
Not applicable.

\noindent\textbf{Data availability}
The datasets generated and analyzed during the current study are available from the corresponding author upon reasonable request.

\noindent\textbf{Materials availability}
Information about the experimental setup and materials is provided in the manuscript. Additional details are available from the corresponding author upon reasonable request.

\noindent\textbf{Code availability}
Custom codes used for data processing and analysis are available from the corresponding author upon reasonable request.

\noindent\textbf{Author contributions}
Conceptualization: Ewe-Wei Saw; Methodology: Ewe-Wei Saw, Linli Fu, Jun Feng, XiaoHui Meng; 
Investigation: Linli Fu, Ewe-Wei Saw, FanXi Gong, YaTing Chen, Jun Feng; Data analysis: Linli Fu, YaTing Chen, FanXi Gong, XiaoHui Meng; 
Writing – original draft: Linli Fu; 
Writing – review \& editing: Ewe-Wei Saw.

\begin{appendices}
	\section{Examples of events resulting in false stereo-matching induced spurious particle (FMIS)
	}\label{secA1}
	To survey the occurrence of FMIS, particles from 15 different time instances were randomly selected from the dataset at $St_{\scriptscriptstyle\mathrm{S}} = 0.28$. The selection spanned different time periods to ensure representative sampling. We identify three different families (type) FMIS events:\\ 
	Type I: One of the implicated particle is temporarily outside the field of view of a single camera, while the other particle remains visible in all three cameras.\\  
	Type II: All three cameras detect both particles, but in at least one camera, the projected distance between particles falls below the pre-set tolerance threshold, producing spurious stereo matches.\\ 
	Type III: All cameras detect both particles, but in one camera the particles trajects along a nearly collinear path relative to the camera's optical axis.
	
	The occurrence statistics of these cases are summarized in Table~\ref{tab:FMIS_statistics}. Type I events occurred in 7 out of 15 sampled times (47\%). The image sequence showcasing this type is shown in Sec.~\ref{subsec3.1} of the main text and will not be repeated here. Here, we provide examples for the Type II and Type III events, along with quantitative information on spurious particle generation and track fragmentation.
   
   \begin{table}[htbp]
   	\centering
   	\caption{Occurrence statistics of FMIS types from 15 randomly selected time at $St_{\scriptscriptstyle\mathrm{S}} = 0.28$.}
   	\begin{tabular}{lccc}
   		\hline
   		FMIS Type & Description & Occurrence & Probability \\
   		\hline
   		Type I & One particle absent in one camera  & 7 & 47\% \\
   		Type II & Full multi-view visibility & 3 & 20\% \\
   		Type III & Particle projection overlap & 5 & 33\% \\
   		\hline
   	\end{tabular}
   	\label{tab:FMIS_statistics}
   \end{table}
   
	\subsection{Type II: FMIS with full multi-view visibility}\label{sec:type2}
\begin{figure}[htbp]
		\centering
		\begin{subfigure}[b]{0.32\textwidth}
			\centering
			\includegraphics[width=\textwidth]{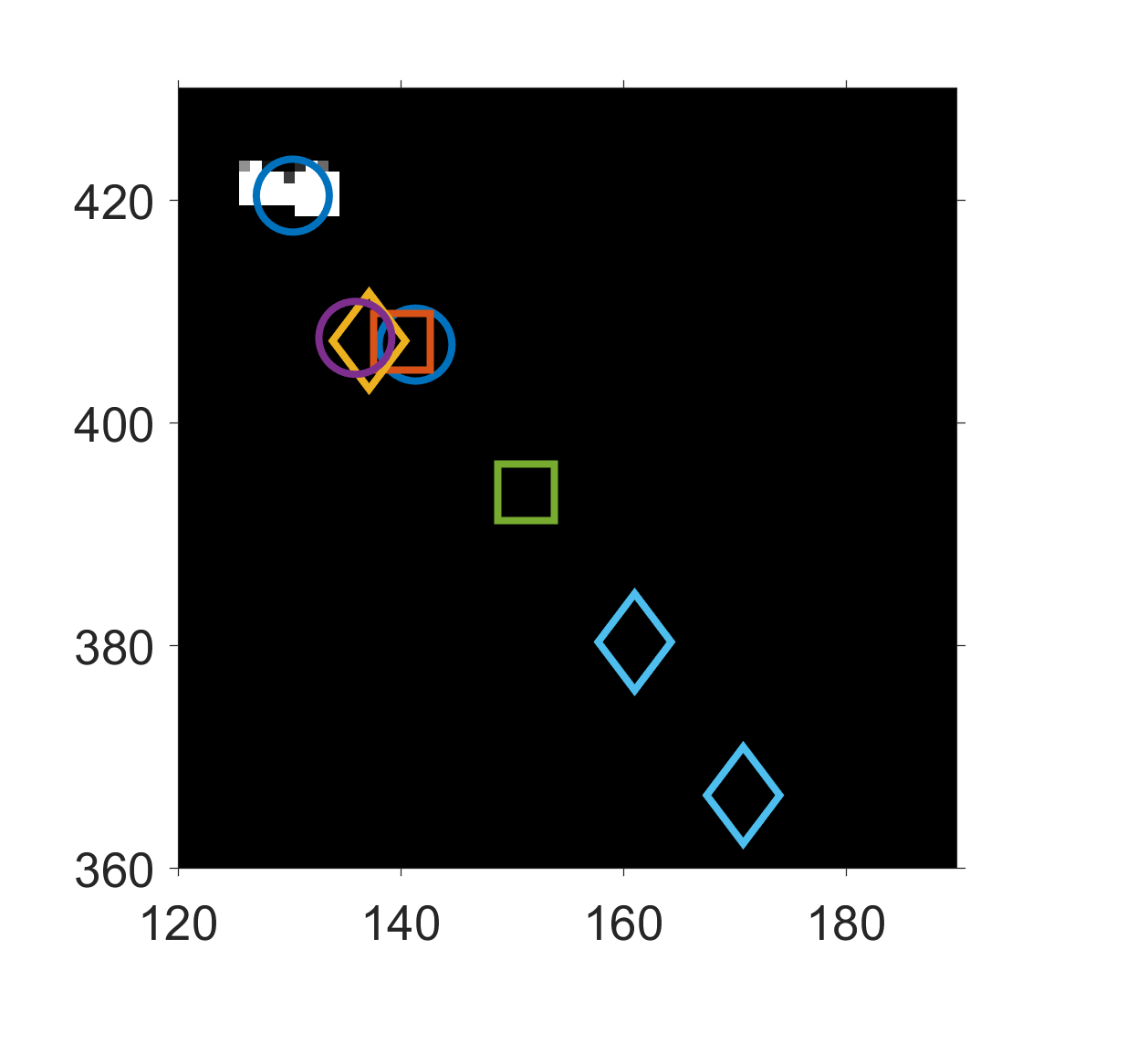}
			\caption{Camera\,1 at $T - 1\Delta t$}
			\label{fig:Cam1 at T - 1t_4808}
		\end{subfigure}
		\hfill
		\begin{subfigure}[b]{0.32\textwidth}
			\centering
			\includegraphics[width=\textwidth]{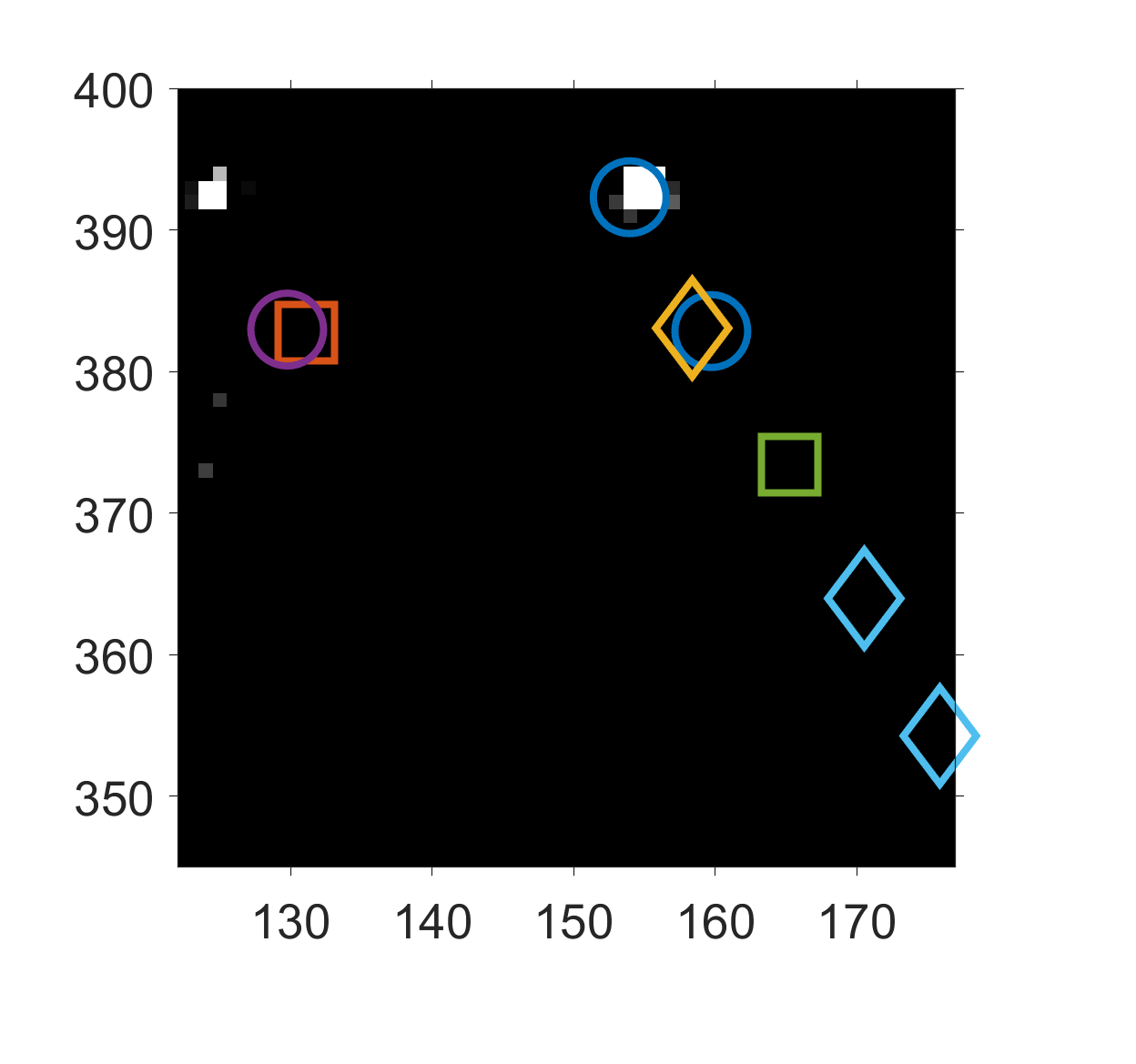}
			\caption{Camera\,2 at $T - 1\Delta t$}
			\label{fig:Cam2 at T - 1t_4808}
		\end{subfigure}
		\hfill
		\begin{subfigure}[b]{0.32\textwidth}
			\centering
			\includegraphics[width=\textwidth]{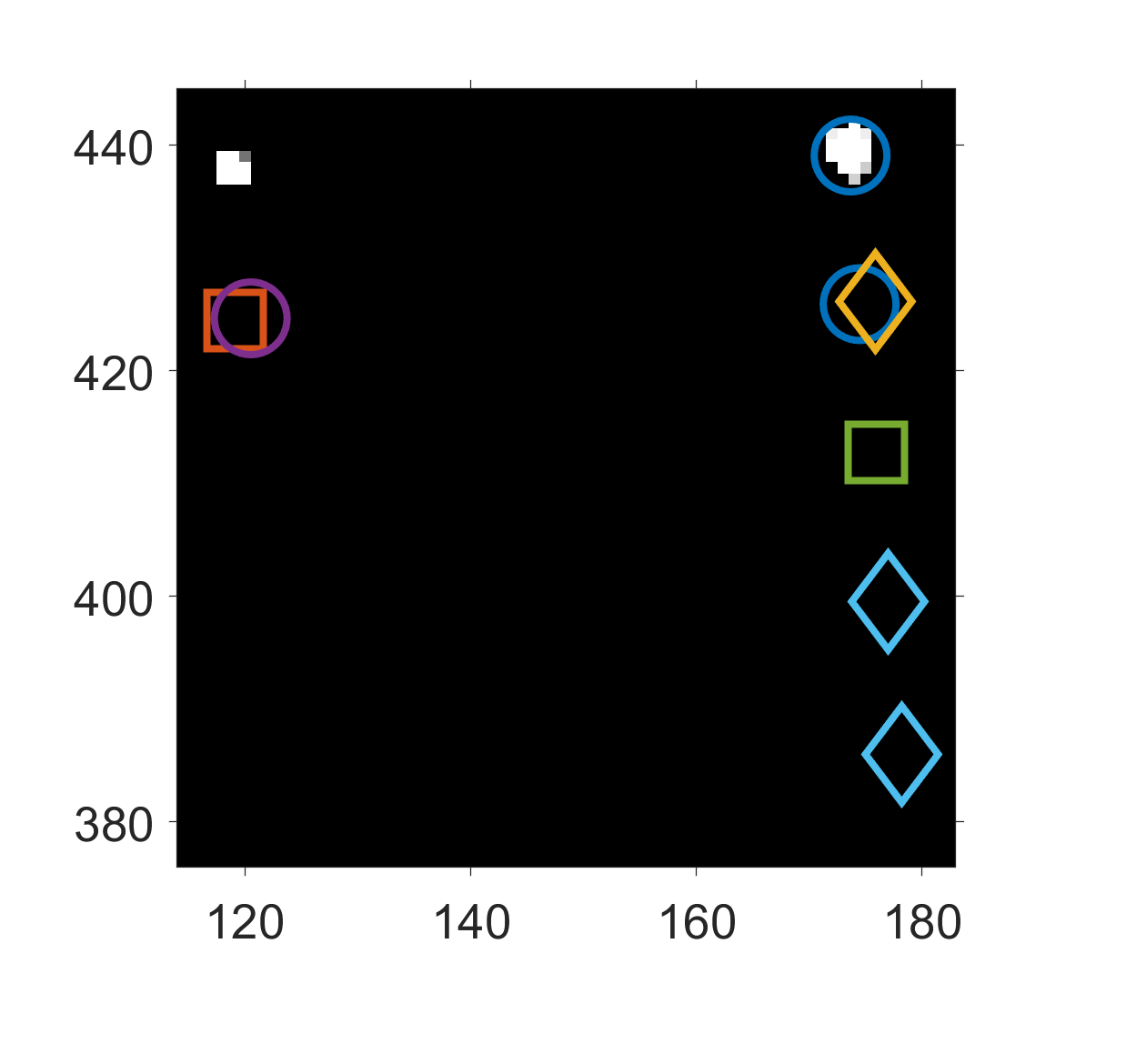}
			\caption{Camera\,3 at $T - 1\Delta t$}
			\label{fig:Cam3 at T - 1t_4808}
		\end{subfigure}
		\hspace{0.05\textwidth}
		\begin{subfigure}[b]{0.32\textwidth}
			\centering
			\includegraphics[width=\textwidth]{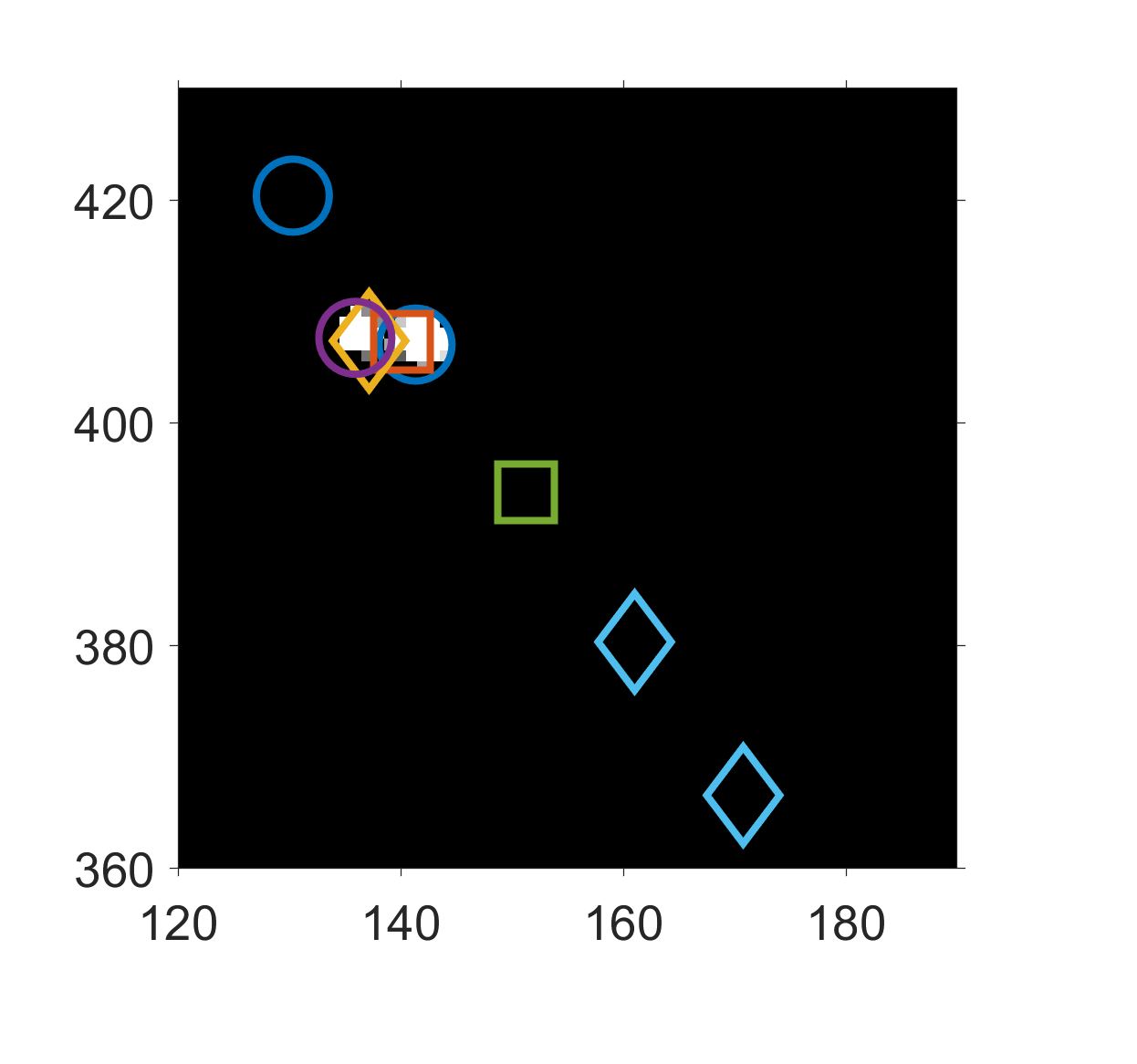}
			\caption{Camera\,1 at $T$}
			\label{fig:Cam1 at T_4808}
		\end{subfigure}
		\hfill
		\begin{subfigure}[b]{0.32\textwidth}
			\centering
			\includegraphics[width=\textwidth]{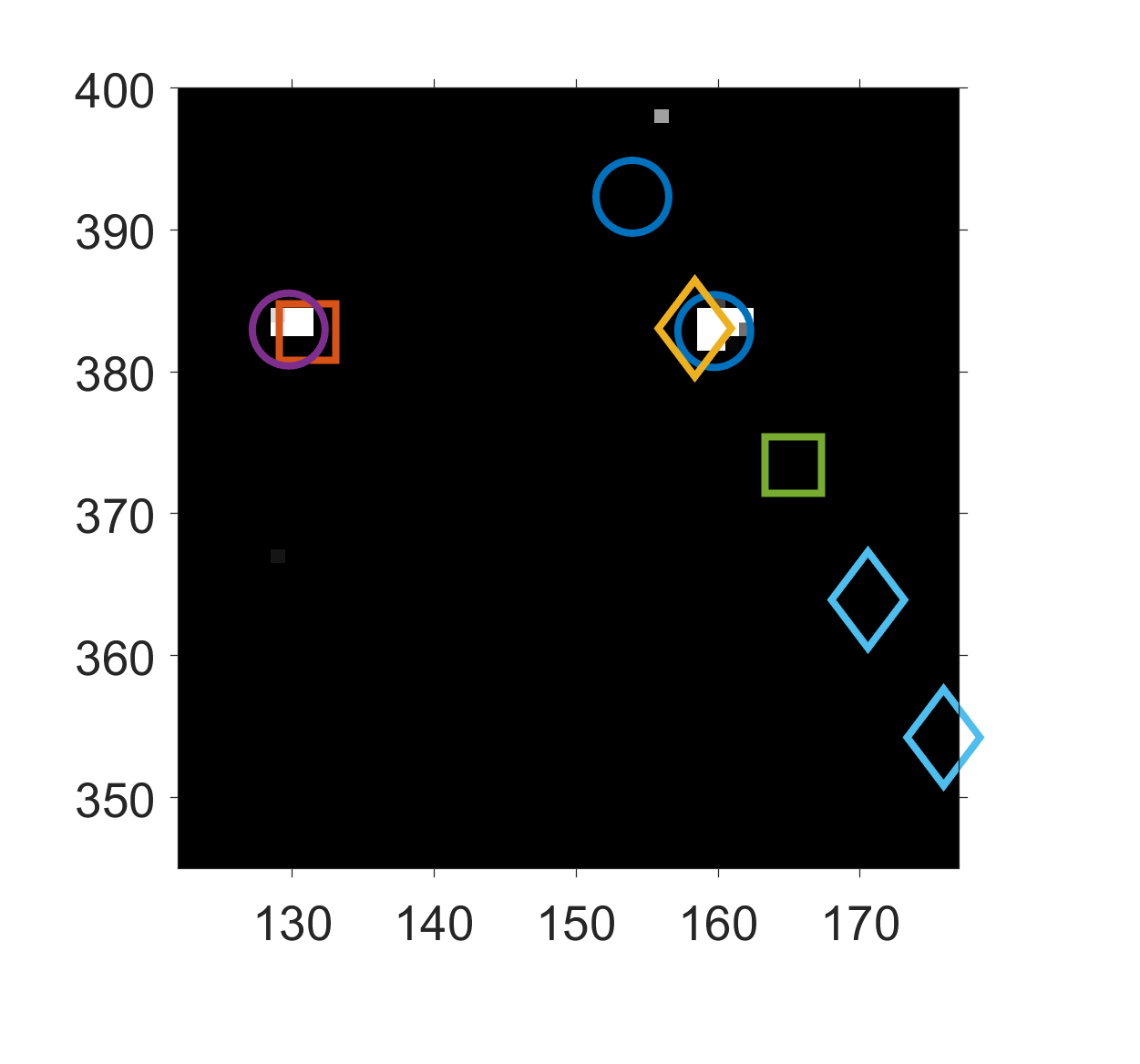}
			\caption{Camera\,2 at $T$}
			\label{fig:Cam2 at T_4808}
		\end{subfigure}
		\hfill
		\begin{subfigure}[b]{0.32\textwidth}
			\centering
			\includegraphics[width=\textwidth]{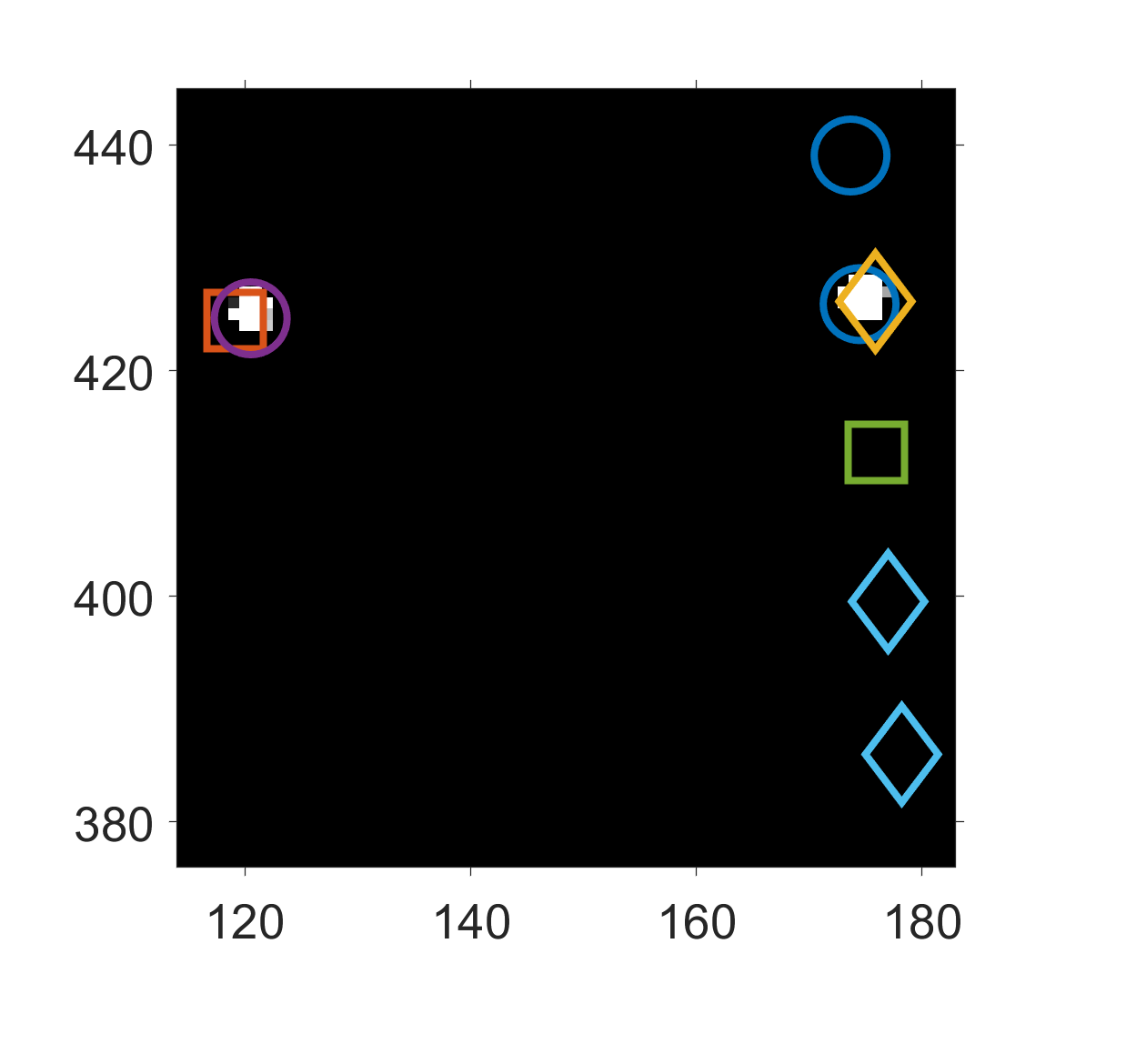}
			\caption{Camera\,3 at $T$}
			\label{fig:Cam3 at T_4808}
		\end{subfigure}
		\hspace{0.05\textwidth}
		\begin{subfigure}[b]{0.32\textwidth}
			\centering
			\includegraphics[width=\textwidth]{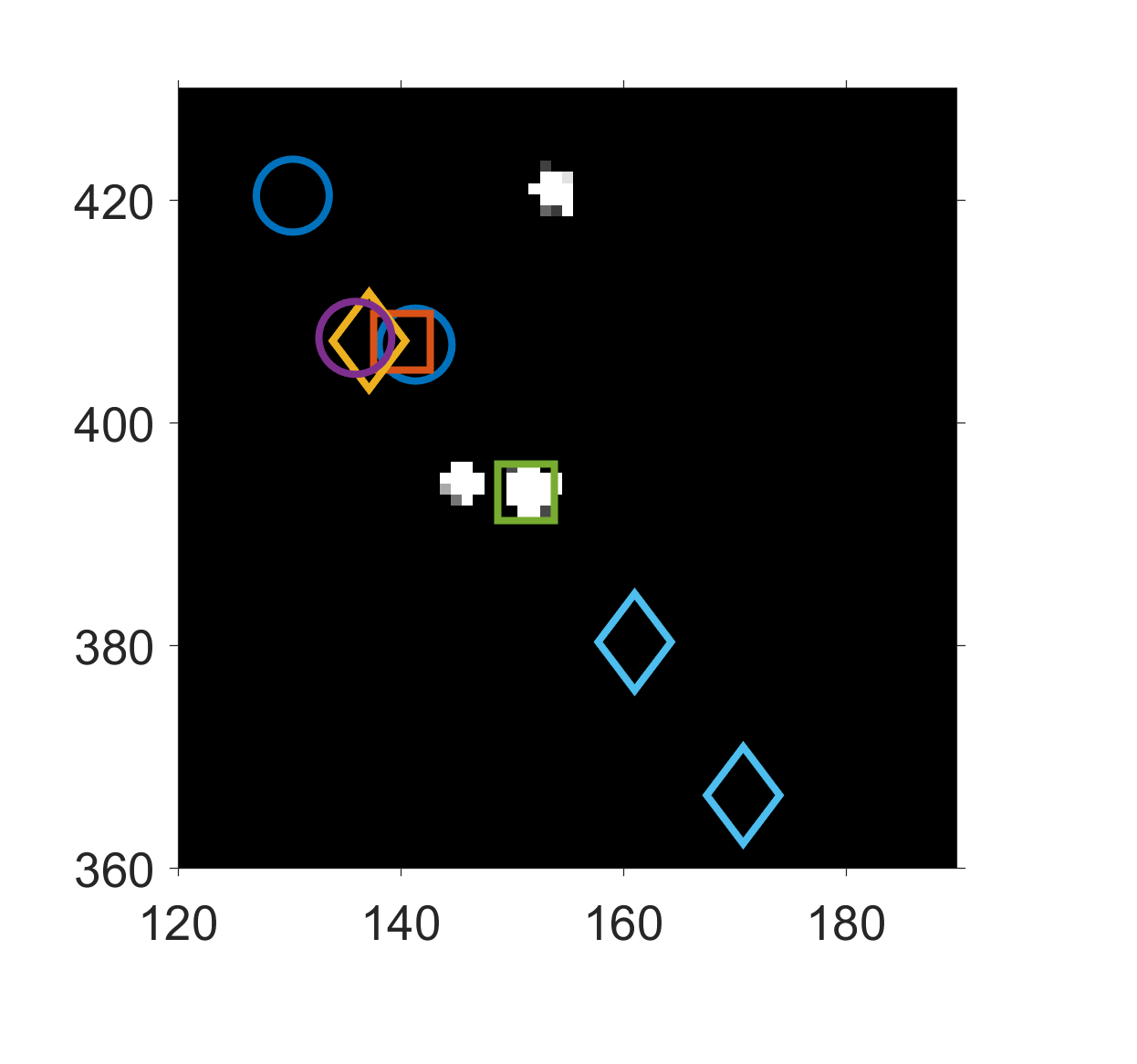}
			\caption{Camera\,1 at $T + 1\Delta t$}
			\label{fig:Cam1 at T + 1t_4808}
		\end{subfigure}
		\hfill
		\begin{subfigure}[b]{0.32\textwidth}
			\centering
			\includegraphics[width=\textwidth]{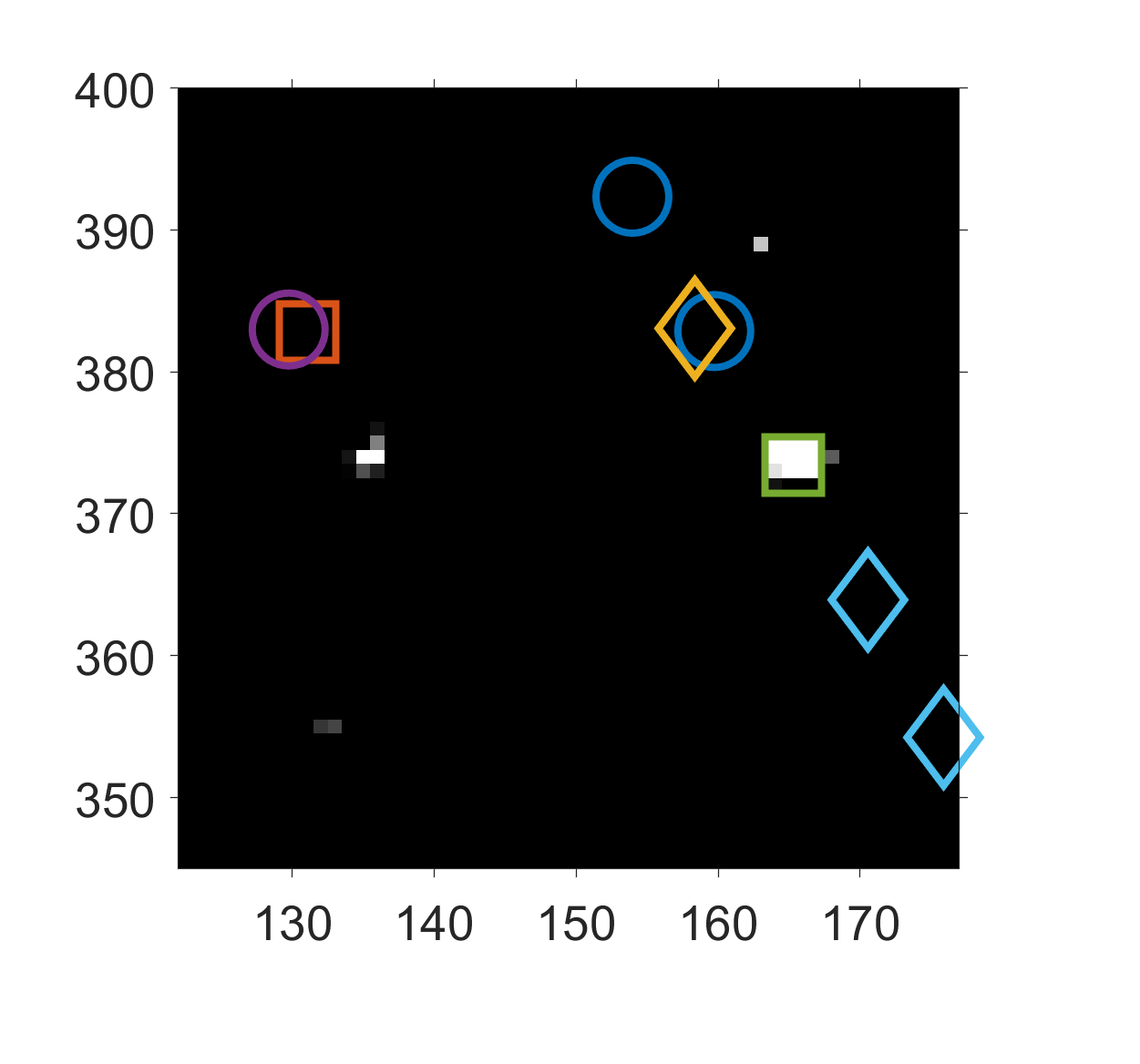}
			\caption{Camera\,2 at $T + 1\Delta t$}
			\label{fig:Cam2 at T + 1t_4808}
		\end{subfigure}
		\hfill
		\begin{subfigure}[b]{0.32\textwidth}
			\centering
			\includegraphics[width=\textwidth]{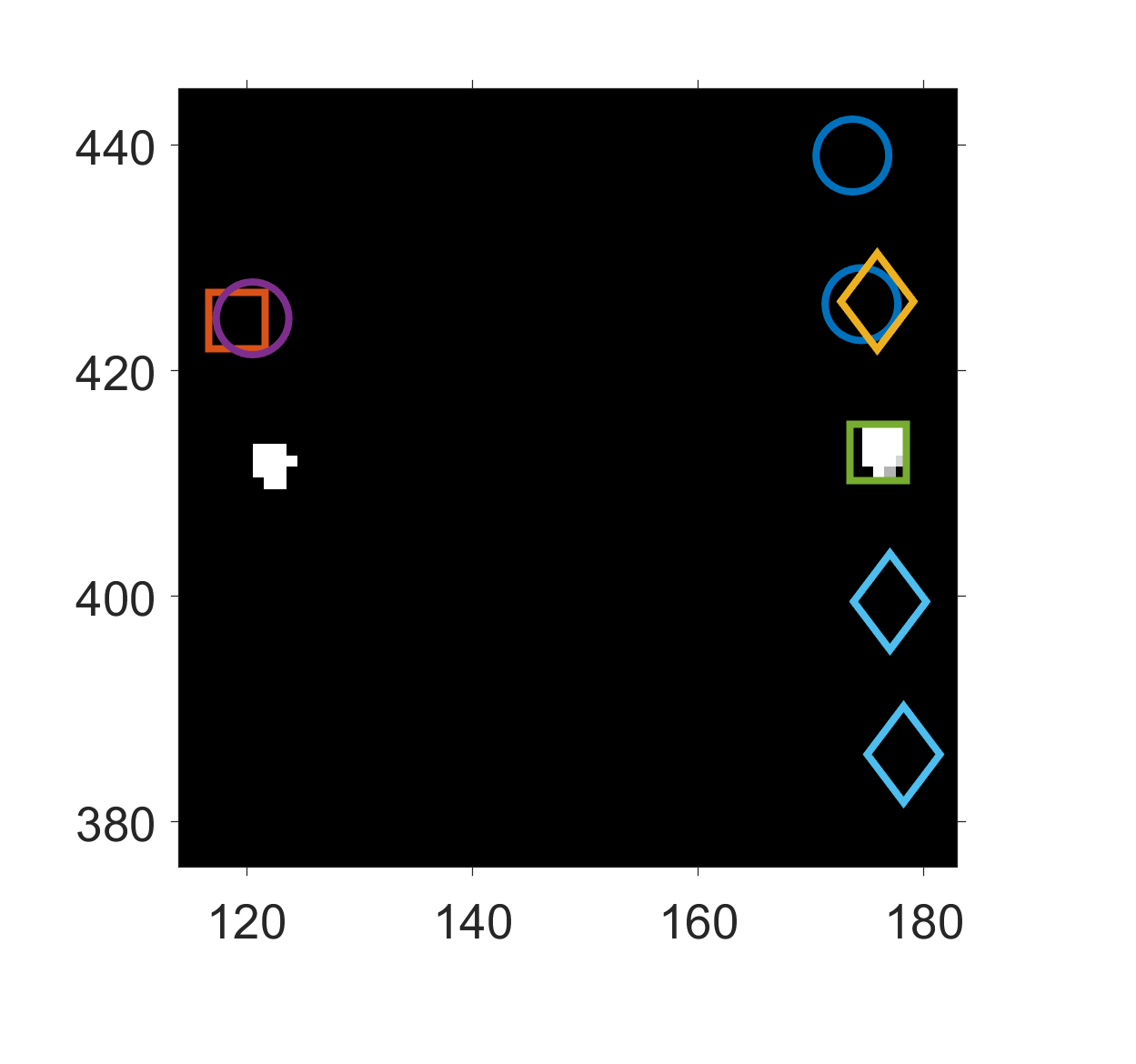}
			\caption{Camera\,3 at $T + 1\Delta t$}
			\label{fig:Cam3 at T + 1t_4808}
		\end{subfigure}
		\hfill
		\begin{subfigure}[b]{0.32\textwidth}
			\centering
			\includegraphics[width=\textwidth]{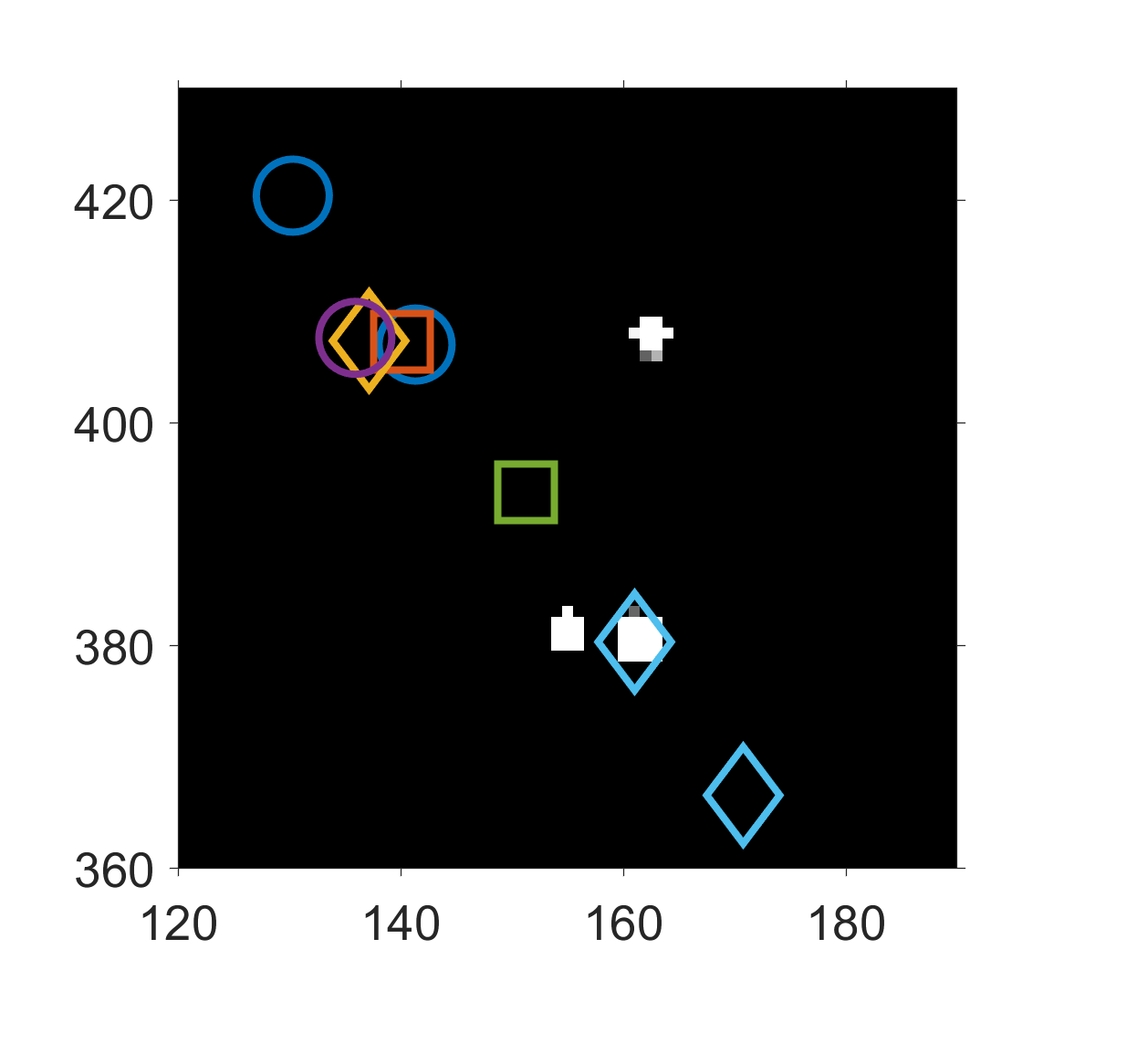}
			\caption{Camera\,1 at $T + 2\Delta t$}
			\label{fig:Cam1 at T + 2t_4808}
		\end{subfigure}
		\hfill
		\begin{subfigure}[b]{0.32\textwidth}
			\centering
			\includegraphics[width=\textwidth]{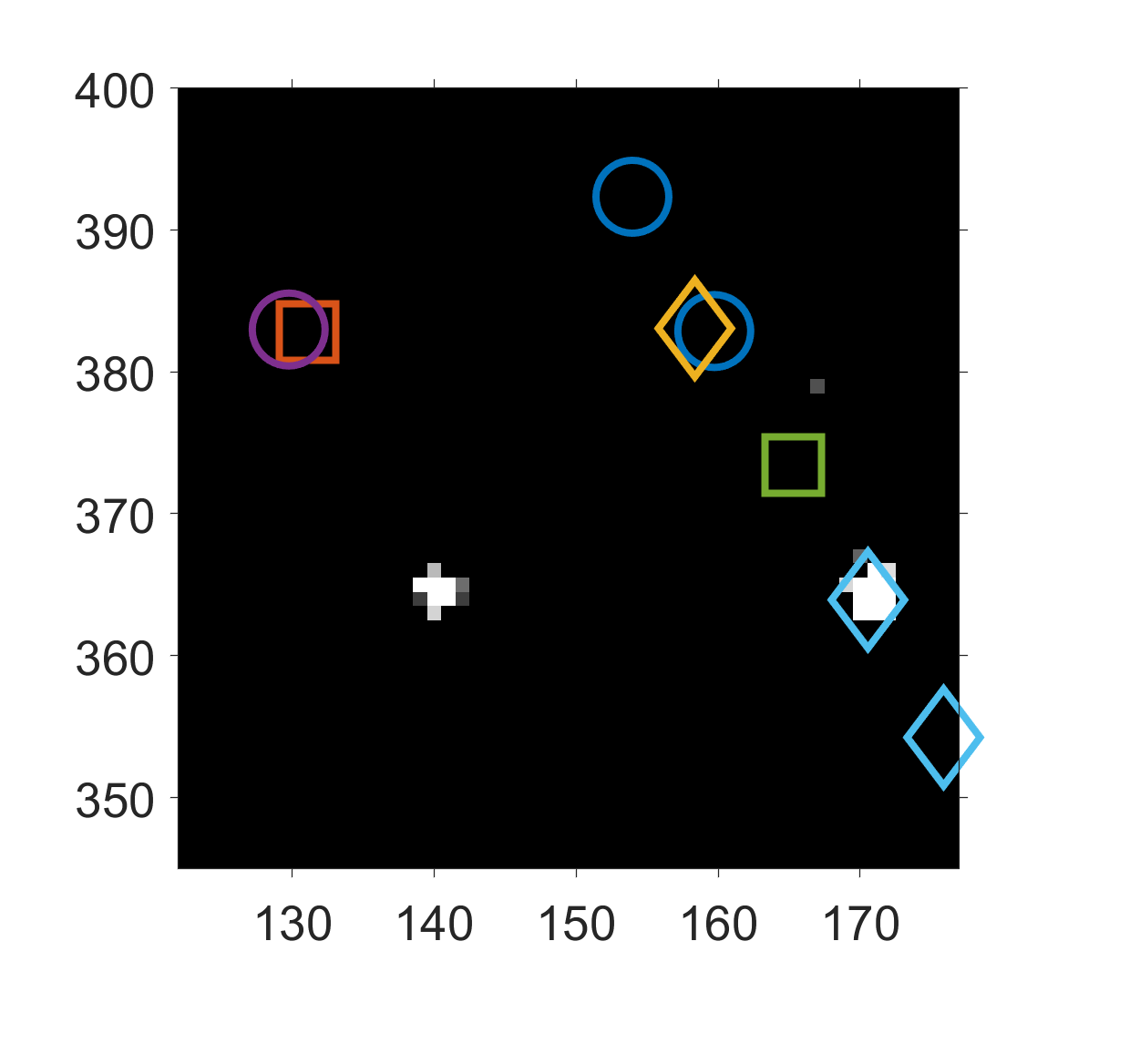}
			\caption{Camera\,2 at $T + 2\Delta t$}
			\label{fig:Cam2 at T + 2t_4808}
		\end{subfigure}
		\hfill
		\begin{subfigure}[b]{0.32\textwidth}
			\centering
			\includegraphics[width=\textwidth]{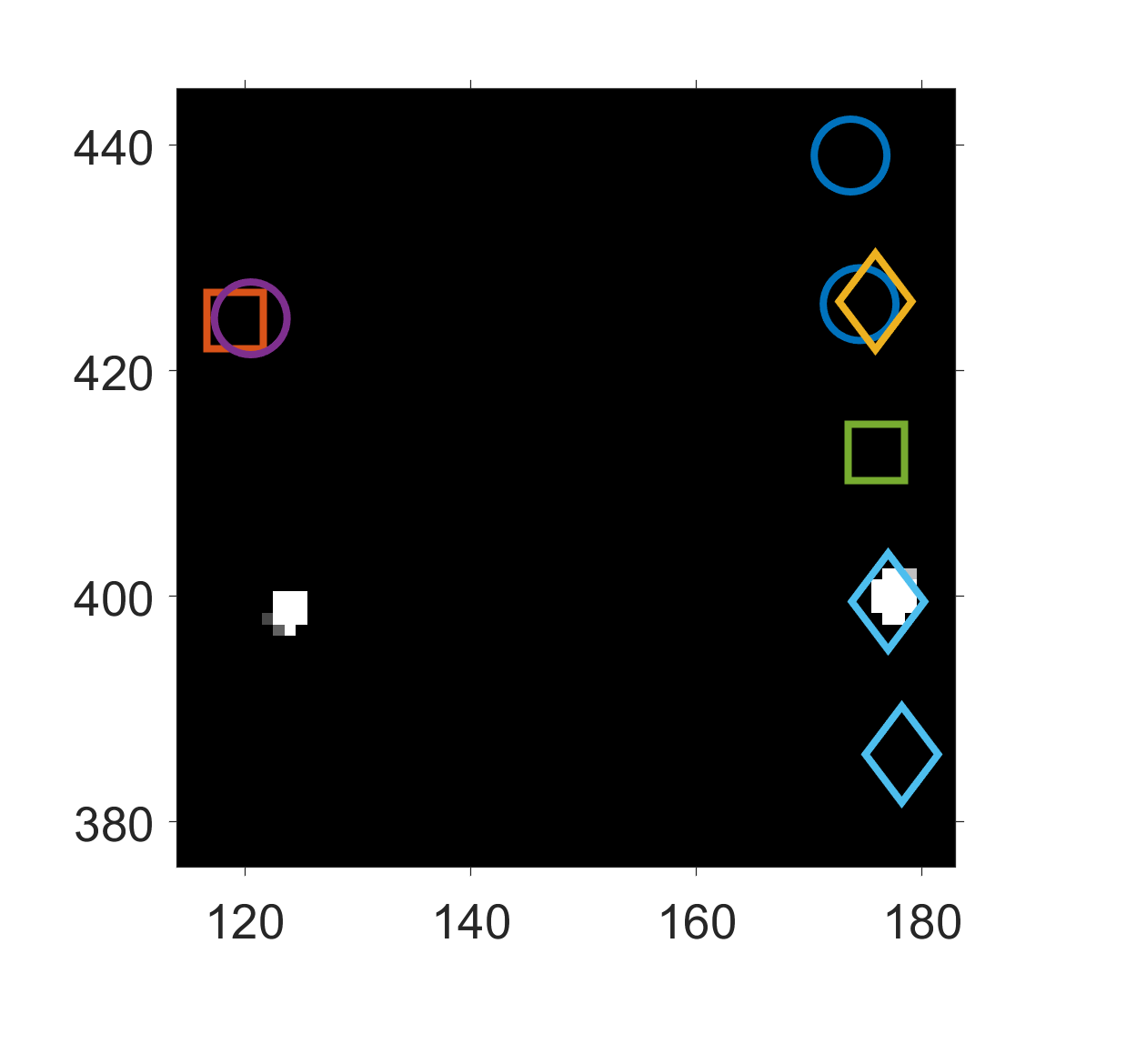}
			\caption{Camera\,3 at $T + 2\Delta t$}
			\label{fig:Cam3 at T + 2t_4808}
		\end{subfigure}
		\caption{Type II FMIS: all cameras fully resolve both particles. Panels (a)--(c), (d)--(f), (g)--(i), and (j)--(l) correspond to $T-\Delta t$, $T$, $T+\Delta t$, and $T+2\Delta t$, respectively. Colors and markers distinguish different particle trajectories. Red square and yellow rhombus indicate spurious particles generated by the stereo-matching algorithm at the frame $T$. This also results in track fragmentation until at $T+2\Delta t$ a new continuous track (cyan rhombuses) is generated.}
		\label{fig:tracking_error_demo_4808}
	\end{figure}

	Type II FMIS events occur when all cameras fully resolve both implicated particles, but the projection is too close in one or more views generates ambiguous ray intersections. Fig.~\ref{fig:tracking_error_demo_4808} illustrates a typical example. Initially (at $T - X \Delta t$, not shown), a particle pair is well-separated in all three camera views. As they approach each other, they reach a state of minimum separation (below tolerance) in Camera 1 at frame $T$. In this example, we observe that at $T$, the separation between the two particle's centroids in the image plane of Camera 1 reduces to $\Delta \approx 1.2$ pixels. This separation is significantly smaller than the average particle image diameter of $ \approx 4.4$ pixels (corresponding to $d = 17.6\,\mu\text{m}$ in physical space). 
	
	This creates a geometric configuration where multiple ray-combinations extrapolated from particle images satisfy the matching tolerance. Consequently, the stereo-matching algorithm erroneously reconstructs two additional spurious particles (red square and yellow rhombus at $T$). These spurious particles also effectively ``hijack'' the matching process, leading to the termination of the original Lagrangian track (blue circle) at $T+\Delta t$. (The algorithm is then forced to starts a new track, i.e.,  the green square, in place of the original track.)
	
	While the tracking stabilizes and resumes a continuous path by $T+2\Delta t$, the fragmenting of the trajectories and the temporary surge in particle count (from two to four) directly impact the small-scale statistics. 
	This contribute to an unphysical enhancement of the RDF at very small separation scales $r$, as the spurious particles are always placed in close proximity to the true particles.
	
	\subsection{Type III: FMIS with particle projection overlap}
\begin{figure}
		\centering
		\begin{subfigure}[b]{0.32\textwidth}
			\centering
			\includegraphics[width=\textwidth]{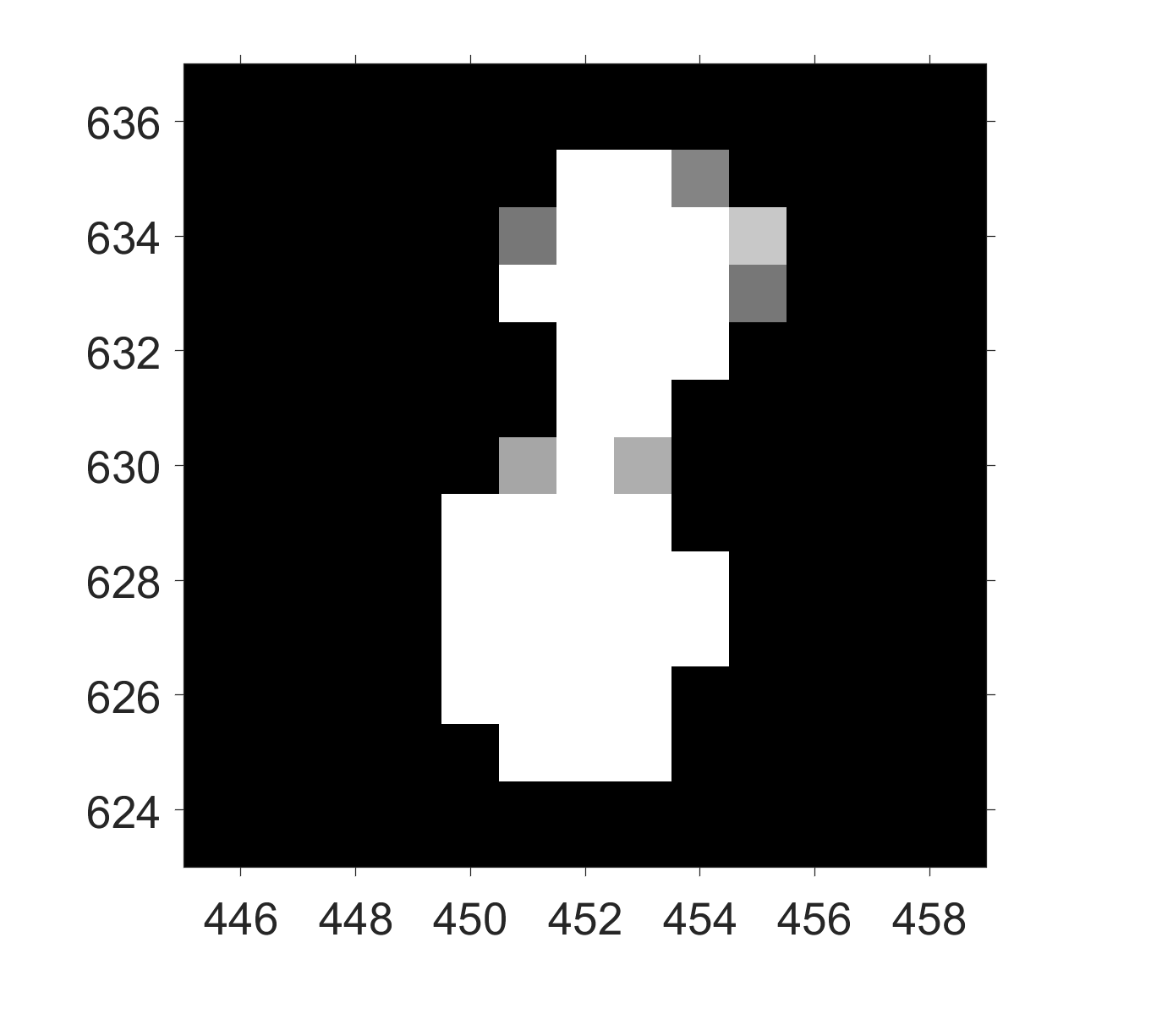}
			\caption{Camera\,1 at $T - 32\Delta t$}
			\label{fig:Camera1 at T - 32t}
		\end{subfigure}
		\hfill
		\begin{subfigure}[b]{0.32\textwidth}
			\centering
			\includegraphics[width=\textwidth]{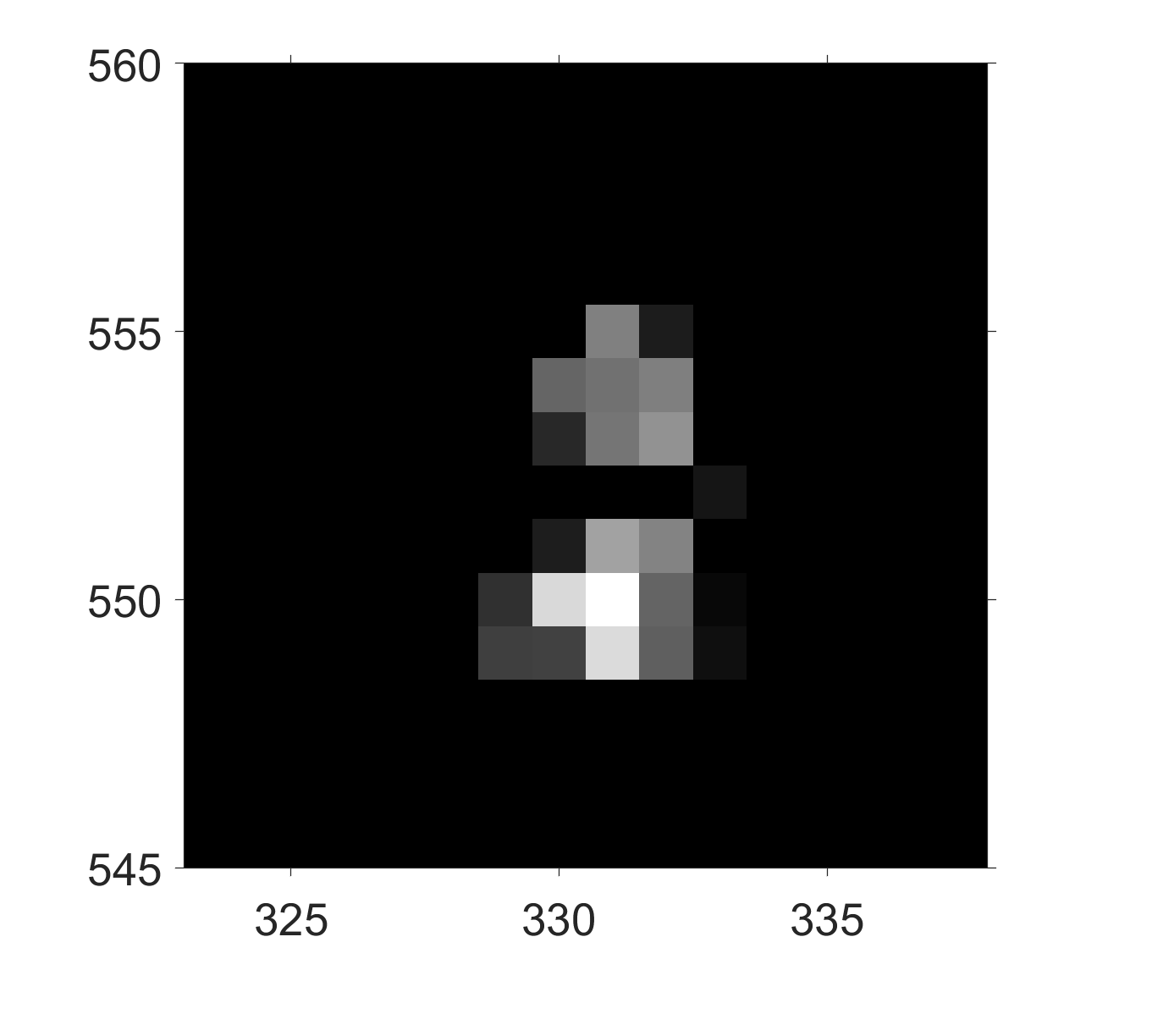}
			\caption{Camera\,2 at $T - 33\Delta t$}
			\label{fig:Camera2 at T - 33t}
		\end{subfigure}
		\hfill
		\begin{subfigure}[b]{0.32\textwidth}
			\centering
			\includegraphics[width=\textwidth]{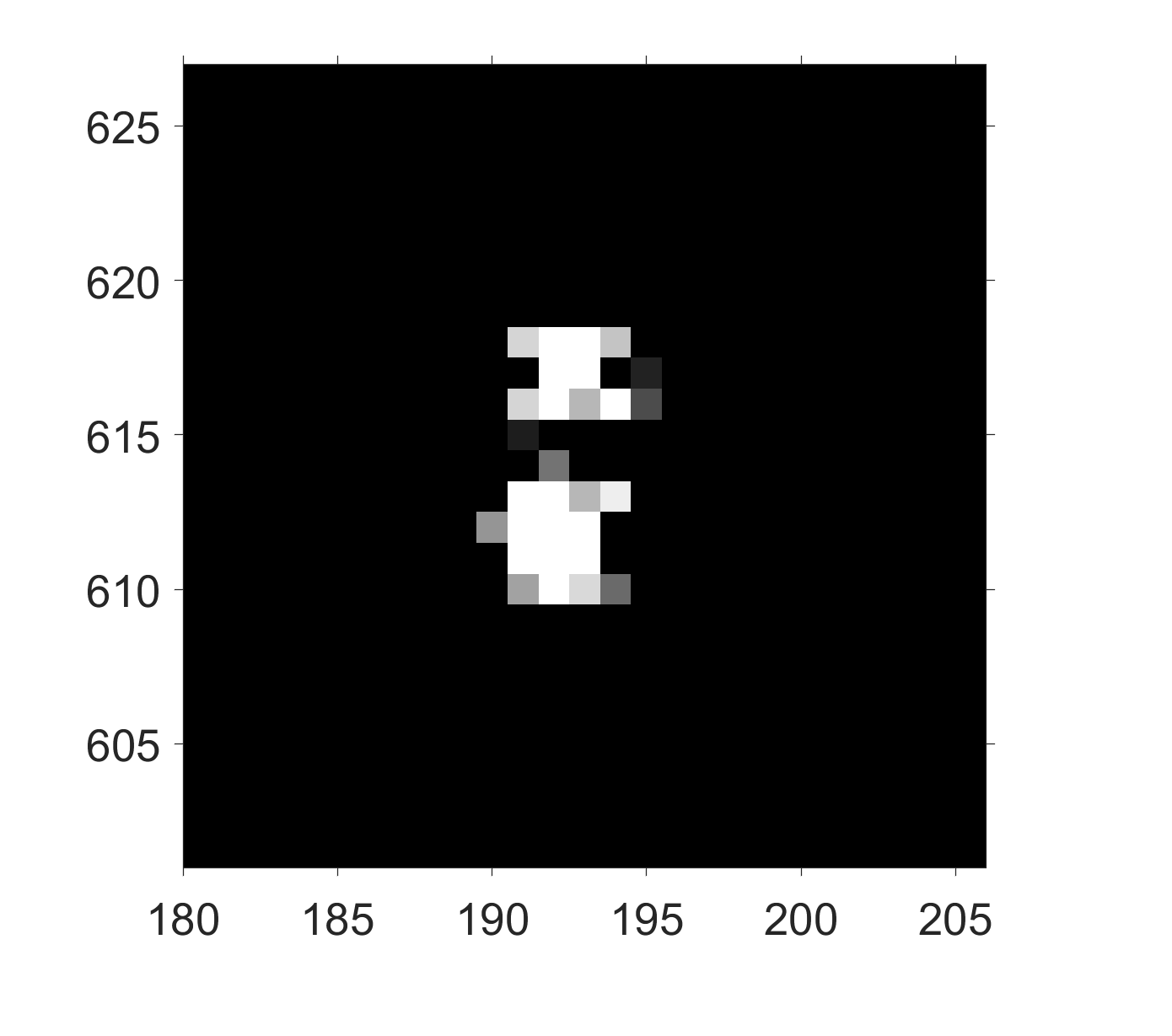}
			\caption{Camera\,3 at $T - 27\Delta t$}
			\label{fig:Camera3 at T - 27t}
		\end{subfigure}
		\hspace{0.05\textwidth}
		\begin{subfigure}[b]{0.32\textwidth}
			\centering
			\includegraphics[width=\textwidth]{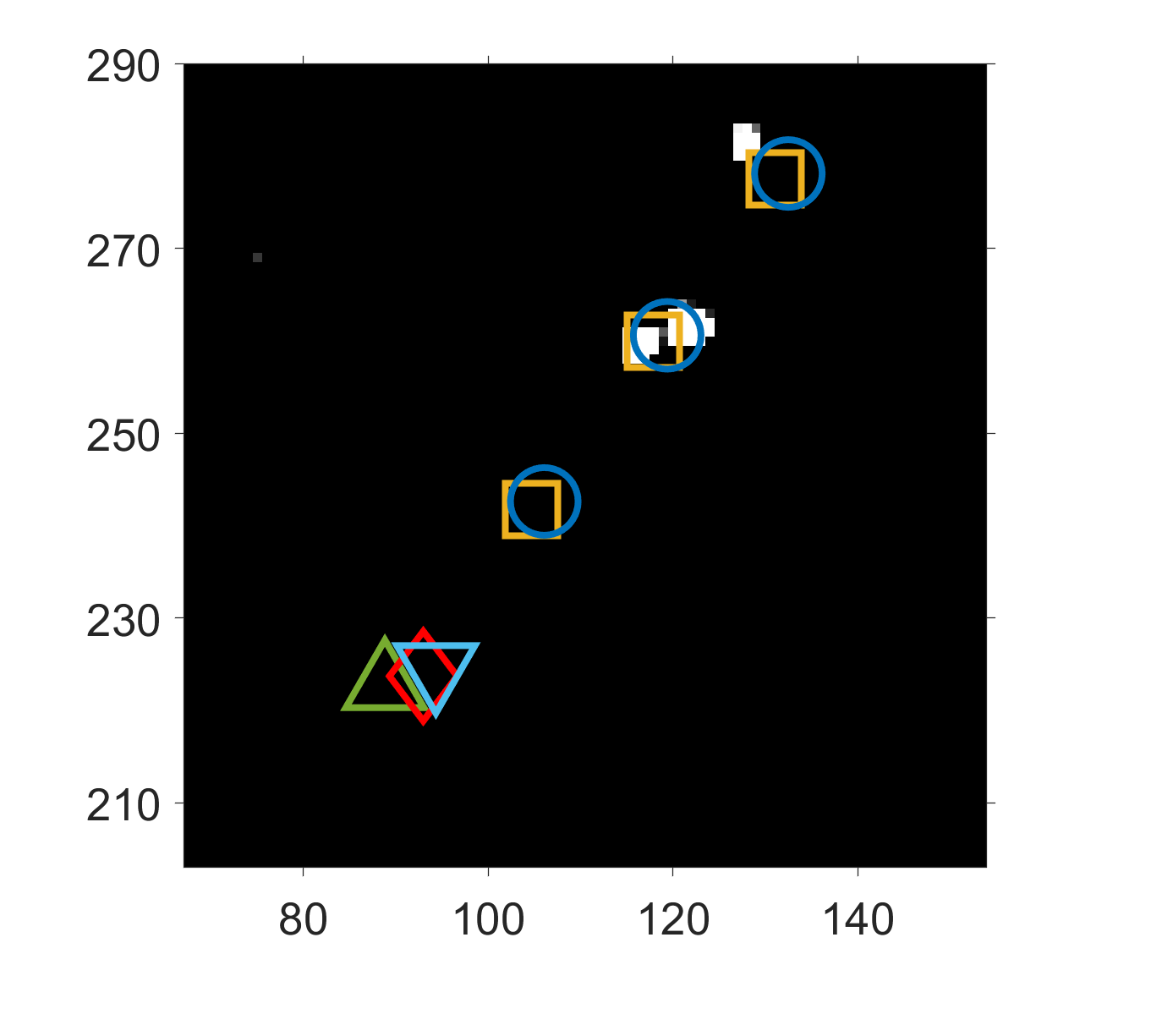}
			\caption{Camera\,1 at $T - 2\Delta t$}
			\label{fig:Camera1 at T - 2t_83742}
		\end{subfigure}
		\hfill
		\begin{subfigure}[b]{0.32\textwidth}
			\centering
			\includegraphics[width=\textwidth]{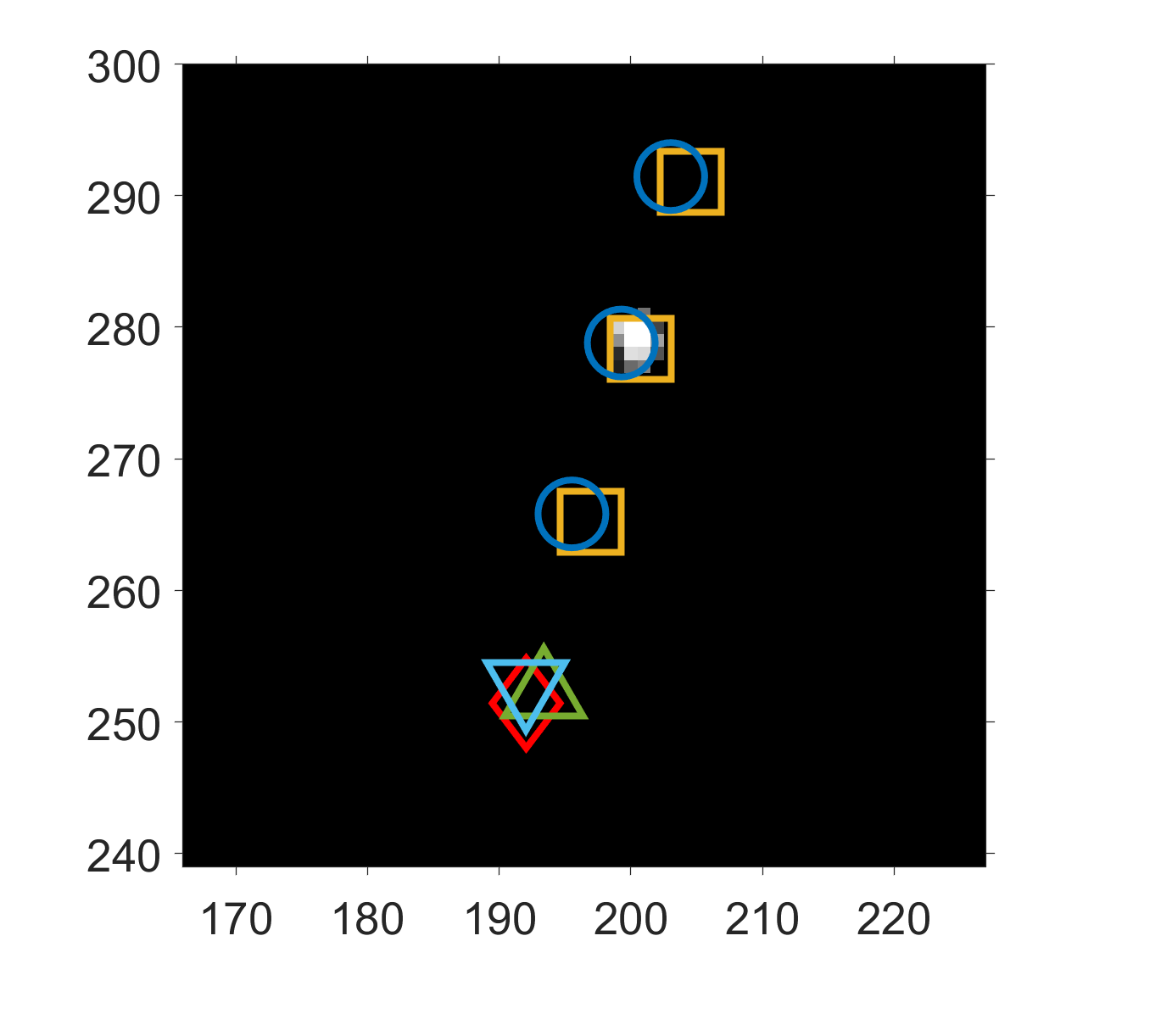}
			\caption{Camera\,2 at $T - 2\Delta t$}
			\label{fig:Camera2 at T - 2t_83742}
		\end{subfigure}
		\hfill
		\begin{subfigure}[b]{0.32\textwidth}
			\centering
			\includegraphics[width=\textwidth]{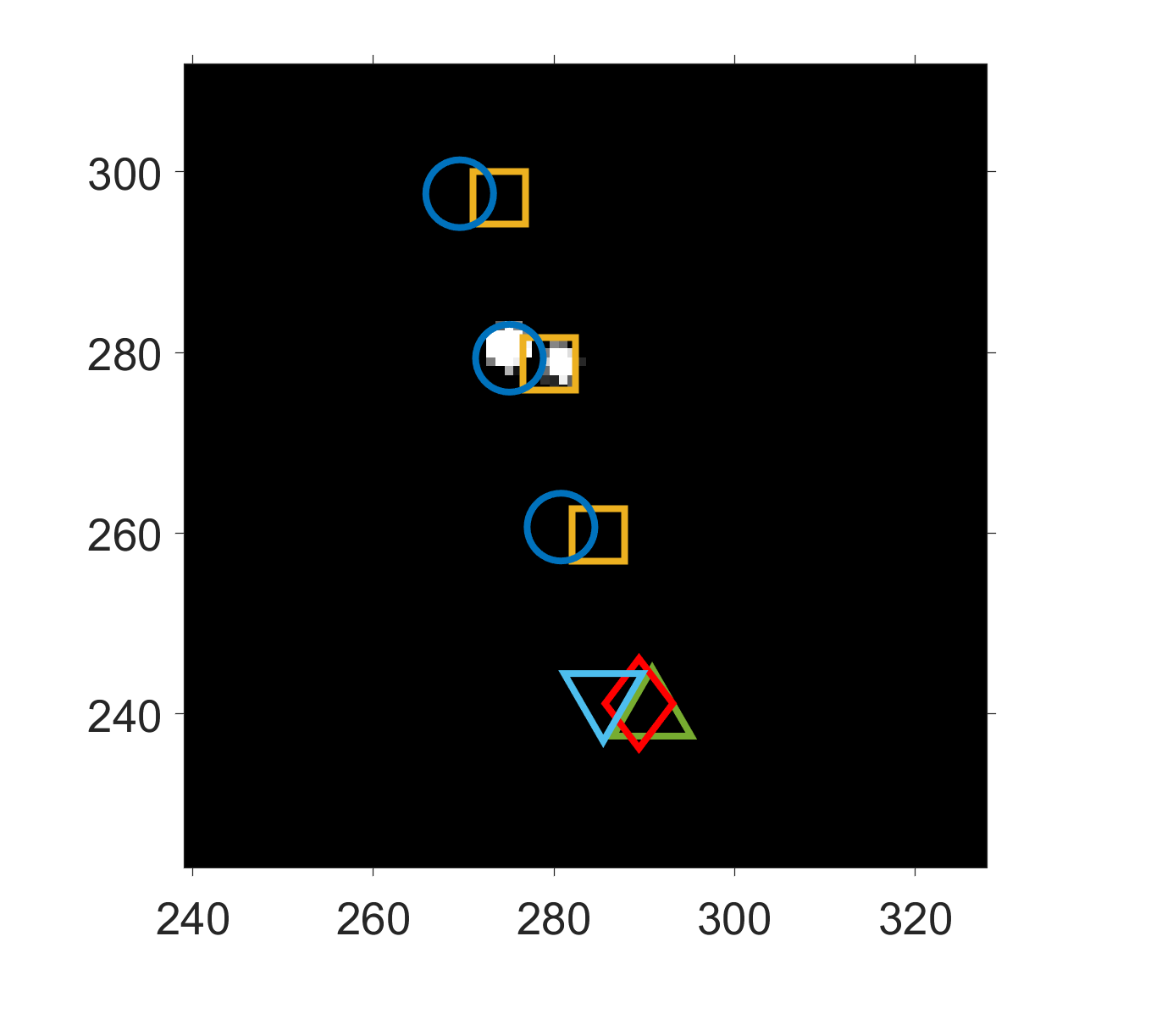}
			\caption{Camera\,3 at $T - 2\Delta t$}
			\label{fig:Camera3 at T - 2t_83742}
		\end{subfigure}
		\hspace{0.05\textwidth}
		\begin{subfigure}[b]{0.32\textwidth}
			\centering
			\includegraphics[width=\textwidth]{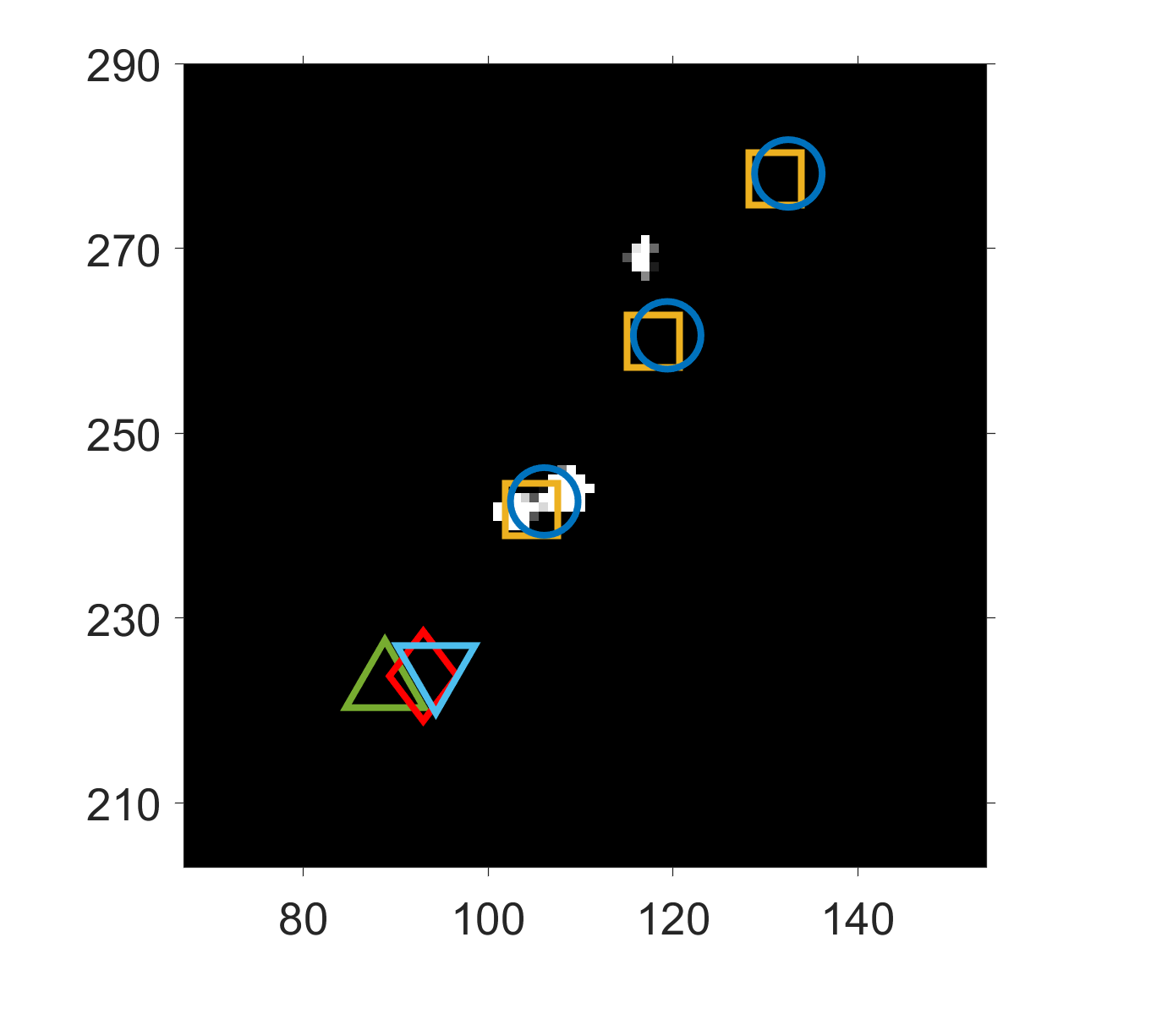}
			\caption{Camera\,1 at $T - 1\Delta t$}
			\label{fig:Camera1 at T - 1t_83742}
		\end{subfigure}
		\hfill
		\begin{subfigure}[b]{0.32\textwidth}
			\centering
			\includegraphics[width=\textwidth]{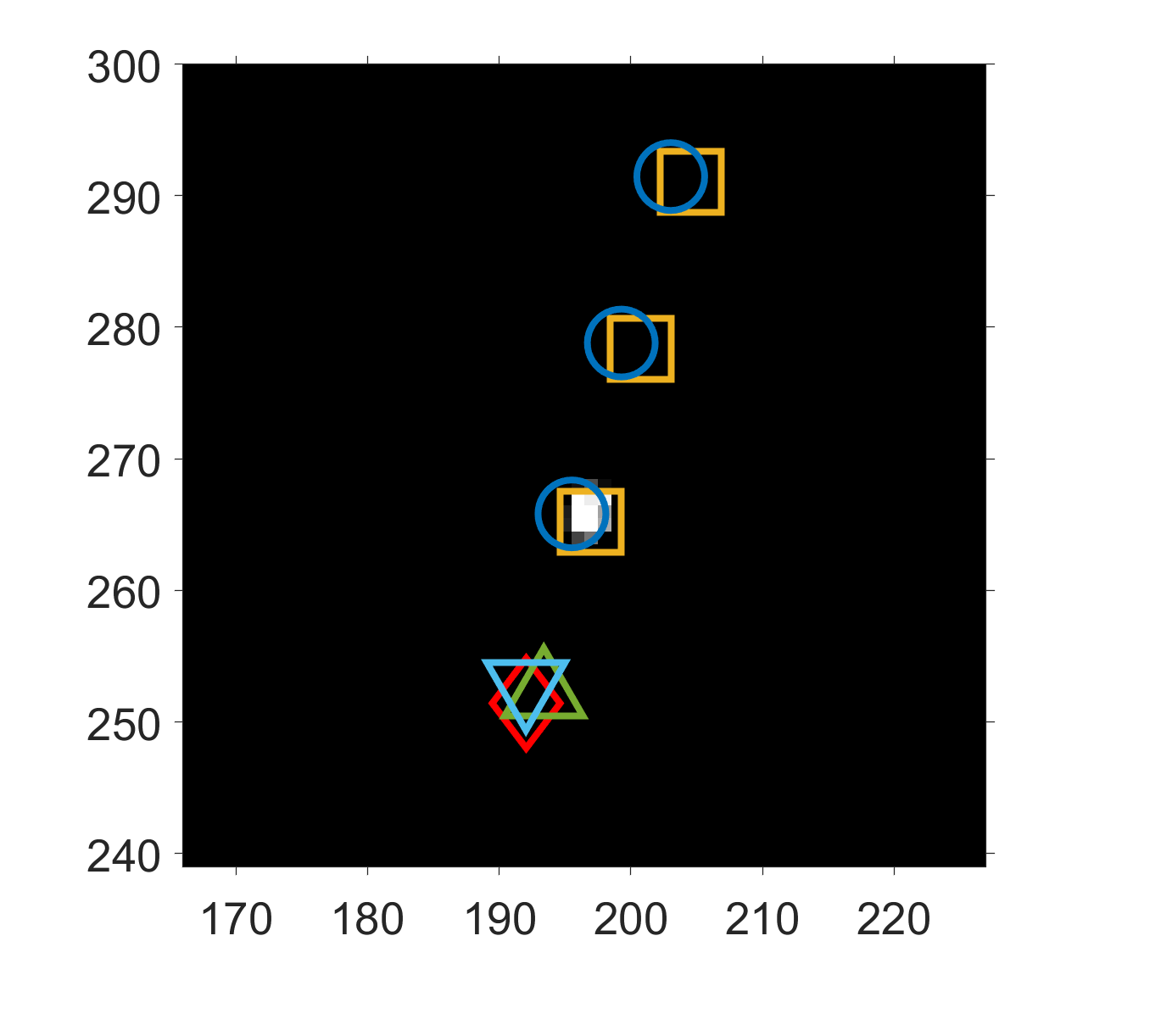}
			\caption{Camera\,2 at $T - 1\Delta t$}
			\label{fig:Camera2 at T - 1t_83742}
		\end{subfigure}
		\hfill
		\begin{subfigure}[b]{0.32\textwidth}
			\centering
			\includegraphics[width=\textwidth]{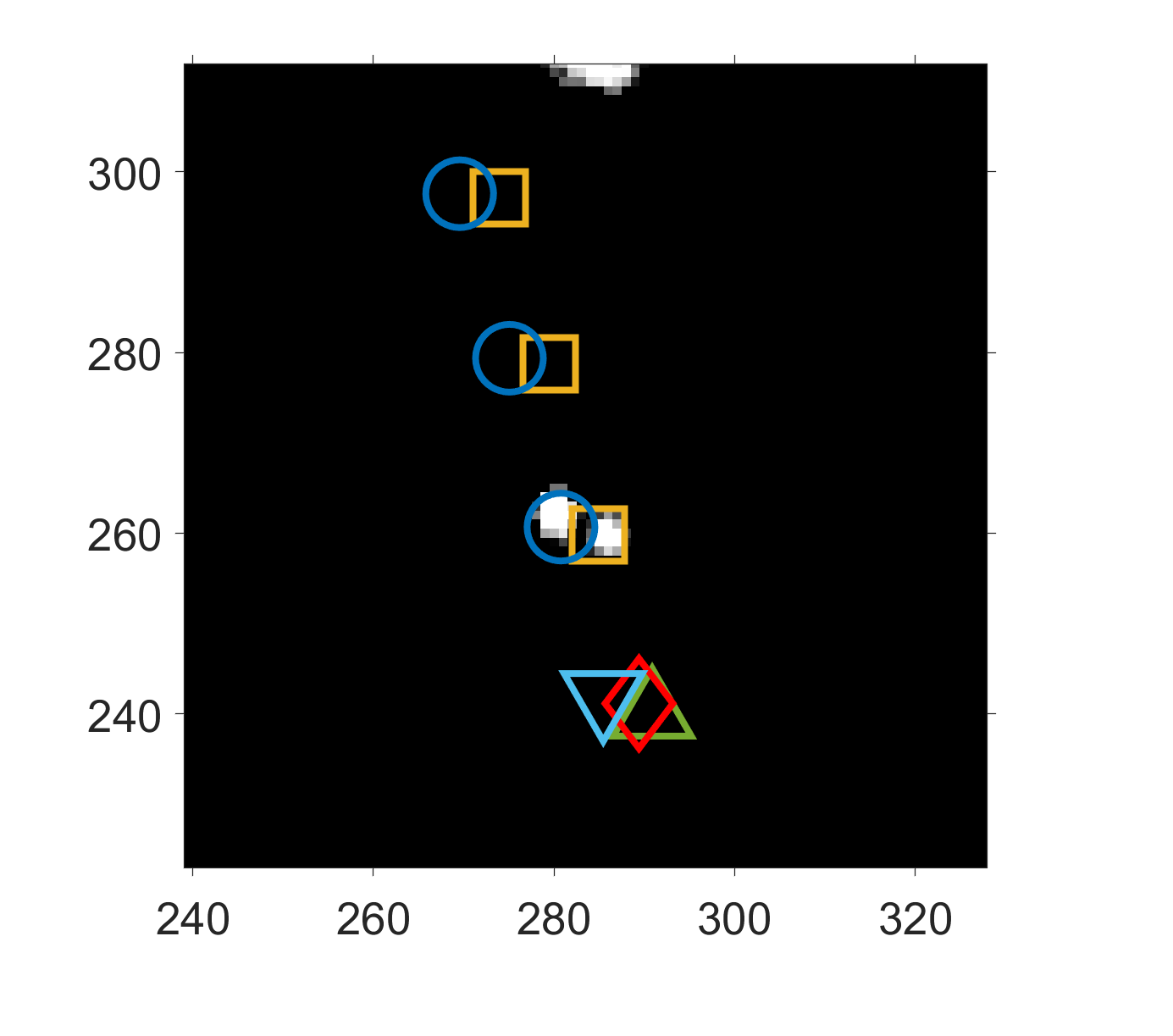}
			\caption{Camera\,3 at $T - 1\Delta t$}
			\label{fig:Camera3 at T - 1t_83742}
		\end{subfigure}
		\hfill
		\begin{subfigure}[b]{0.32\textwidth}
			\centering
			\includegraphics[width=\textwidth]{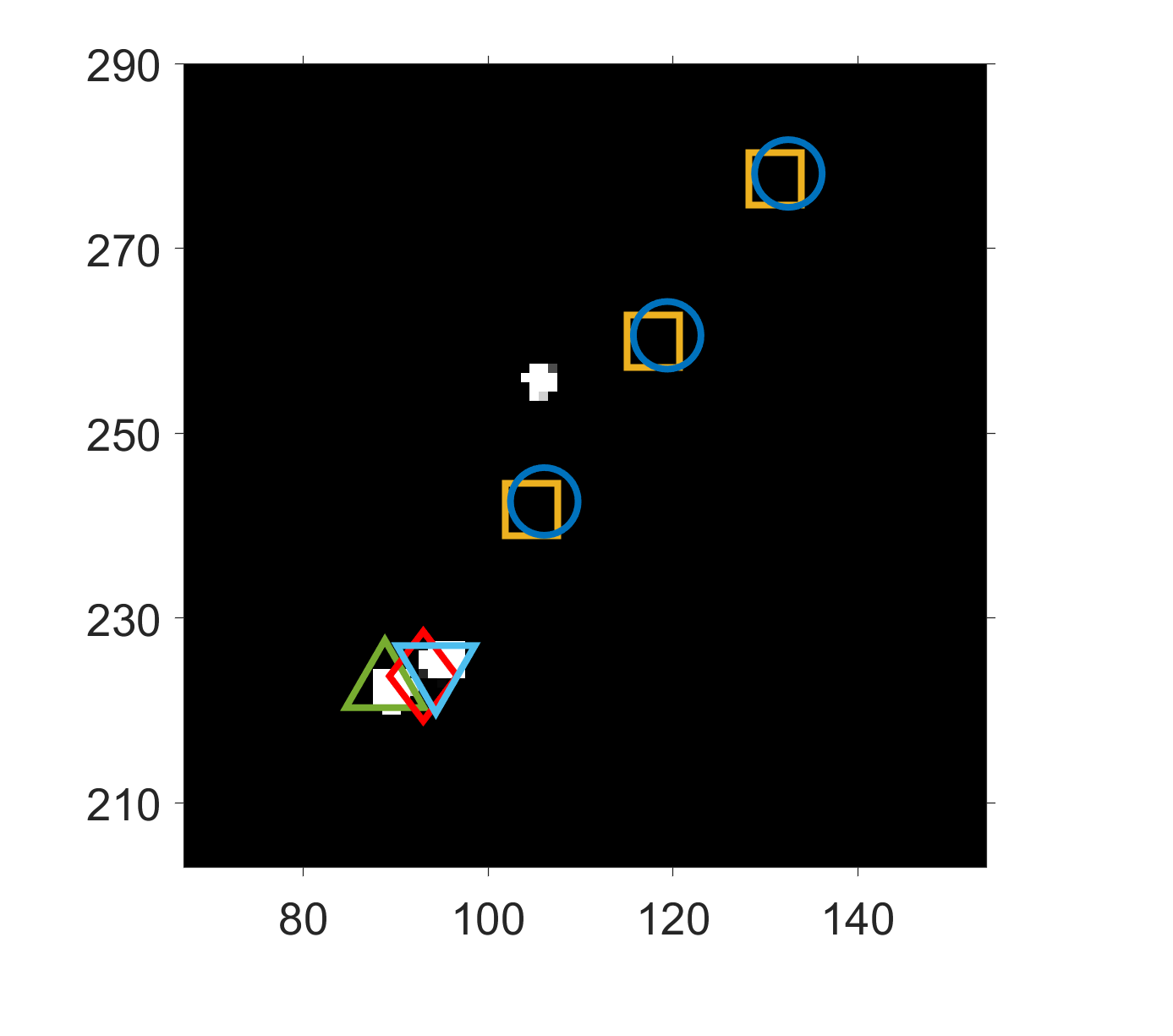}
			\caption{Camera\,1 at $T$}
			\label{fig:Camera0 at T_83742}
		\end{subfigure}
		\hfill
		\begin{subfigure}[b]{0.32\textwidth}
			\centering
			\includegraphics[width=\textwidth]{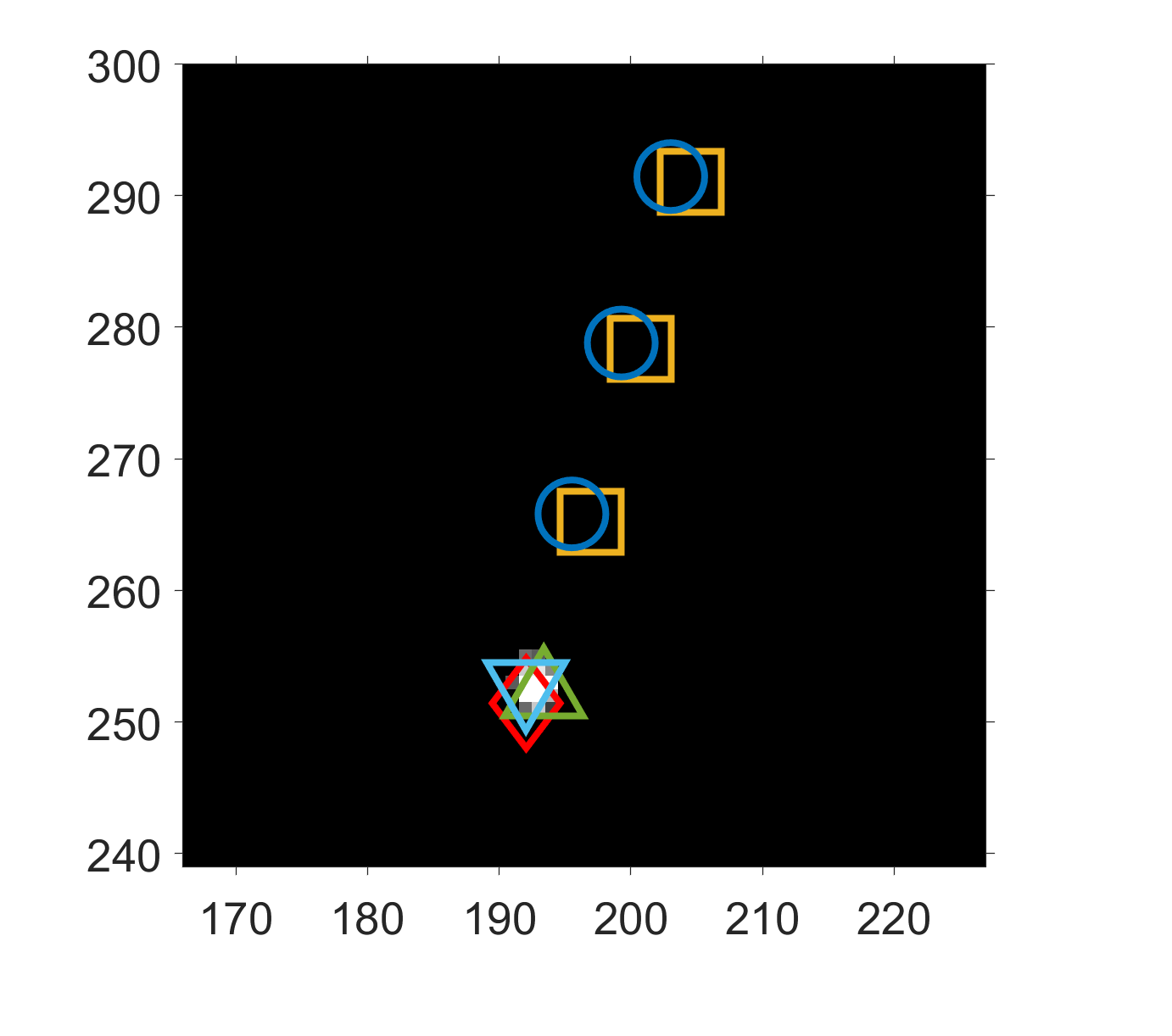}
			\caption{Camera\,2 at $T$}
			\label{fig:Camera2 at T_83742}
		\end{subfigure}
		\hfill
		\begin{subfigure}[b]{0.32\textwidth}
			\centering
			\includegraphics[width=\textwidth]{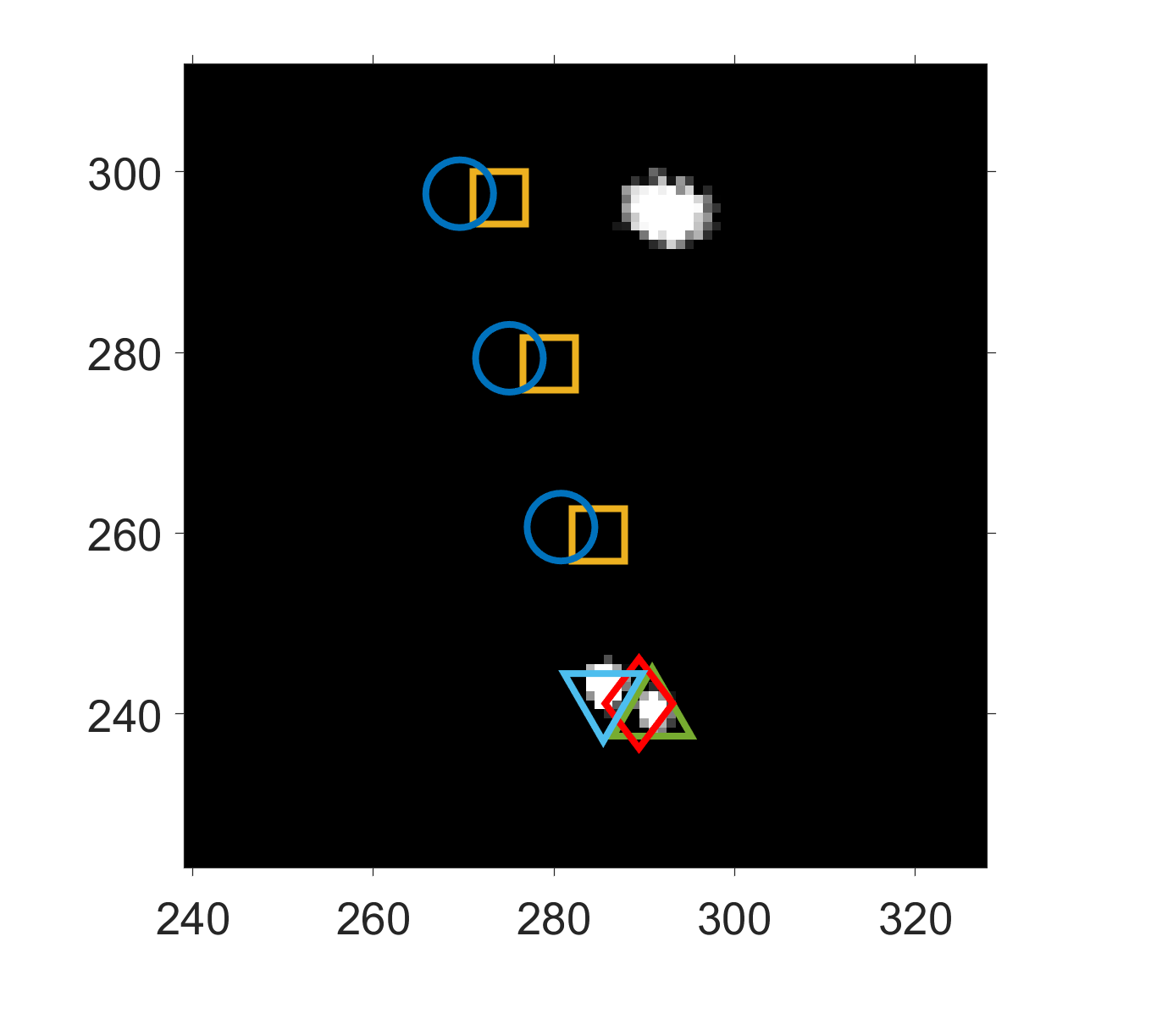}
			\caption{Camera\,3 at $T$}
			\label{fig:Camera3 at T_83742}
		\end{subfigure}
		\caption{Type III FMIS: Partial overlap of particles in a single camera. Different colors and markers indicate distinct detected particle tracks. At the initial time, two closely spaced particles are observed in all cameras. At $T-2\Delta t$ and $T-\Delta t$, even though the particle images in Camera\,2 increasingly overlap, the system successfully resolves and tracks the two physical trajectories. At $T$, the two particle images are too close such that multiple duplicated matchings occurs resulting in the appearance of the spurious particle (red diamond) at $T$. This illustrates how FMIS inflates number of closely separated particles.}
		\label{fig:tracking_error_demo_83742}
	\end{figure}

	Type III FMIS arises from partial particle projection overlap in a single camera. Fig.~\ref{fig:tracking_error_demo_83742} illustrates an FMIS event caused by the partial overlap of two particles in Camera 2. Initially (at $T-27\Delta t$ or earlier), both particles are clearly resolved as distinct intensity peaks in all cameras, as shown in the zoomed-in views in Fig.~\ref{fig:tracking_error_demo_83742}(a)--(c).  
	
	At $T-2\Delta t$ and $T-1\Delta t$, although the particle images begin to overlap in Camera 2, the stereo-matching algorithm successfully maintains two distinct trajectories (blue circle and yellow square). However, at frame $T$, multiple ray-combinations satisfy the matching tolerance, the system erroneously generates a spurious third particle (red triangle) that mathematically satisfies the the matching tolerance. 
    
	The occurrence of this FMIS artifact not only generates spurious particles but also artificially fragments Lagrangian trajectories. 
	
\end{appendices}

\bibliography{sn-bibliography}

\end{document}